\documentclass[english,floatfix,11pt,superscriptaddress,aps,prd,reprint,showkeys,nofootinbib]{revtex4-2}
\usepackage{amsmath}
\usepackage{amssymb}
\usepackage{amsbsy}
\usepackage{amsfonts}
\usepackage{amsopn}
\usepackage{amstext}
\usepackage{graphicx}
\usepackage[english]{babel}
\usepackage{color}
\usepackage{slashed}
\usepackage{esint}
\usepackage{float}
\usepackage{units}
\usepackage{textcomp}
\usepackage{wasysym}
\usepackage{hyperref}
\usepackage{orcidlink}

\graphicspath{{./}}

\newcommand{\lmm}{\lambda_{\rm MM}}
\newcommand{\Qeff}{Q_{\rm eff}^2}

\begin{document}

\title{ModMax black hole surrounded by perfect-fluid dark matter in Lorentz-violating Kalb--Ramond gravity}

\author{Fernando M. Belchior\orcidlink{0009-0006-8675-7849}}
\email{fernandobelcks7@gmail.com}
\affiliation{Departamento de Física, Universidade Federal da Paraíba, Centro de Ciências Exatas e da Natureza, 58051-970, João Pessoa, Paraíba, Brazil}

\author{Faizuddin Ahmed\orcidlink{0000-0003-2196-9622}}
\email{faizuddinahmed15@gmail.com}
\affiliation{Department of Physics, The Assam Royal Global University, Guwahati-781035, Assam, India}

\author{Edilberto O. Silva\orcidlink{0000-0002-0297-5747}}
\email{edilberto.silva@ufma.br (Corresponding author)}
\affiliation{Programa de P\'os-Gradua\c c\~ao em F\'{\i}sica \& Coordena\c c\~ao do Curso de F\'isica -- Bacharelado, Universidade Federal do Maranh\~{a}o, 65085-580 S\~{a}o Lu\'is, Maranh\~{a}o, Brazil}

\begin{abstract}
We investigate a ModMax black hole surrounded by perfect-fluid dark matter within the framework of Lorentz-violating Kalb-Ramond gravity. The model combines three physically distinct contributions: nonlinear electrodynamic corrections from the ModMax sector, Lorentz-symmetry-breaking effects induced by the background Kalb-Ramond field, and environmental modifications associated with the surrounding dark matter fluid. We obtain the corresponding static and spherically symmetric black hole geometry and analyze how the charge, ModMax parameter, Kalb-Ramond coupling, and dark matter parameter affect the horizon structure and thermodynamic behavior. In particular, we study the Hawking temperature, entropy, heat capacity, and Helmholtz free energy, showing that the combined effects of nonlinear electrodynamics and Lorentz violation may shift the extremal configuration, modify the thermal stability regions, and generate nontrivial phase behavior. The perfect fluid dark matter contribution introduces an additional logarithmic correction to the geometry, becoming especially relevant at intermediate radial scales. Our results indicate that ModMax electrodynamics can effectively screen the electric sector, while the Kalb-Ramond parameter amplifies the geometric deformation and changes the thermodynamic response of the system. These features suggest that black holes in Lorentz-violating backgrounds surrounded by dark matter provide a useful arena for probing deviations from standard charged black-hole thermodynamics.

\end{abstract}


\maketitle

\section{Introduction}

Black holes occupy a central position in the gravitational context and high-energy astrophysics. Since Schwarzschild obtained the first exact vacuum solution of Einstein's field equations \cite{Schwarzschild}, and Reissner and Nordstr\"om introduced the charged generalization \cite{Reissner,Nordstrom}, black holes have become standard backgrounds for testing the interplay between geometry, matter fields, and fundamental interactions. The classical analysis of horizon structure and relativistic motion was substantially developed by Carter and by the later formulation of black-hole mechanics due to Bardeen, Carter, and Hawking \cite{Carter:1968rr,Bardeen:1973gs}. The thermodynamic interpretation initiated by Bekenstein and Hawking, and further developed by Gibbons and Hawking, established the deep connection between horizon area, temperature, entropy, and quantum emission \cite{Bekenstein:1973ur,Hawking:1974sw,Gibbons:1977mu}. These developments are part of the broader causal and geometrical framework summarized by Hawking and Ellis \cite{Hawking:1973uf}, and they also provide the basis for modern Noether-charge derivations of black-hole entropy by Wald and by Iyer and Wald \cite{Wald:1993nt,Iyer:1994ys}. On the observational side, the LIGO Scientific and Virgo collaborations detected gravitational waves from binary black-hole mergers \cite{LIGOScientific:2016aoc}, while the Event Horizon Telescope collaboration obtained horizon-scale images of M87$^*$ and Sagittarius A$^*$ \cite{EventHorizonTelescope:2019dse,EventHorizonTelescope:2019ggy,EventHorizonTelescope:2022wkp}. These observations have transformed black-hole physics into a precision field in which modified gravity, nonlinear fields, and environmental matter distributions can be constrained through strong-field phenomena.

Perturbations, scattering, absorption, and shadows provide complementary probes of black-hole geometries. The theory of black-hole perturbations began with the Regge--Wheeler and Zerilli equations \cite{Regge:1957td,Zerilli:1970se} and was generalized to rotating backgrounds through Teukolsky's formalism \cite{Teukolsky:1973ha}. The quasinormal-mode program, reviewed for instance by Kokkotas and Schmidt, Berti, Cardoso, and Starinets, and Konoplya and Zhidenko, connects the damped response of compact objects with the parameters of the underlying spacetime \cite{Kokkotas:1999bd,Berti:2009kk,Konoplya:2011qq}. In parallel, the absorption and greybody spectra of black holes have been investigated since the works of Page and Sanchez \cite{Page:1976df,Sanchez:1976xm}. Low-frequency and string-inspired greybody factors were further studied by Maldacena and Strominger, Klebanov, and Das, Gibbons, and Mathur \cite{Maldacena:1996ix,Klebanov:1997cx,Das:1996we}. Rigorous greybody-factor bounds were later developed by Visser and collaborators and applied to dirty or modified black-hole backgrounds \cite{Visser:1998ke,Boonserm:2021owk}. The optical appearance of black holes also has a long history, from Synge's calculation of the Schwarzschild shadow and the Kerr-ray-tracing analysis of Bardeen to recent reviews by Cunha and Herdeiro and by Perlick and Tsupko \cite{Synge:1966okc,Bardeen:1973tla,Cunha:2018acu,Perlick:2021aok}. These tools are essential for distinguishing modifications of the near-horizon geometry from changes in the effective potential outside the horizon.

Among the possible extensions of general relativity, Lorentz-symmetry-violating gravity has attracted considerable attention. Spontaneous Lorentz violation can arise naturally in field theories with tensor vacuum expectation values, including scenarios inspired by string theory \cite{Kostelecky:1988zi,Kostelecky:2003fs,Bailey:2006fd,Bluhm:2004ep}. Antisymmetric tensor fields of Kalb--Ramond type were originally introduced in string-motivated contexts by Kalb and Ramond \cite{Kalb:1974yc}, and they offer a natural setting in which a background two-form can break local Lorentz symmetry spontaneously. In the gravitational sector, Altschul, Bailey, and Kosteleck\'y analyzed Lorentz violation with an antisymmetric tensor \cite{Altschul:2009ae}, while Aashish and Panda studied quantum aspects of antisymmetric tensor fields with spontaneous Lorentz violation \cite{Aashish:2019ykb}. More recently, Lessa, Silva, Maluf, and Almeida obtained black-hole solutions with a background Kalb--Ramond field \cite{Maluf:2018jwc}; Yang, Chen, Duan, and Zhao derived static and spherically symmetric black holes in this framework \cite{Yang:2023wtu}; and Duan, Zhao, and Yang extended the construction to electrically charged configurations \cite{Duan:2023gng}. Static neutral and optical properties of related Kalb--Ramond black holes have also been explored in recent works \cite{Liu:2024oas,Junior:2024ety,AhmedFathiSilva2026ChargedKRPFDMPublished}. These studies show that the Kalb--Ramond parameter can affect horizons, photon spheres, shadows, orbital motion, and thermodynamic response, making it a useful phenomenological probe of Lorentz-violating gravity.

Nonlinear electrodynamics provides another important route beyond the standard Einstein--Maxwell system. Born and Infeld introduced the first celebrated nonlinear theory of electrodynamics to regularize the self-energy of point charges \cite{Born:1934gh}, and later studies by Gibbons and Rasheed clarified the role of electric-magnetic duality in nonlinear electromagnetic theories \cite{Gibbons:1995cv}. ModMax electrodynamics, introduced by Bandos, Lechner, Sorokin, and Townsend, is especially interesting because it is a nonlinear deformation of Maxwell theory that preserves conformal invariance and electromagnetic duality \cite{Bandos:2020jsw}. When coupled to gravity, ModMax-type fields lead to charged black-hole solutions in which the electromagnetic contribution is effectively screened by the nonlinear parameter, thereby modifying the horizon structure, thermodynamics, shadows, and wave propagation. Dyonic ModMax black holes and their shadows, lensing, quasinormal modes, greybody bounds, and neutrino propagation were studied by Pantig, Mastrototaro, Lambiase, and \"Ovg\"un \cite{Pantig:2022gih}. Related Einstein--ModMax and generalized ModMax black-hole configurations have been analyzed by Barrientos, Cisterna, Kubiz\v{n}\'ak, and Oliva and by Kruglov \cite{Barrientos:2024umq,Kruglov:2022zsk}. Scattering and absorption by ModMax black holes have also been studied recently by Campos and collaborators \cite{Campos:2025modmax}, and ModMax black holes surrounded by perfect fluid dark matter have been considered in AdS thermodynamics by Shaukat and collaborators \cite{Shaukat:2025mvk}. These results motivate the use of ModMax electrodynamics as a controlled nonlinear electromagnetic sector in compact-object physics.

A further motivation for the present study comes from the fact that astrophysical black holes are not isolated objects. They may be embedded in accretion flows, matter distributions, dark matter halos, or effective fluids. Kiselev's construction of black holes surrounded by fluid-like matter provides a useful phenomenological template \cite{Kiselev:2002dx}, and several authors have used perfect fluid dark matter (PFDM) to model environmental corrections to the black-hole metric. Xu, Hou, Wang, and Liao investigated the effect of PFDM on the thermodynamics and phase transition of Reissner--Nordstr\"om--AdS black holes \cite{Xu:2016jod}, while Hou, Xu, and Wang analyzed the shadow and deflection angle of rotating black holes in PFDM \cite{Hou:2018avu}. Thermodynamic geometry, optical properties, and extensions involving nonlinear electrodynamics or Euler--Heisenberg corrections have also been explored in PFDM backgrounds \cite{Li:2024euler}. In this context, the PFDM contribution is relevant because it introduces a logarithmic correction to the metric, which can remain significant at intermediate radial scales and therefore influence horizons, photon spheres, thermodynamic stability, greybody factors, and absorption cross sections.

Motivated by these developments, in this work, we study a ModMax black hole surrounded by perfect fluid dark matter within Lorentz-violating Kalb--Ramond gravity. The model combines three physically distinct ingredients. The first is the Kalb--Ramond background, which encodes Lorentz-violating gravitational effects through the parameter $\alpha$. The second is the ModMax electromagnetic sector, whose purely electric branch modifies the charge contribution through the screened combination $\Qeff=Q^2e^{-\lmm}$. The third is the PFDM distribution, controlled by the parameter $\beta$, which contributes a logarithmic matter correction to the geometry. The simultaneous presence of these sectors leads to a rich structure in the horizon equation, photon sphere, optical radius, Hawking temperature, heat capacity, Helmholtz free energy, Hawking-radiation sparsity, greybody factor, and scalar absorption cross section.

It is important to distinguish the present work from the closely related charged KR+PFDM black-hole analysis recently posted as Ref.~\cite{AhmedFathiSilva2026ChargedKRPFDMPublished}. That work considers the linear charged sector in a Kalb--Ramond background surrounded by PFDM and develops a broad optical and dynamical phenomenology including photon spheres, shadows, epicyclic frequencies, QPO models, Bayesian parameter estimation, thermodynamics, and Hawking-radiation sparsity. Here we study a different electromagnetic source: the charge sector is generated by ModMax nonlinear electrodynamics and enters the metric through the screened combination $\Qeff=Q^2e^{-\lmm}$. Consequently, the present solution reduces to the charged KR+PFDM geometry only in the Maxwell limit $\lmm=0$ with the corresponding parameter normalization, whereas for $\lmm\neq0$ it represents a distinct nonlinear-electrodynamic deformation. The main physical question addressed here is therefore not a repetition of the charged KR+PFDM analysis, but how ModMax screening competes with the Kalb--Ramond deformation and the PFDM logarithmic term in the horizon structure, optical radius, thermal response, Hawking-radiation sparsity, greybody factor, and scalar absorption cross section. In particular, the scalar perturbation and absorption sector developed below provides a clear phenomenological discriminator of the ModMax extension.

To achieve the goal described above, we organize the paper as follows. In Sec.~\ref{s2}, we present the theoretical setup and obtain the black-hole solution in Kalb--Ramond gravity coupled to ModMax electrodynamics and perfect fluid dark matter. In Sec.~\ref{s3}, we investigate the horizon and geodesic structure along with the motion of test particles and discuss the role played by each physical parameter. In Sec.~\ref{s4}, we derive the main thermodynamic quantities and analyze the thermal stability of the system. In Sec.~\ref{s5}, we investigate the sparsity of the Hawking radiation. In Sec.~\ref{s6}, we investigate the scattering properties of massless scalar perturbations by calculating the greybody factor and the absorption cross section. Finally, in Sec.~\ref{s7}, we summarize our results and comment on possible extensions of the present work.

\section{Black hole solution}\label{s2}

In this section, we will discuss a static and spherically symmetric black hole solution with ModMax electrodynamics and surrounded by PFDM in a LV Kalb-Ramond gravity background. Within this gravitational framework, LV effects are induced by a rank-two antisymmetric KR field, which acquires a nonzero VEV. On the other hand, the dark matter halo is modeled through a PFDM energy-momentum tensor. We also consider ModMax electrodynamics. In this sense, the resulting geometry may be viewed as a Reissner–Nordström-like black hole dressed simultaneously by a KR background, ModMax electrodynamics, and a PFDM halo.

Let us begin with the four-dimensional action \cite{Altschul:2009ae, Maluf:2018jwc, Aashish:2019ykb}
\begin{align}
S=\int d^4x\,\sqrt{-g}\bigg[
\frac{1}{2\kappa}\left(R-2\Lambda+\epsilon\,B^{\mu\lambda}B^{\nu}{}_{\lambda}R_{\mu\nu}\right)
\nonumber\\-\frac{1}{12}H_{\lambda\mu\nu}H^{\lambda\mu\nu}
-V\!\left(B_{\mu\nu}B^{\mu\nu}\pm b^2\right)
+\mathcal{L}_{\rm MM}+\mathcal{L}_{\rm dm}
\bigg].
\end{align}
Above, we have considered $\kappa=8\pi G$, while $\epsilon$ is the nonminimal coupling between gravity and the KR field, $b^2>0$ is a constant, $\Lambda$ is the cosmological constant, and $H_{\mu\nu\rho}=\partial_{[\mu}B_{\nu\rho]}$ is the KR field strength. The potential $V\!\left(B_{\mu\nu}B^{\mu\nu}\pm b^2\right)$ enforces spontaneous LSB through the nonzero VEV $\langle B_{\mu\nu}\rangle=b_{\mu\nu}$ and $b_{\mu\nu}b^{\mu\nu}=\mp b^2$. The variation of the action with respect to the metric provides the following equation
\begin{equation}
R_{\mu\nu}-\frac{1}{2}g_{\mu\nu}R+\Lambda g_{\mu\nu}
=\kappa\left(T_{\mu\nu}^{\rm KR}+T_{\mu\nu}^{\rm (MM)}+T_{\mu\nu}^{\rm (DM)}\right).
\end{equation}
The PFDM source is taken as follows
\begin{align}
T^{\nu}{}_{\mu}{}^{\rm (DM)}
&=\mathrm{diag}[-\rho,p,p,p]
\nonumber\\&=\mathrm{diag}\!\left[
\frac{\beta}{8\pi r^3},
\frac{\beta}{8\pi r^3},
-\frac{\beta}{16\pi r^3},
-\frac{\beta}{16\pi r^3}
\right],
\end{align}
where $\beta$ is the PFDM parameter. The KR contribution can be written as follows
\begin{align}
\kappa T_{\mu\nu}^{\rm KR}
&=\frac{1}{2}H_{\mu\alpha\beta}H_{\nu}{}^{\alpha\beta}
-\frac{1}{12}g_{\mu\nu}H_{\alpha\beta\rho}H^{\alpha\beta\rho}
\nonumber\\
&+2V'(X)\,B_{\alpha\mu}B^{\alpha}{}_{\nu}
-g_{\mu\nu}V(X)
\nonumber\\
&+\epsilon\Bigg[
\frac{1}{2}g_{\mu\nu}B^{\rho\sigma}B^{\chi}{}_{\sigma}R_{\rho\chi}
-B^{\rho}{}_{\mu}B^{\chi}{}_{\nu}R_{\rho\chi}
\nonumber\\
&-B^{\alpha\beta}B_{\nu\beta}R_{\mu\alpha}
-B^{\alpha\beta}B_{\mu\beta}R_{\nu\alpha}
\nonumber\\
&+\frac{1}{2}\nabla_{\alpha}\nabla_{\mu}\!\left(B^{\alpha\beta}B_{\nu\beta}\right)
+\frac{1}{2}\nabla_{\alpha}\nabla_{\nu}\!\left(B^{\alpha\beta}B_{\mu\beta}\right)
\nonumber\\
&-\frac{1}{2}\nabla^{\rho}\nabla_{\rho}\!\left(B_{\mu}{}^{\sigma}B_{\nu\sigma}\right)
-\frac{1}{2}g_{\mu\nu}\nabla_{\rho}\nabla_{\chi}
\!\left(B^{\rho\sigma}B^{\chi}{}_{\sigma}\right)
\Bigg],
\end{align}
with $X=B_{\mu\nu}B^{\mu\nu}$. In what follows, we set $\Lambda=0$ and adopt the following line element
\begin{equation}
ds^2=-A(r)\,dt^2+B(r)\,dr^2+r^2d\theta^2+r^2\sin^2\theta\,d\phi^2 .
\end{equation}
For this KR background, let us consider a pseudoelectric configuration in which the only nonvanishing components are $b_{01}$ and $b_{10}$. Then, the constant-norm condition is given by
\begin{equation}
b_{01}=-b_{10}=|b|\,\sqrt{\frac{A(r)B(r)}{2}}.
\end{equation}
With this, one assumes the KR field is frozen at its VEV, so that the modified Einstein equations reduce to a coupled system for $A(r)$ and $B(r)$. A particularly useful combination of the independent equations is
\begin{equation}
B(r)A'(r)+A(r)B'(r)=0,
\end{equation}
which immediately implies
\begin{equation}
B(r)=\frac{1}{A(r)}.
\end{equation}

At this point, we conveniently define the LV parameter given by $\alpha=\frac{\epsilon b^2}{2}$. To proceed further, let us substitute $B(r)=A(r)^{-1}$ back into the field equations, thereby obtaining the exact black hole geometry given by
\begin{equation}
ds^2=-A(r)\,dt^2+\frac{dr^2}{A(r)}+r^2d\theta^2+r^2\sin^2\theta\,d\phi^2,
\end{equation}
with metric function
\begin{equation}
A(r)=\frac{1}{1-\alpha}
-\frac{2M}{r}
+\frac{\beta}{(1-\alpha)\,r}\,
\log\!\left(\frac{r}{|\beta|}\right).
\label{eq:A_KRPFDM}
\end{equation}

In equation \eqref{eq:A_KRPFDM}, we see that the physical role of the two deformation parameters is transparent, where $\alpha$ encodes the LV effect of the KR vacuum, while $\beta$ measures the influence of the PFDM distribution. In the limit $\beta\to 0$, one recovers the static KR black hole without PFDM. A further limit $\alpha\to 0$ reduces the solution to the Schwarzschild spacetime. The event horizon is determined by the largest positive root of $A(r_h)=0$. In the notation employed in the original derivation, the horizon can be written as
\begin{equation}
r_h=\beta\,W\!\left(
e^{\frac{2(1-\alpha)M}{\beta}}
\right),
\end{equation}
where $W$ denotes the Lambert-$W$ function or $\mathrm{ProductLog}$. In the Schwarzschild limit, $\alpha,\beta\rightarrow 0$, one has $r_h\rightarrow 2M$. In addition to PFDM, we can introduce ModMax electrodynamics, which is a nonlinear, conformally invariant deformation of Maxwell theory controlled by a real parameter $\lmm\ge 0$. To obtain the black hole solution, it is convenient to write the two electromagnetic invariants given by
\begin{align}
S \equiv -\frac14\,F_{\mu\nu}F^{\mu\nu},\\
P \equiv -\frac14\,F_{\mu\nu}\tilde F^{\mu\nu},
\end{align}
where $\tilde F^{\mu\nu}\equiv\frac12\,\varepsilon^{\mu\nu\alpha\beta}F_{\alpha\beta}$. With this, Maxwell theory corresponds to $\mathcal L_{\rm Mxw}=S$. With this, one expresses the ModMax Lagrangian density as follows
\begin{align}
\mathcal L_{\rm MM}(S,P;\lmm)=
S\,\cosh\lmm+\sqrt{S^2+P^2}\,\sinh\lmm,
\label{eq:modmax_lagrangian}
\end{align}
which is analytic for $(S,P)\neq (0,0)$, reduces to Maxwell at $\lmm=0$, and preserves electric-magnetic duality with a deformation of the constitutive relations. In the static and spherically symmetric case, the line element is commonly written as
\begin{align}
ds^2=-f(r)\,dt^2+\frac{dr^2}{f(r)}+r^2(d\theta^2+\sin^2\theta\,d\phi^2),
\end{align}
where the metric function $f(r)$ contains the gravitational mass contribution and the effective electric term generated by the ModMax sector. In Einstein gravity, the metric function for the ModMax black hole is given by
\begin{align}
    f(r)=1-\frac{2M}{r}+\frac{Q^2\,e^{-\lmm}}{r^2}
\end{align}

It is worth noting that the ModMax black hole preserves the basic Reissner-Nordstr\"om-type structure. On the other hand, the charge sector is effectively dressed by nonlinear electrodynamic effects, making the horizon structure, thermodynamics, and wave absorption explicitly dependent on the parameter $\lmm$. 

For the purely electric branch considered in this work, the ModMax field equations preserve the Reissner--Nordstr\"om-type radial dependence of the charge term, while replacing the Maxwell contribution by the screened combination $\Qeff=Q^2e^{-\lmm}$. We therefore work with the screened-charge extension of the charged KR+PFDM geometry. This construction is understood within the standard static, spherically symmetric, purely electric ModMax sector: it recovers the charged KR+PFDM black hole in the Maxwell limit $\lmm=0$, the uncharged KR+PFDM solution when $Q=0$, and the Reissner--Nordstr\"om--ModMax form when the KR and PFDM deformations are removed. In this sector, the static and spherically symmetric geometry is described by the following line element
\begin{align}
ds^2=-A(r)\,dt^2+\frac{1}{A(r)}\,dr^2+r^2(d\theta^2+\sin^2\theta\,d\phi^2),
\label{eq:sss_metric}
\end{align}
where
\begin{equation}
A(r)=\frac{1}{1-\alpha}
-\frac{2M}{r}+\frac{Q^2\,e^{-\lmm}}{(1-\alpha)^2\, r^2}
+\frac{\beta}{(1-\alpha)\,r}\,
\log\!\left(\frac{r}{|\beta|}\right).
\label{BH sol}
\end{equation}

\begin{figure*}[tbhp]
\centering

\begin{minipage}[t]{0.48\textwidth}
    \centering
    \includegraphics[height=4.8cm]{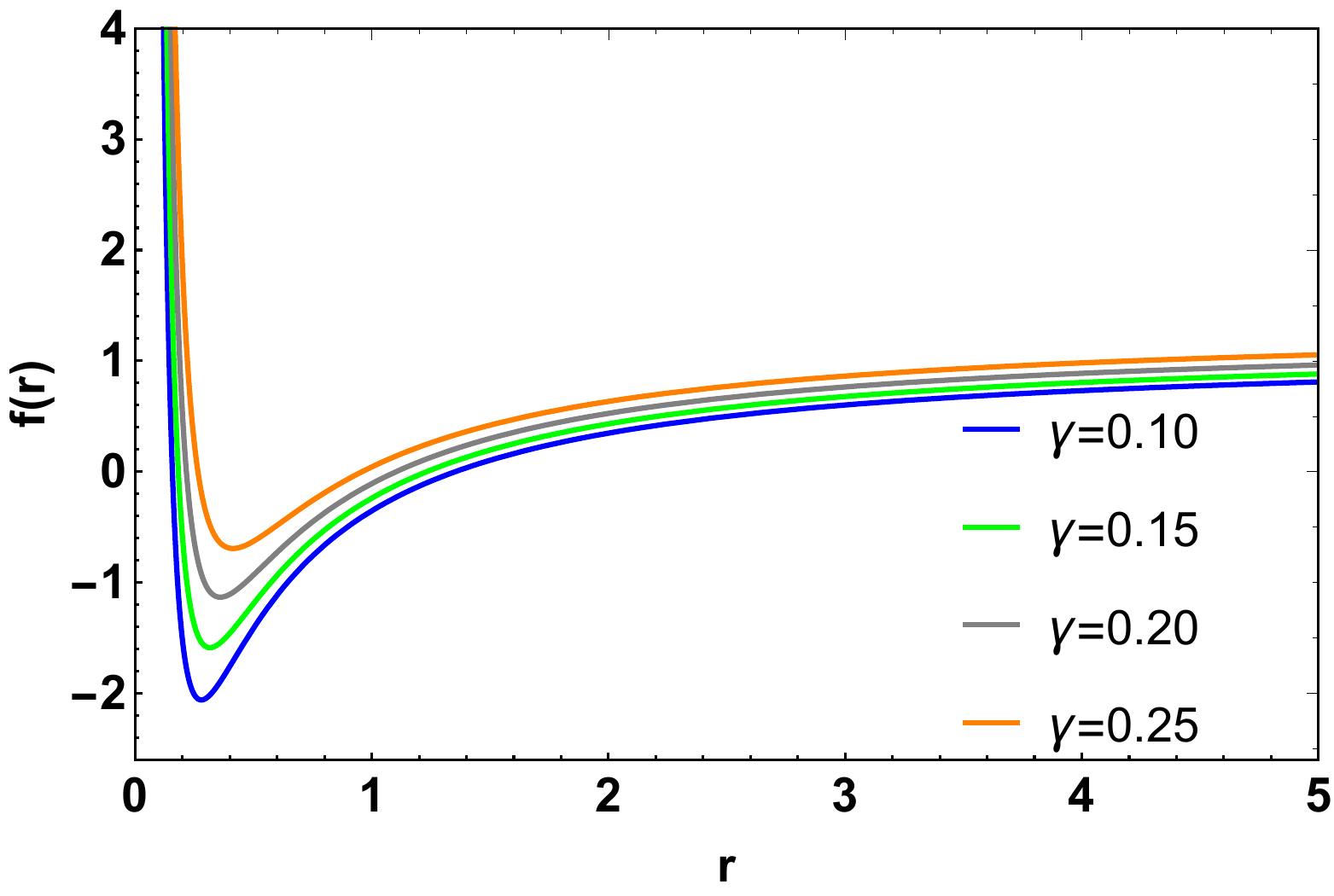}

    \vspace{0.15cm}
    \parbox{0.95\textwidth}{\centering\footnotesize
    (a) Variation of the KR/LV parameter, with the remaining parameters fixed.}
\end{minipage}
\hfill
\begin{minipage}[t]{0.48\textwidth}
    \centering
    \includegraphics[height=4.8cm]{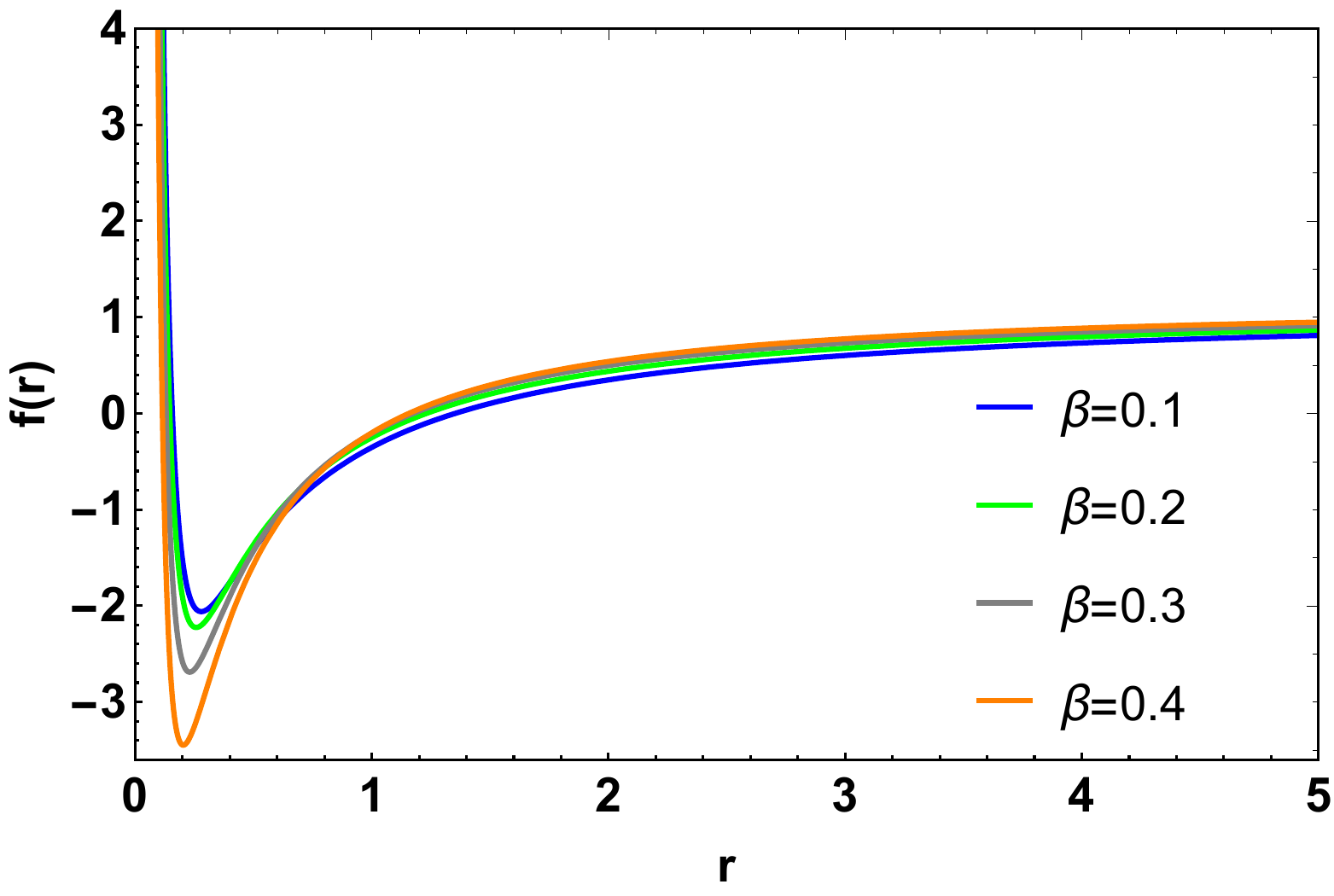}

    \vspace{0.15cm}
    \parbox{0.95\textwidth}{\centering\footnotesize
    (b) Variation of the PFDM parameter, with the remaining parameters fixed.}
\end{minipage}

\vspace{0.45cm}

\begin{minipage}[t]{0.48\textwidth}
    \centering
    \includegraphics[height=4.8cm]{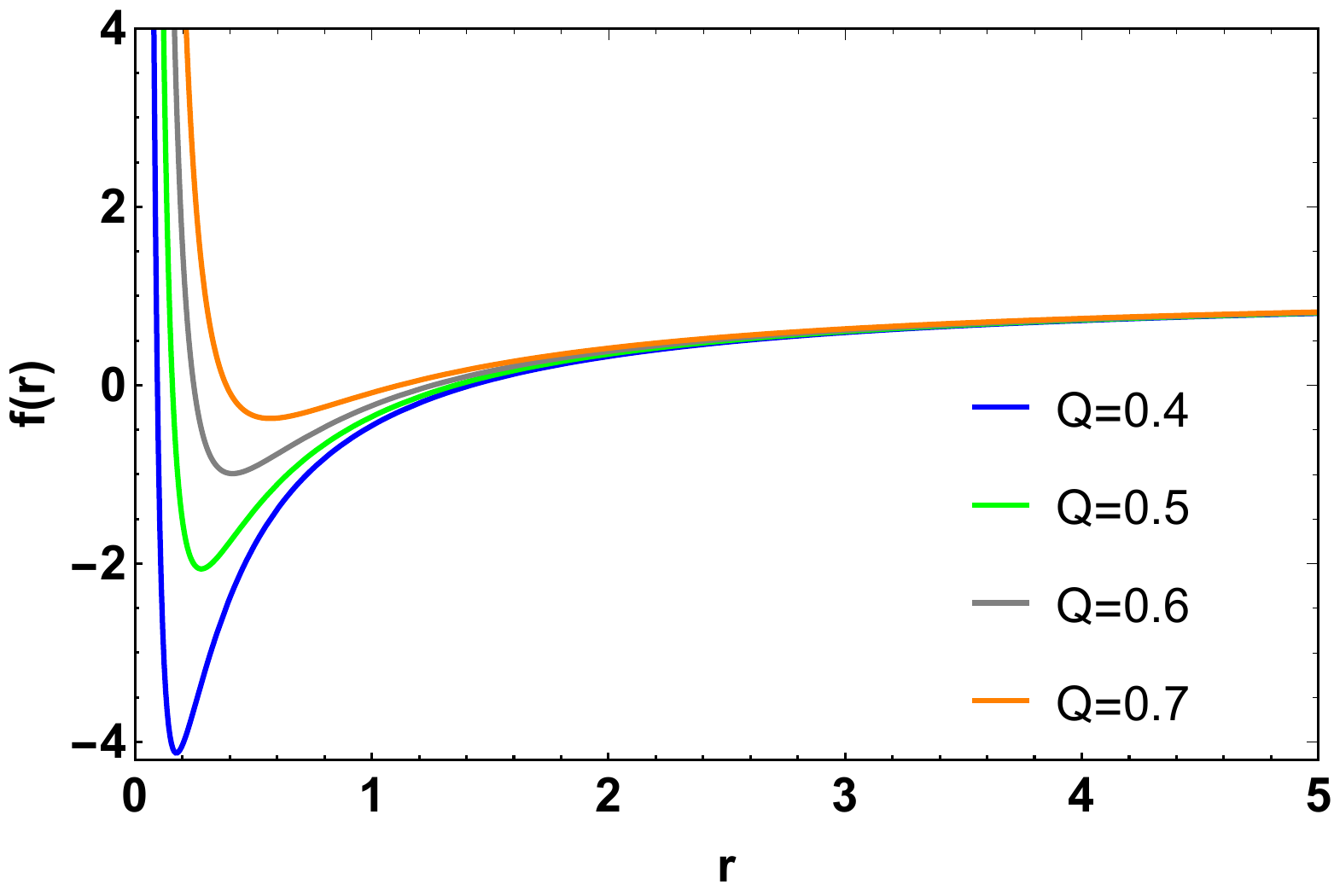}

    \vspace{0.15cm}
    \parbox{0.95\textwidth}{\centering\footnotesize
    (c) Variation of the electric charge, with the remaining parameters fixed.}
\end{minipage}
\hfill
\begin{minipage}[t]{0.48\textwidth}
    \centering
    \includegraphics[height=4.8cm]{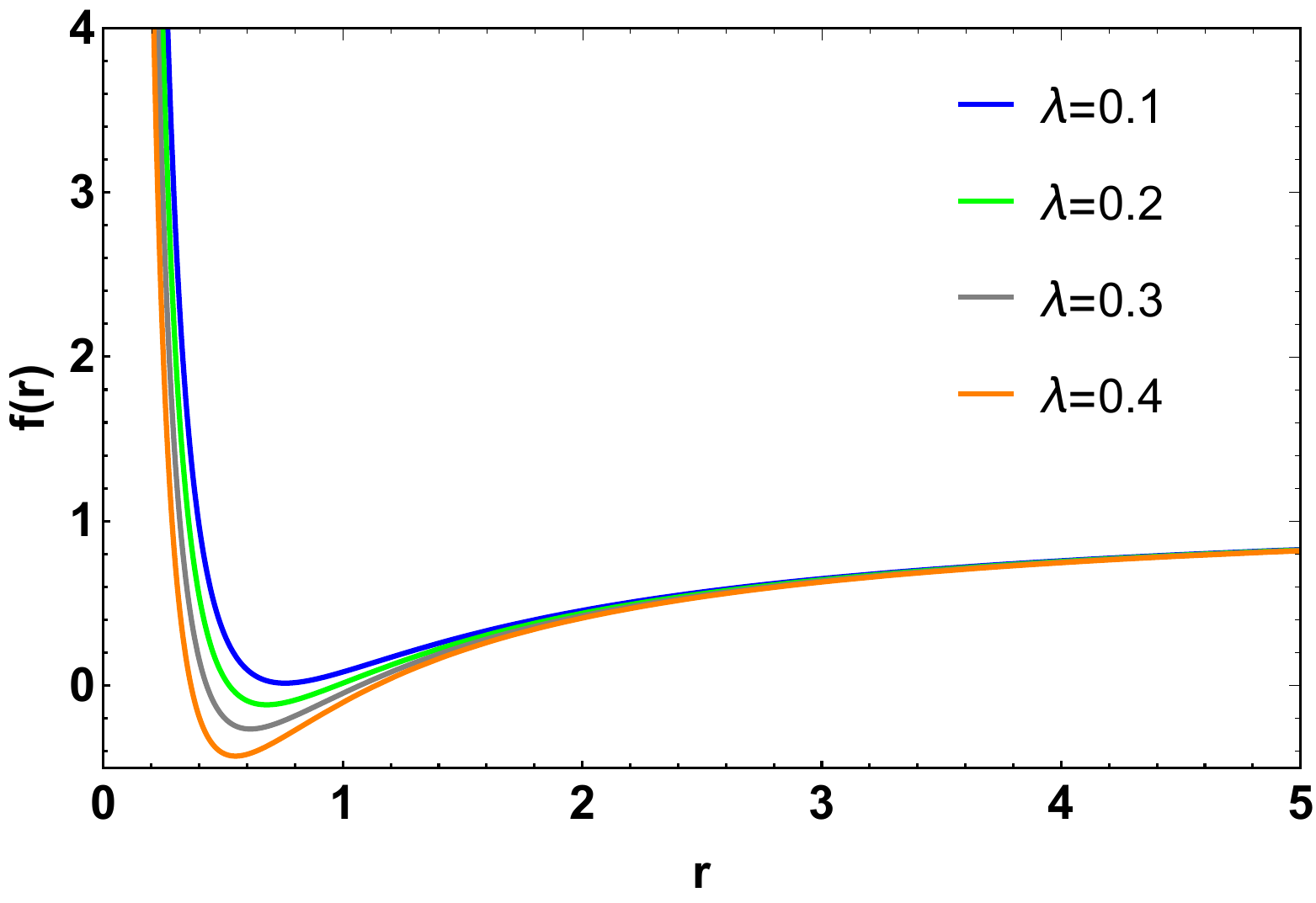}

    \vspace{0.15cm}
    \parbox{0.95\textwidth}{\centering\footnotesize
    (d) Variation of the ModMax parameter, with the remaining parameters fixed.}
\end{minipage}

\vspace{-0.2cm}

\caption{Behavior of the metric function $f(r)$ under variations of the model parameters. Panel (a) displays the effect of the KR/LV parameter, panel (b) the effect of the PFDM parameter, panel (c) the effect of the electric charge, and panel (d) the effect of the ModMax parameter. In the graphical legends, $\gamma\equiv\alpha$ and $\lambda\equiv\lambda_{\rm MM}$.}
\label{f}
\end{figure*}

From this point on, we use the notations $f(r)$ and $A(r)$ interchangeably, with $f(r)\equiv A(r)$, whenever this makes contact with the standard black-hole thermodynamics and perturbation literature.

We note that the metric function in Eq.~\eqref{BH sol} has a Reissner-Nordstr\"om-like core, because the charge contribution behaves as $Q^2/r^2$, but it is dressed by both the LV factor $(1-\alpha)^{-1}$ and the ModMax exponential $e^{-\lmm}$. The logarithmic term proportional to $\beta$ represents the surrounding PFDM distribution. Since this term decays only as $\ln(r/\beta)/r$, it is subleading at infinity but it remains more persistent than a pure $1/r^2$ correction.

Moreover, the event horizon is obtained from $f(r_h)=0$. Equivalently, one may write the mass parameter as $M=M(r_h)$. This form is useful because it shows how the external parameters compete when $M$ is kept fixed. Increasing $\alpha$ amplifies all terms weighted by $(1-\alpha)^{-1}$ or $(1-\alpha)^{-2}$. In the parameter range used in the plots, the outer horizon moves inward as $\alpha$ increases. The same inward shift is found when $Q$ or $\beta$ is increased. By contrast, increasing $\lmm$ suppresses the effective charge contribution $\Qeff$ and therefore moves the outer horizon outward. This is one of the cleanest physical effects of the ModMax sector, where for positive $\lmm$, the electric contribution is screened.

The plots of $f(r)$ in Fig.~\ref{f} reflect these facts directly. For fixed $M,Q,\beta,\lmm$, larger $\alpha$ raises the asymptotic value $f(\infty)=1/(1-\alpha)$ and changes the radial position where $f(r)$ crosses zero. Increasing $\beta$ strengthens the logarithmic exterior contribution in the plotted region, where $r>\beta$, and also shifts the zero of $f(r)$. Increasing $Q$ enhances the near-horizon repulsive $1/r^2$ term, while increasing $\lmm$ weakens it. Hence the $\lmm$ dependence goes in the opposite direction to the $Q$ dependence, as expected from the screened combination $\Qeff$.

To avoid any ambiguity in the graphical presentation, we note that the symbols displayed inside the figure legends follow the notation used when the plots were produced: the symbol $\gamma$ denotes the same Lorentz-violating Kalb--Ramond parameter denoted by $\alpha$ in the analytical discussion, while the symbol $\lambda$ denotes the ModMax parameter $\lambda_{\rm MM}$. The panel descriptions below are therefore written in terms of the physical parameter varied, and the internal graphical labels should be read with the identifications $\gamma\equiv\alpha$ and $\lambda\equiv\lambda_{\rm MM}$.

\section{Geodesic structure and black-hole shadow}\label{s3}

In this section, we will delve into the analysis of the motion of test particles and photons in the spacetime described by \eqref{BH sol}. In order to examine the geodesic motion and the optical appearance of the black hole, it is useful to introduce the compact notation
\begin{align}
&\delta\equiv 1-\alpha,\\
&\Qeff\equiv Q^2e^{-\lmm},
\label{eq:delta_Qeff_shadow}
\end{align}
so that, we write
\begin{equation}
A(r)=\frac{1}{\delta}
-\frac{2M}{r}
+\frac{\Qeff}{\delta^2r^2}
+\frac{\beta}{\delta r}
\log\!\left(\frac{r}{|\beta|}\right).
\label{eq:A_delta_shadow}
\end{equation}
In what follows, let us assume $\delta>0$, or equivalently $\alpha<1$, in order to preserve the correct asymptotic signature. The event horizon is located at the largest positive root of $A(r_h)=0$, namely
\begin{equation}
r_h^2
-2M\delta r_h
+\frac{\Qeff}{\delta}
+\beta r_h\log\!\left(\frac{r_h}{|\beta|}\right)
=0.
\label{eq:horizon_equation_shadow}
\end{equation}
Because of the logarithmic contribution produced by the perfect fluid dark matter, the horizon equation is not algebraic and, in general, must be solved numerically.

The geodesic motion can be derived from the Lagrangian
\begin{equation}
2\mathcal{L}
=-A(r)\dot{t}^{\,2}
+\frac{\dot{r}^{\,2}}{A(r)}
+r^2\dot{\theta}^{\,2}
+r^2\sin^2\theta\,\dot{\phi}^{\,2},
\label{eq:lagrangian_ABinverse}
\end{equation}
where the dot denotes differentiation with respect to an affine parameter. Due to the spherical symmetry, we may restrict the motion to the equatorial plane,
\begin{equation}
\theta=\frac{\pi}{2},
\qquad
\dot{\theta}=0.
\label{eq:equatorial_shadow}
\end{equation}
The energy and angular momentum per unit mass are conserved quantities and are given by
\begin{equation}
E=A(r)\dot{t},
\qquad
L=r^2\dot{\phi}.
\label{eq:EL_shadow}
\end{equation}
The normalization condition reads
\begin{equation}
g_{\mu\nu}\dot{x}^{\mu}\dot{x}^{\nu}=-\varepsilon,
\label{eq:normalization_shadow}
\end{equation}
where $\varepsilon=1$ for massive particles and $\varepsilon=0$ for photons. Substituting Eq.~\eqref{eq:EL_shadow} into Eq.~\eqref{eq:normalization_shadow}, one obtains the radial equation
\begin{equation}
\dot{r}^{\,2}
=
E^2
-
A(r)\left(\varepsilon+\frac{L^2}{r^2}\right).
\label{eq:radial_equation_ABinverse}
\end{equation}
Therefore, the effective potential is
\begin{equation}
V_{\rm eff}(r)
=
A(r)\left(\varepsilon+\frac{L^2}{r^2}\right),
\label{eq:Veff_general_shadow}
\end{equation}
and the allowed region of motion is determined by
\begin{equation}
E^2\geq V_{\rm eff}(r).
\label{eq:allowed_motion_shadow}
\end{equation}

For massive particles, one has
\begin{equation}
V_{\rm eff}^{(\rm timelike)}(r)
=
A(r)\left(1+\frac{L^2}{r^2}\right),
\label{eq:Veff_timelike_shadow}
\end{equation}
whereas for photons,
\begin{equation}
V_{\rm eff}^{(\rm null)}(r)
=
\frac{A(r)L^2}{r^2}.
\label{eq:Veff_null_shadow}
\end{equation}
Thus, the null geodesic structure is governed by the ratio $A(r)/r^2$.

\begin{figure*}[tbhp]
\centering

\begin{minipage}[t]{0.48\textwidth}
    \centering
    \includegraphics[height=4.8cm]{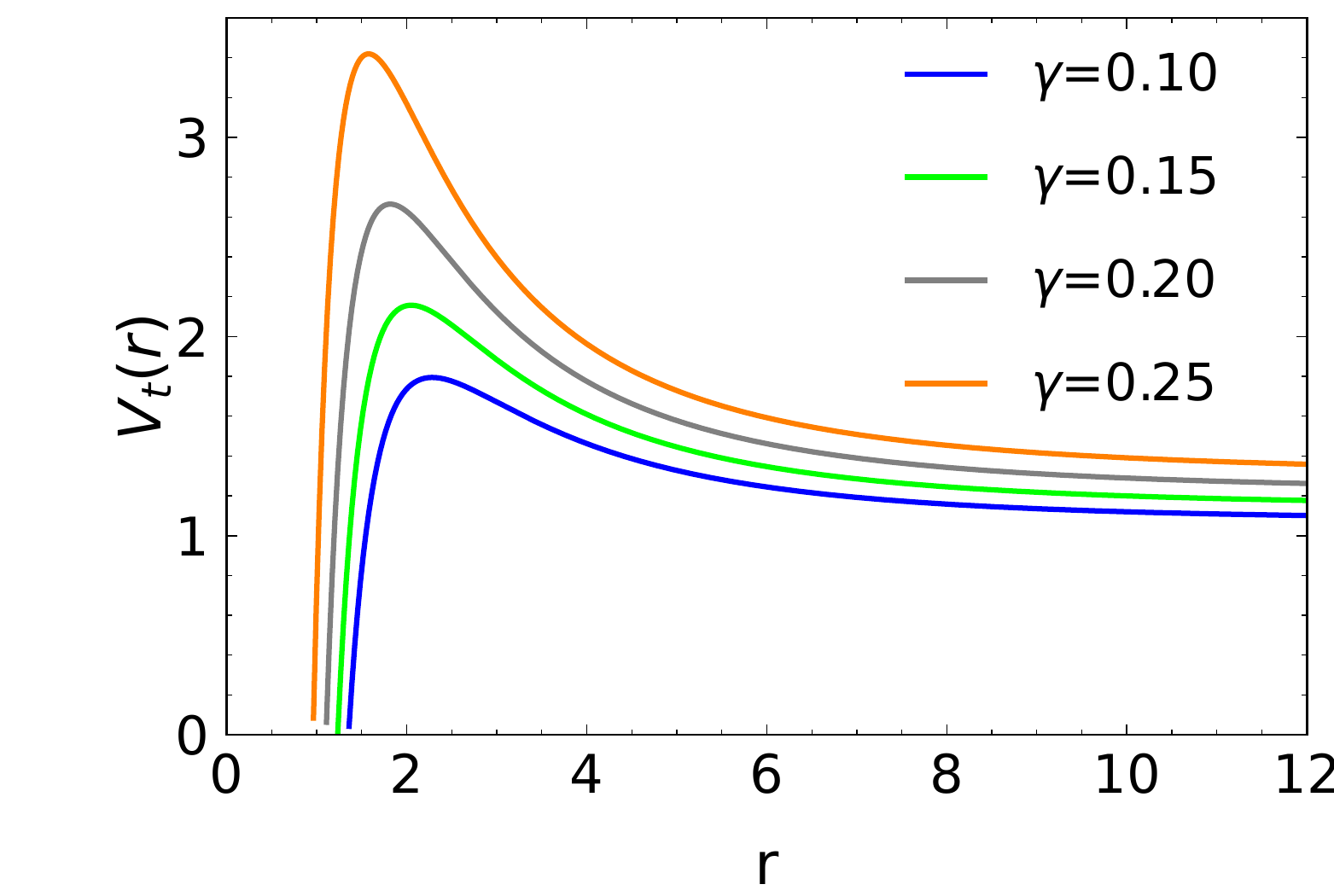}

    \vspace{0.15cm}
    \parbox{0.95\textwidth}{\centering\footnotesize
    (a) Variation of the KR/LV parameter, with the remaining parameters fixed.}
\end{minipage}
\hfill
\begin{minipage}[t]{0.48\textwidth}
    \centering
    \includegraphics[height=4.8cm]{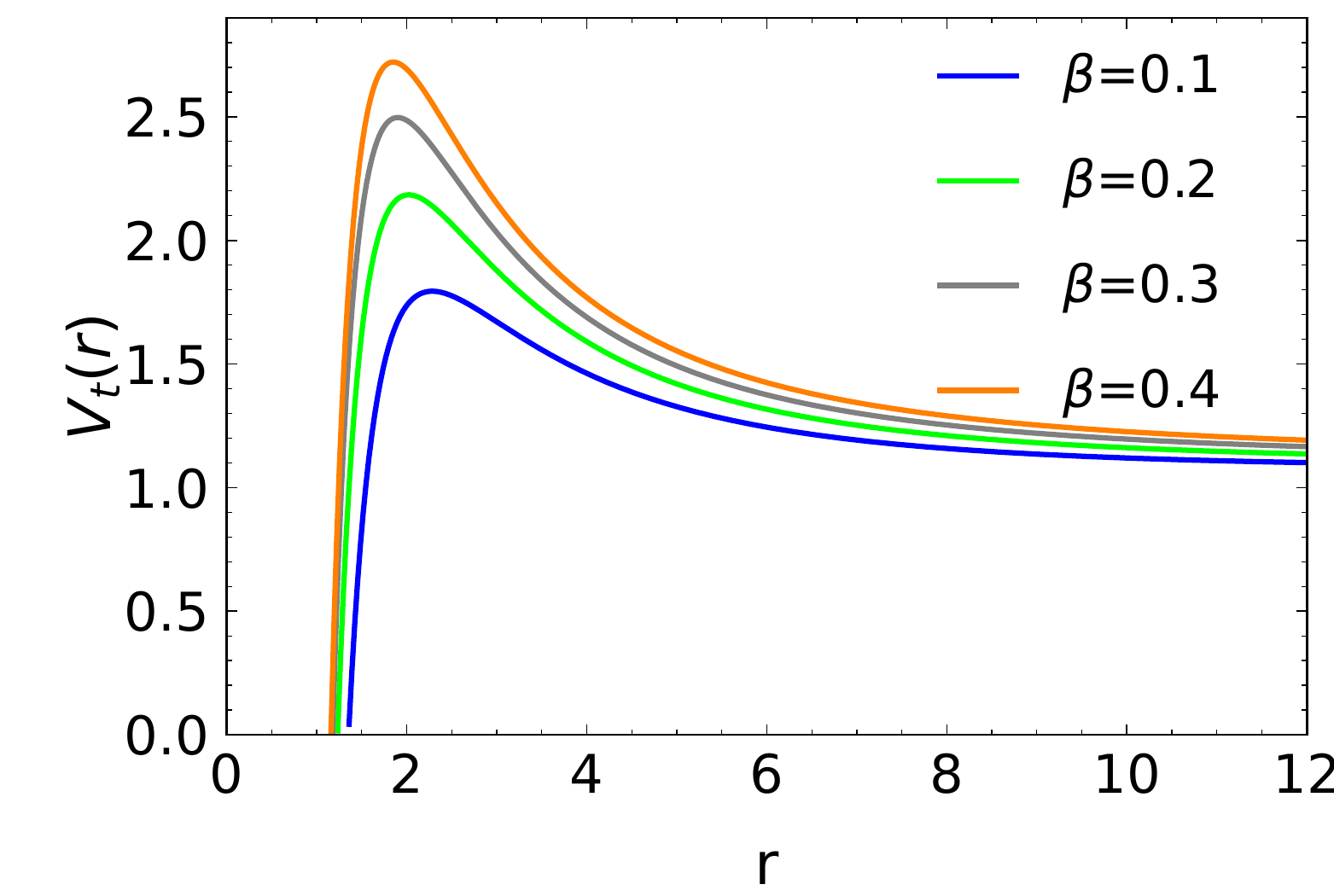}

    \vspace{0.15cm}
    \parbox{0.95\textwidth}{\centering\footnotesize
    (b) Variation of the PFDM parameter, with the remaining parameters fixed.}
\end{minipage}

\vspace{0.35cm}

\begin{minipage}[t]{0.48\textwidth}
    \centering
    \includegraphics[height=4.8cm]{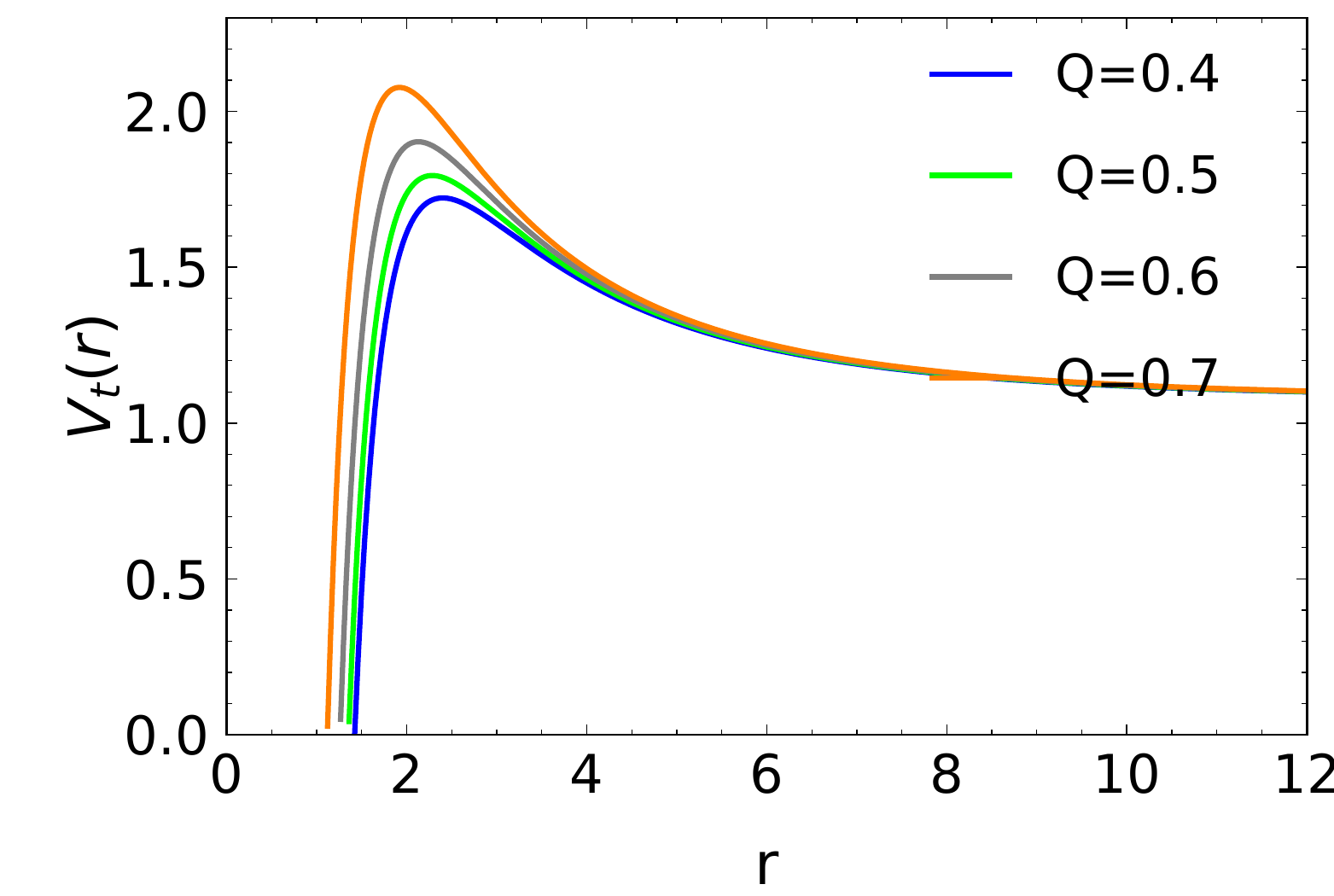}

    \vspace{0.15cm}
    \parbox{0.95\textwidth}{\centering\footnotesize
    (c) Variation of the electric charge, with the remaining parameters fixed.}
\end{minipage}
\hfill
\begin{minipage}[t]{0.48\textwidth}
    \centering
    \includegraphics[height=4.8cm]{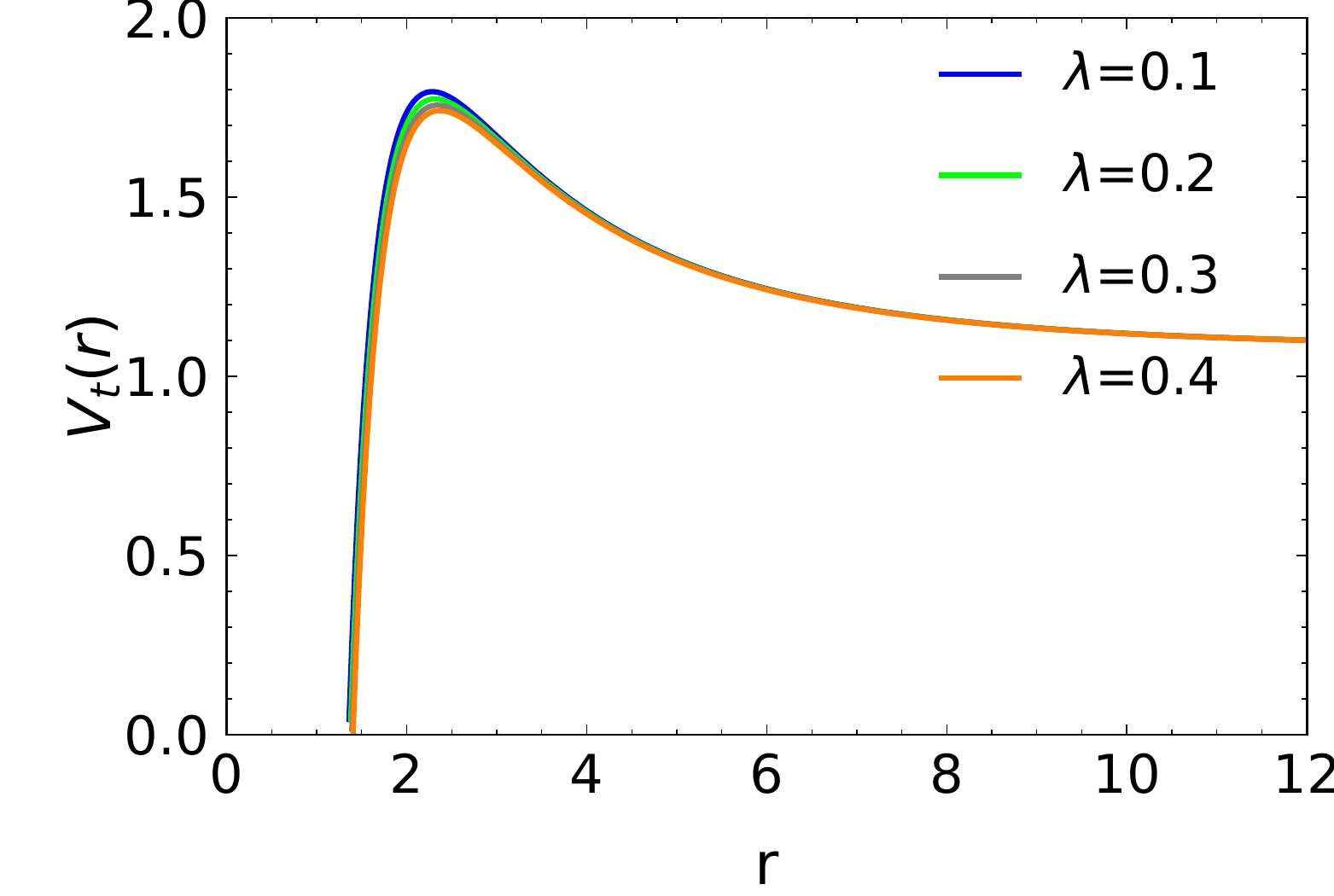}

    \vspace{0.15cm}
    \parbox{0.95\textwidth}{\centering\footnotesize
    (d) Variation of the ModMax parameter, with the remaining parameters fixed.}
\end{minipage}

\vspace{-0.2cm}

\caption{Behavior of the timelike effective potential $V_t(r)\equiv V_{\rm eff}^{(\rm timelike)}(r)$ for $L=4$ under variations of the model parameters. Panel (a) displays the effect of the KR/LV parameter, panel (b) the effect of the PFDM parameter, panel (c) the effect of the electric charge, and panel (d) the effect of the ModMax parameter. In the graphical legends, $\gamma\equiv\alpha$ and $\lambda\equiv\lambda_{\rm MM}$.}
\label{Vtimelike}
\end{figure*}

\begin{figure*}[tbhp]
\centering

\begin{minipage}[t]{0.48\textwidth}
    \centering
    \includegraphics[height=4.8cm]{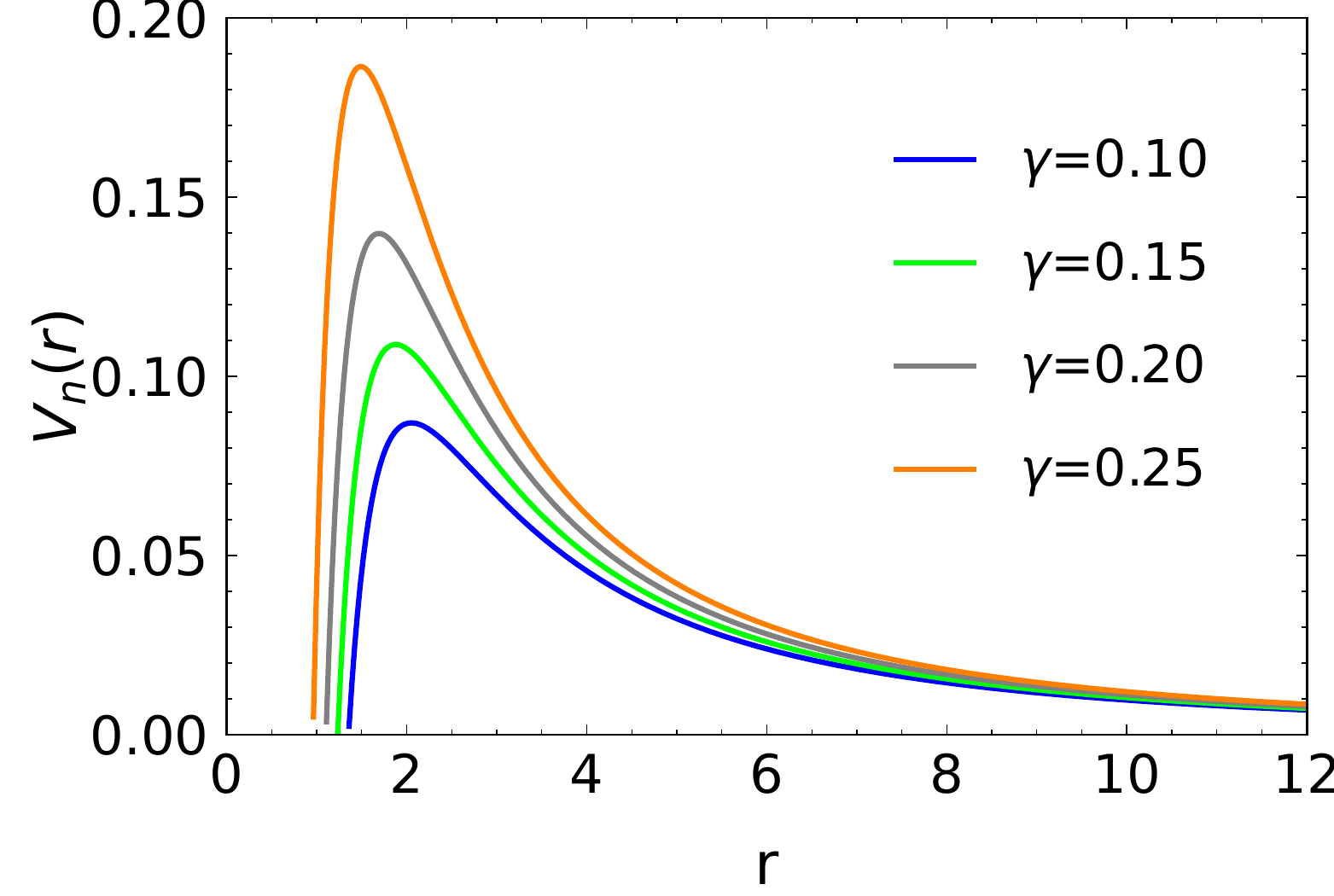}

    \vspace{0.15cm}
    \parbox{0.95\textwidth}{\centering\footnotesize
    (a) Variation of the KR/LV parameter, with the remaining parameters fixed.}
\end{minipage}
\hfill
\begin{minipage}[t]{0.48\textwidth}
    \centering
    \includegraphics[height=4.8cm]{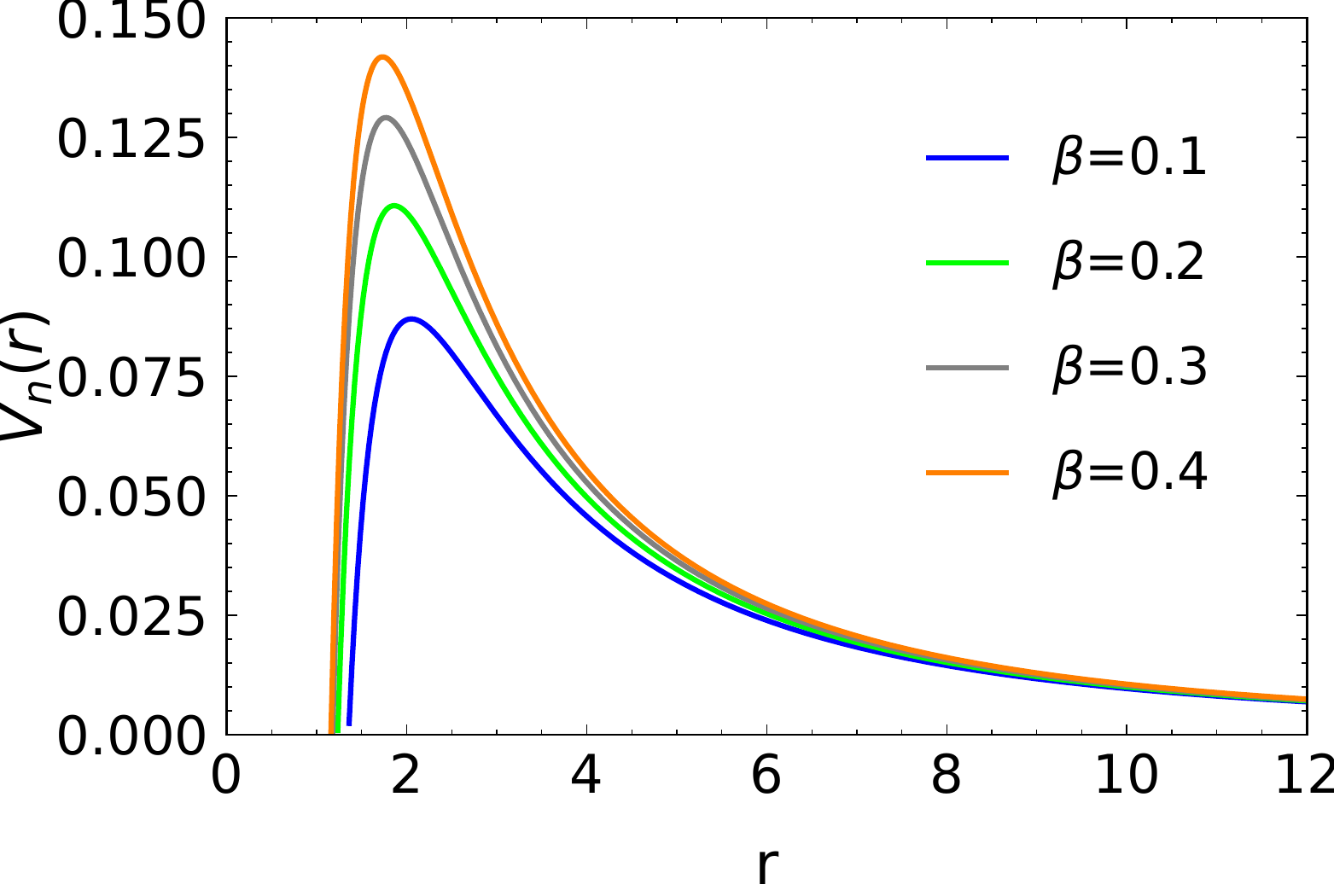}

    \vspace{0.15cm}
    \parbox{0.95\textwidth}{\centering\footnotesize
    (b) Variation of the PFDM parameter, with the remaining parameters fixed.}
\end{minipage}

\vspace{0.35cm}

\begin{minipage}[t]{0.48\textwidth}
    \centering
    \includegraphics[height=4.8cm]{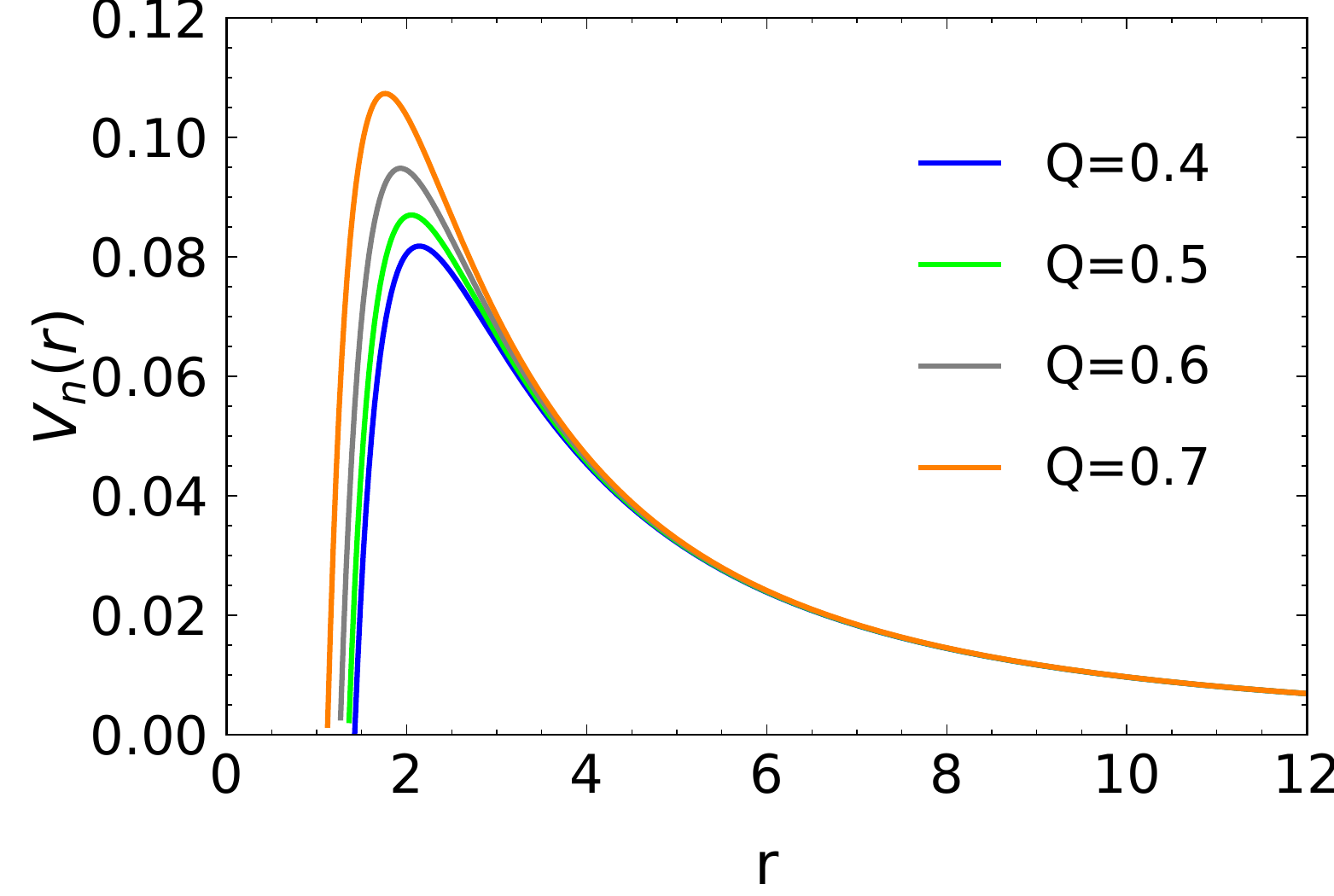}

    \vspace{0.15cm}
    \parbox{0.95\textwidth}{\centering\footnotesize
    (c) Variation of the electric charge, with the remaining parameters fixed.}
\end{minipage}
\hfill
\begin{minipage}[t]{0.48\textwidth}
    \centering
    \includegraphics[height=4.8cm]{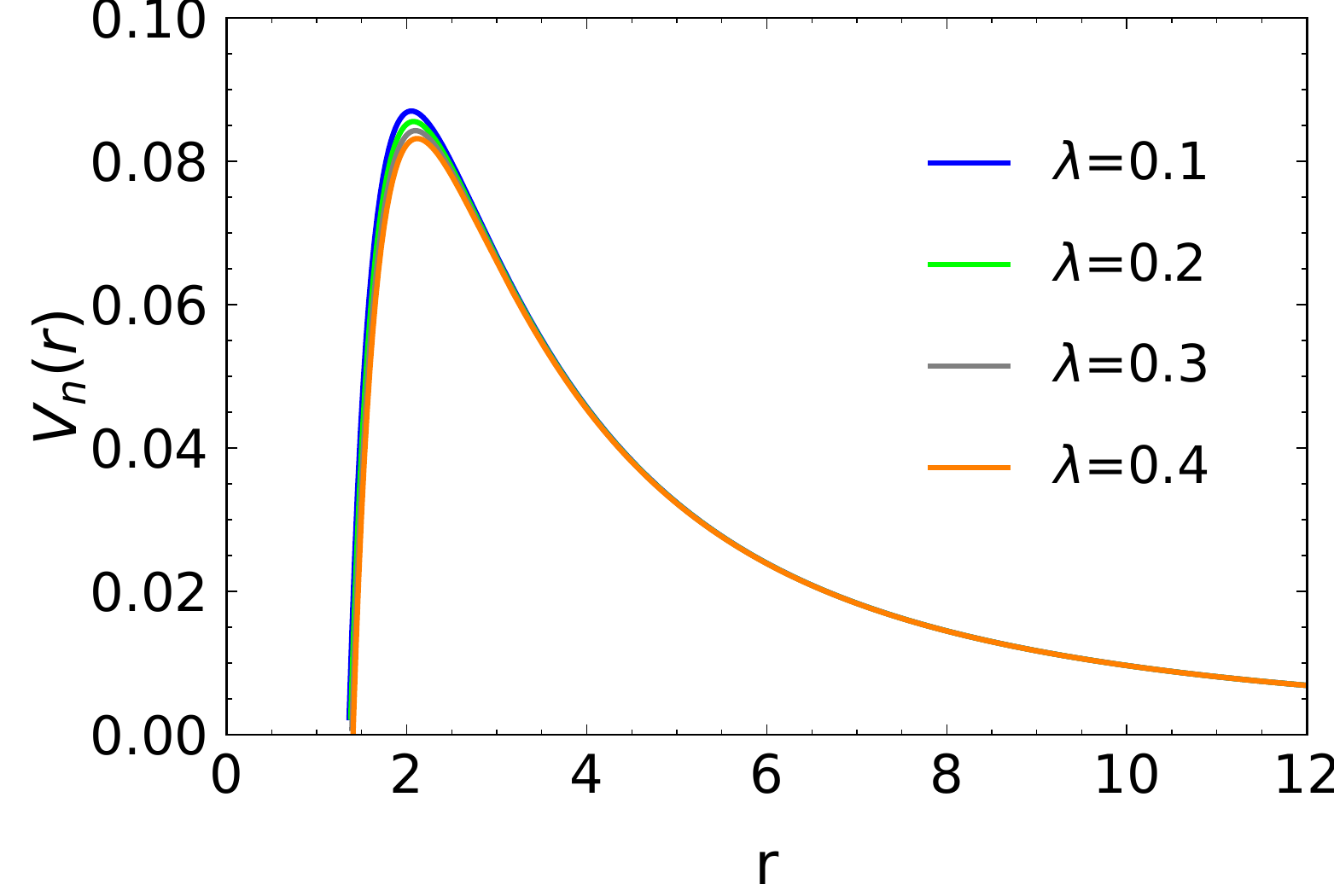}

    \vspace{0.15cm}
    \parbox{0.95\textwidth}{\centering\footnotesize
    (d) Variation of the ModMax parameter, with the remaining parameters fixed.}
\end{minipage}

\vspace{-0.2cm}

\caption{Behavior of the null effective potential $V_n(r)\equiv V_{\rm eff}^{(\rm null)}(r)$ for $L=1$ under variations of the model parameters. Panel (a) displays the effect of the KR/LV parameter, panel (b) the effect of the PFDM parameter, panel (c) the effect of the electric charge, and panel (d) the effect of the ModMax parameter. In the graphical legends, $\gamma\equiv\alpha$ and $\lambda\equiv\lambda_{\rm MM}$.}
\label{Vnull}
\end{figure*}

The first and second derivatives of the metric function are
\begin{equation}
A'(r)
=
\frac{2M}{r^2}
-\frac{2\Qeff}{\delta^2r^3}
+\frac{\beta}{\delta r^2}
\left[
1-\log\!\left(\frac{r}{|\beta|}\right)
\right],
\label{eq:A_prime_shadow}
\end{equation}
and
\begin{equation}
A''(r)
=
-\frac{4M}{r^3}
+\frac{6\Qeff}{\delta^2r^4}
+\frac{\beta}{\delta r^3}
\left[
2\log\!\left(\frac{r}{|\beta|}\right)-3
\right].
\label{eq:A_second_shadow}
\end{equation}

Circular timelike orbits are defined by
\begin{equation}
\dot{r}=0,
\qquad
\frac{dV_{\rm eff}^{(\rm timelike)}}{dr}=0.
\label{eq:circular_timelike_conditions_shadow}
\end{equation}
These conditions lead to
\begin{equation}
E_c^2
=
\frac{2A(r_c)^2}{2A(r_c)-r_cA'(r_c)},
\label{eq:Ec_shadow}
\end{equation}
and
\begin{equation}
L_c^2
=
\frac{r_c^3A'(r_c)}{2A(r_c)-r_cA'(r_c)}.
\label{eq:Lc_shadow}
\end{equation}
The circular orbit is physically admissible only if
\begin{equation}
2A(r_c)-r_cA'(r_c)>0.
\label{eq:circular_orbit_condition_shadow}
\end{equation}
The stability of the orbit is governed by the sign of
\begin{equation}
\left.
\frac{d^2V_{\rm eff}^{(\rm timelike)}}{dr^2}
\right|_{r=r_c}.
\label{eq:second_derivative_stability_shadow}
\end{equation}
Stable circular orbits satisfy
\begin{equation}
\left.
\frac{d^2V_{\rm eff}^{(\rm timelike)}}{dr^2}
\right|_{r=r_c}>0,
\label{eq:stable_orbit_shadow}
\end{equation}
while unstable circular orbits satisfy
\begin{equation}
\left.
\frac{d^2V_{\rm eff}^{(\rm timelike)}}{dr^2}
\right|_{r=r_c}<0.
\label{eq:unstable_orbit_shadow}
\end{equation}
The innermost stable circular orbit, ISCO, is obtained from the marginal stability condition
\begin{equation}
\left.
\frac{d^2V_{\rm eff}^{(\rm timelike)}}{dr^2}
\right|_{r=r_{\rm ISCO}}=0.
\label{eq:isco_definition_shadow}
\end{equation}
Equivalently, for the metric \eqref{eq:sss_metric}, the ISCO radius satisfies
\begin{align}
A(r_{\rm ISCO})A''(r_{\rm ISCO})
&-2A'(r_{\rm ISCO})^2
\notag\\&+\frac{3}{r_{\rm ISCO}}A(r_{\rm ISCO})A'(r_{\rm ISCO})
=0.
\label{eq:isco_condition_shadow}
\end{align}
Since $A(r)$ contains a logarithmic perfect-fluid-dark-matter correction, Eq.~\eqref{eq:isco_condition_shadow} is generally solved numerically.

For photons, $\varepsilon=0$, and the radial equation becomes
\begin{equation}
\dot{r}^{\,2}
=E^2-\frac{A(r)L^2}{r^2}.
\label{eq:null_radial_shadow}
\end{equation}
Introducing the impact parameter
\begin{equation}
b\equiv \frac{L}{E},
\label{eq:impact_parameter_shadow}
\end{equation}
one finds
\begin{equation}
\dot{r}^{\,2}
=
E^2
\left[
1-\frac{A(r)b^2}{r^2}
\right].
\label{eq:null_radial_impact_shadow}
\end{equation}
The turning point of a photon trajectory is determined by
\begin{equation}
b^2=\frac{r^2}{A(r)}.
\label{eq:turning_point_shadow}
\end{equation}
A circular photon orbit occurs when the effective potential reaches an extremum. Therefore, the photon-sphere radius $r_{\rm ph}$ is obtained from
\begin{equation}
\frac{d}{dr}\left(\frac{A(r)}{r^2}\right)=0,
\label{eq:photon_condition_compact_shadow}
\end{equation}
or
\begin{equation}
r_{\rm ph}A'(r_{\rm ph})-2A(r_{\rm ph})=0.
\label{eq:photon_condition_shadow}
\end{equation}
Using the explicit form of $A(r)$, this condition gives
\begin{equation}
2r_{\rm ph}^2
-6M\delta r_{\rm ph}
+\frac{4\Qeff}{\delta}
+\beta r_{\rm ph}
\left[
3\log\!\left(\frac{r_{\rm ph}}{|\beta|}\right)-1
\right]
=0.
\label{eq:photon_sphere_equation_shadow}
\end{equation}
Equivalently, in terms of the original parameters,
\begin{align}
\frac{4Q^2e^{-\lmm}}{1-\alpha}
+\beta r_{\rm ph}&
\left[
3\log\!\left(\frac{r_{\rm ph}}{|\beta|}\right)-1
\right]\notag \\&+2r_{\rm ph}^2
-6M(1-\alpha)r_{\rm ph}
=0.
\label{eq:photon_sphere_original_shadow}
\end{align}
The physical photon sphere corresponds to the outermost positive solution satisfying
\begin{equation}
r_{\rm ph}>r_h,
\qquad
A(r_{\rm ph})>0.
\label{eq:physical_photon_sphere_shadow}
\end{equation}

The instability of the photon sphere follows from
\begin{equation}
\left.
\frac{d^2}{dr^2}
\left(\frac{A(r)}{r^2}\right)
\right|_{r=r_{\rm ph}}
=
\frac{r_{\rm ph}A''(r_{\rm ph})-A'(r_{\rm ph})}{r_{\rm ph}^3}.
\label{eq:photon_instability_general_shadow}
\end{equation}
Thus, the photon sphere is unstable when
\begin{equation}
r_{\rm ph}A''(r_{\rm ph})-A'(r_{\rm ph})<0.
\label{eq:photon_instability_shadow}
\end{equation}
For the present black hole,
\begin{equation}
rA''(r)-A'(r)
=
-\frac{6M}{r^2}
+\frac{8\Qeff}{\delta^2r^3}
+\frac{\beta}{\delta r^2}
\left[
3\log\!\left(\frac{r}{|\beta|}\right)-4
\right].
\label{eq:photon_instability_explicit_shadow}
\end{equation}

For completeness, the orbit equation for null geodesics can be written as
\begin{equation}
\left(\frac{dr}{d\phi}\right)^2
=
r^4
\left[
\frac{1}{b^2}
-\frac{A(r)}{r^2}
\right].
\label{eq:null_orbit_r_shadow}
\end{equation}
In terms of $u=1/r$, this becomes
\begin{equation}
\left(\frac{du}{d\phi}\right)^2
=
\frac{1}{b^2}
-u^2 A\!\left(\frac{1}{u}\right).
\label{eq:null_orbit_u_shadow}
\end{equation}

\subsection{Shadow radius}
\label{subsec:shadow_radius}

The critical impact parameter associated with the unstable photon sphere is
\begin{equation}
b_{\rm ph}^2
=
\frac{r_{\rm ph}^2}{A(r_{\rm ph})},
\label{eq:critical_b_shadow}
\end{equation}
or
\begin{equation}
b_{\rm ph}
=
\frac{r_{\rm ph}}{\sqrt{A(r_{\rm ph})}}.
\label{eq:bph_shadow}
\end{equation}
The local angular radius of the shadow measured by a static observer located at $r=r_o$ is obtained from
\begin{equation}
\sin^2\alpha_{\rm sh}
=
\frac{A(r_o)b_{\rm ph}^2}{r_o^2}.
\label{eq:angular_radius_shadow}
\end{equation}
Thus,
\begin{equation}
\sin\alpha_{\rm sh}
=
\frac{r_{\rm ph}}{r_o}
\sqrt{
\frac{A(r_o)}{A(r_{\rm ph})}
}.
\label{eq:angular_radius_explicit_shadow}
\end{equation}

Since the spacetime satisfies
\begin{equation}
A(r\rightarrow\infty)=\frac{1}{\delta},
\label{eq:asymptotic_A_shadow}
\end{equation}
the asymptotically normalized shadow radius is
\begin{equation}
R_{\rm sh}
=
\lim_{r_o\rightarrow\infty}r_o\sin\alpha_{\rm sh}
=
\frac{r_{\rm ph}}{\sqrt{\delta A(r_{\rm ph})}}.
\label{eq:Rshadow_general}
\end{equation}
Substituting the explicit metric function, one obtains
\begin{equation}
R_{\rm sh}
=
\frac{r_{\rm ph}}
{
\sqrt{
1
-\frac{2M\delta}{r_{\rm ph}}
+\frac{\Qeff}{\delta r_{\rm ph}^2}
+\frac{\beta}{r_{\rm ph}}
\log\!\left(\frac{r_{\rm ph}}{|\beta|}\right)
}
}.
\label{eq:Rshadow_explicit_delta}
\end{equation}
In terms of the original parameters,
\begin{equation}
R_{\rm sh}
=
\frac{r_{\rm ph}}
{
\sqrt{
1
-\frac{2M(1-\alpha)}{r_{\rm ph}}
+\frac{Q^2e^{-\lmm}}{(1-\alpha)r_{\rm ph}^2}
+\frac{\beta}{r_{\rm ph}}
\log\!\left(\frac{r_{\rm ph}}{|\beta|}\right)
}
}.
\label{eq:Rshadow_explicit_original}
\end{equation}
The apparent shadow is circular because the spacetime is spherically symmetric. The celestial coordinates may therefore be written as
\begin{equation}
X^2+Y^2=R_{\rm sh}^2.
\label{eq:celestial_shadow_circle}
\end{equation}

\begin{figure*}[tbhp]
\centering

\begin{minipage}[t]{0.48\textwidth}
    \centering
    \includegraphics[height=5.8cm]{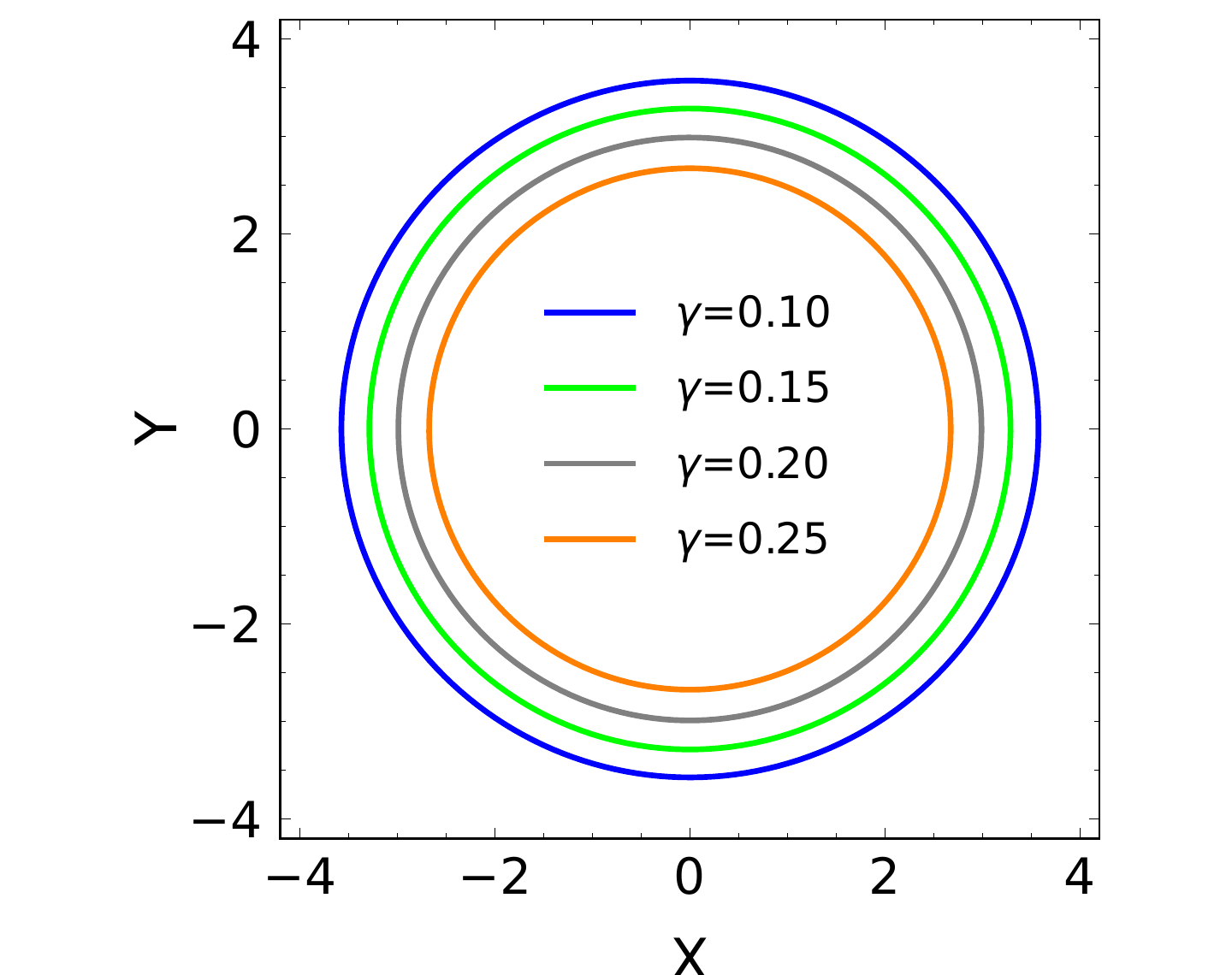}

    \vspace{0.15cm}
    \parbox{0.95\textwidth}{\centering\footnotesize
    (a) Variation of the KR/LV parameter, with the remaining parameters fixed.}
\end{minipage}
\hfill
\begin{minipage}[t]{0.48\textwidth}
    \centering
    \includegraphics[height=5.8cm]{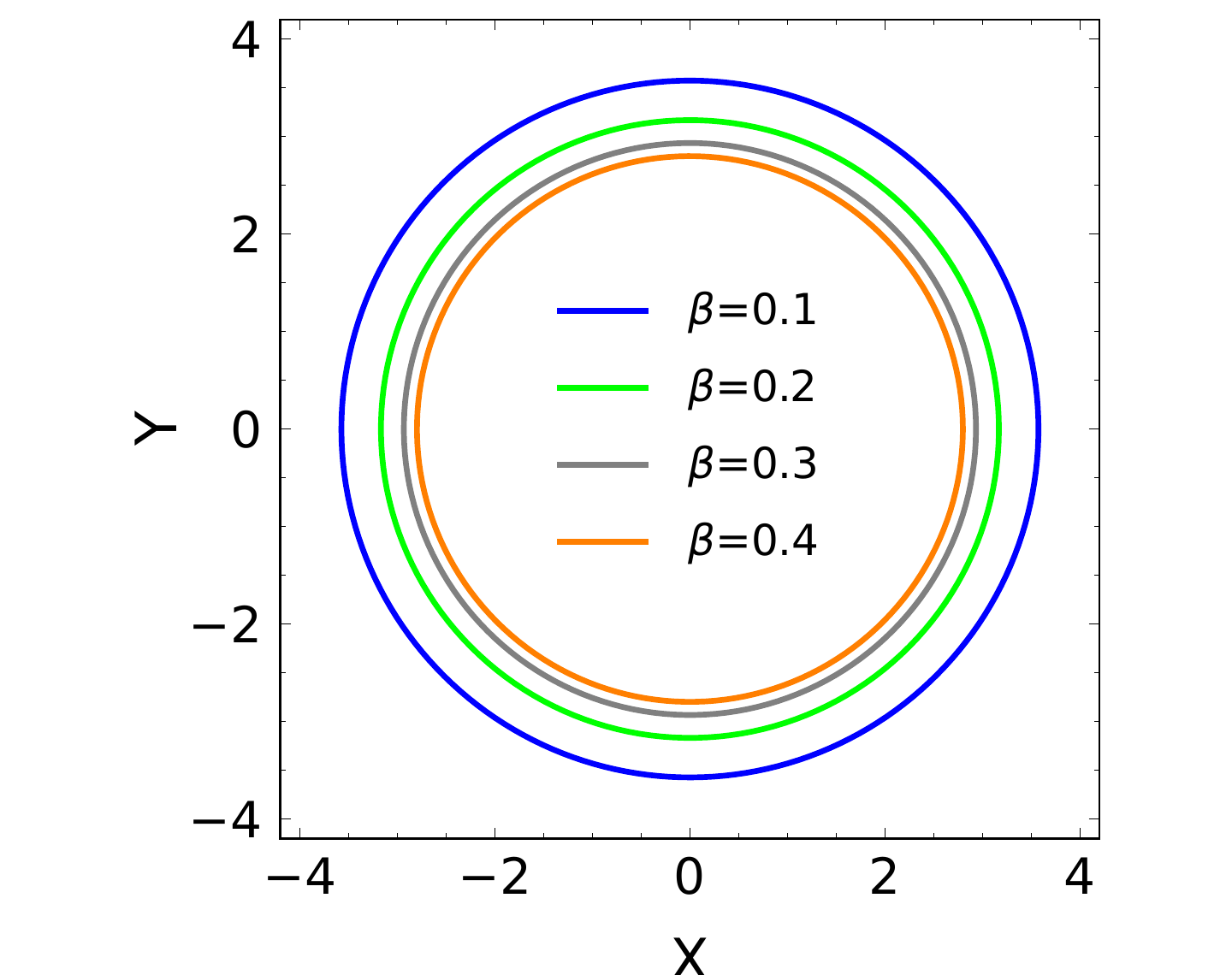}

    \vspace{0.15cm}
    \parbox{0.95\textwidth}{\centering\footnotesize
    (b) Variation of the PFDM parameter, with the remaining parameters fixed.}
\end{minipage}

\vspace{0.35cm}

\begin{minipage}[t]{0.48\textwidth}
    \centering
    \includegraphics[height=5.8cm]{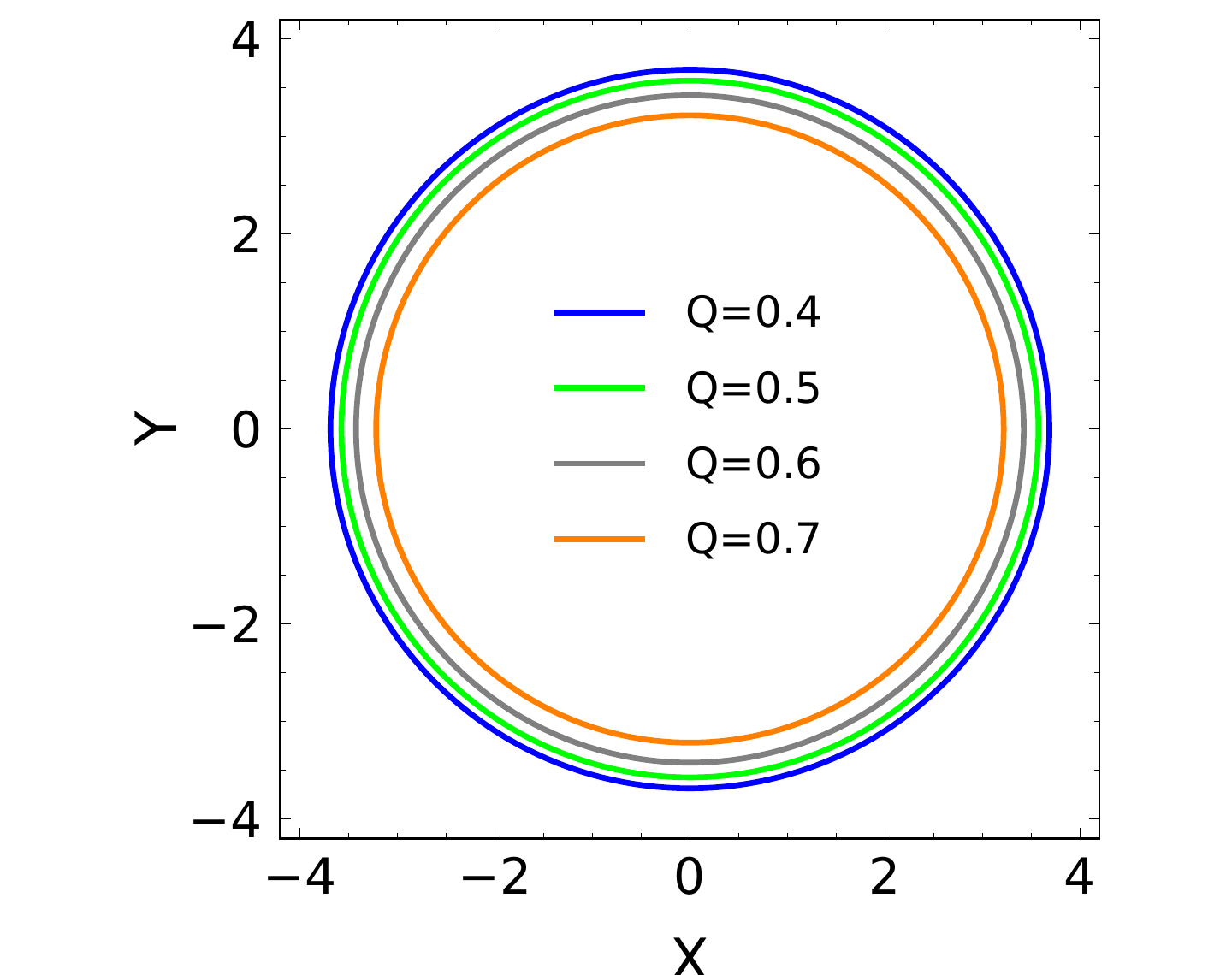}

    \vspace{0.15cm}
    \parbox{0.95\textwidth}{\centering\footnotesize
    (c) Variation of the electric charge, with the remaining parameters fixed.}
\end{minipage}
\hfill
\begin{minipage}[t]{0.48\textwidth}
    \centering
    \includegraphics[height=5.8cm]{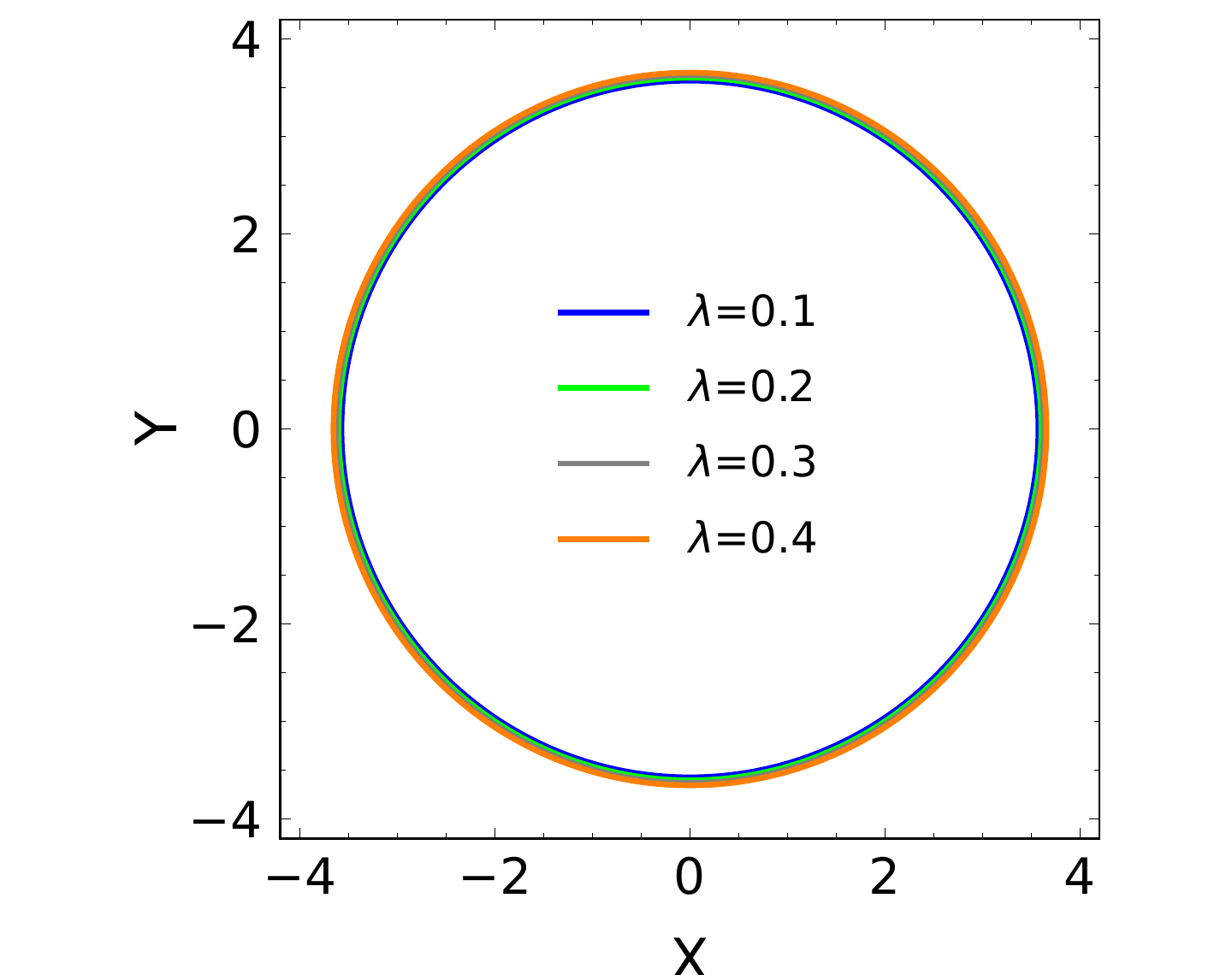}

    \vspace{0.15cm}
    \parbox{0.95\textwidth}{\centering\footnotesize
    (d) Variation of the ModMax parameter, with the remaining parameters fixed.}
\end{minipage}

\vspace{-0.2cm}

\caption{Apparent shadow profiles in the celestial plane $(X,Y)$ under variations of the model parameters. Panel (a) displays the effect of the KR/LV parameter, panel (b) the effect of the PFDM parameter, panel (c) the effect of the electric charge, and panel (d) the effect of the ModMax parameter. In the graphical legends, $\gamma\equiv\alpha$ and $\lambda\equiv\lambda_{\rm MM}$.}
\label{SHADOWprofiles}
\end{figure*}

In the Schwarzschild limit,
\begin{equation}
\alpha=0,
\qquad
Q=0,
\qquad
\beta=0,
\label{eq:schwarzschild_limit_shadow}
\end{equation}
one obtains
\begin{equation}
r_{\rm ph}=3M,
\qquad
R_{\rm sh}=3\sqrt{3}\,M,
\label{eq:schwarzschild_shadow_result}
\end{equation}
as expected. The same null geodesic equations can be used to study light deflection. From Eq.~\eqref{eq:null_radial_impact_shadow}, one obtains
\begin{equation}
\frac{d\phi}{dr}
=
\frac{b}{r^2}
\frac{1}{
\sqrt{
1-\dfrac{A(r)b^2}{r^2}
}
}.
\label{eq:dphi_dr_shadow}
\end{equation}
If $r_0$ denotes the closest approach distance, then
\begin{equation}
b^2=\frac{r_0^2}{A(r_0)}.
\label{eq:b_closest_approach_shadow}
\end{equation}
The total deflection angle is therefore
\begin{equation}
\hat{\alpha}
=
2\int_{r_0}^{\infty}
\frac{b\,dr}{r^2
\sqrt{
1-\dfrac{A(r)b^2}{r^2}
}
}
-\pi.
\label{eq:deflection_angle_shadow}
\end{equation}
The divergence of this integral as $r_0\rightarrow r_{\rm ph}$ signals the strong-deflection regime, where photons wind around the black hole before escaping to infinity.

\subsection{Physical role of the parameters}
\label{subsec:parameter_effects_shadow}

The geodesic structure and the black-hole shadow are controlled by the competition between the Kalb--Ramond parameter $\alpha$, the ModMax parameter $\lmm$, the electric charge $Q$, and the perfect fluid dark matter parameter $\beta$. The nonlinear electromagnetic sector enters through the effective squared charge
\begin{equation}
\Qeff=Q^2e^{-\lmm}.
\label{eq:screened_charge_shadow}
\end{equation}
Therefore, increasing $\lmm$ weakens the effective electric contribution. As a result, the geometry tends to approach the corresponding neutral configuration, and the photon-sphere radius moves toward the neutral Kalb-Ramond-dark-matter value.

The LV parameter appears through $\delta=1-\alpha$. It changes the asymptotic normalization of the metric, modifies the charge contribution, and shifts both the horizon and photon-sphere radii. Consequently, $\alpha$ has a direct influence on the apparent shadow radius through Eq.~\eqref{eq:Rshadow_explicit_original}.

The perfect fluid dark matter parameter $\beta$ introduces the logarithmic correction
\begin{equation}
\frac{\beta}{r}
\log\!\left(\frac{r}{|\beta|}\right),
\label{eq:pfdm_correction_shadow}
\end{equation}
which decays more slowly than the pure charge term. This correction is especially relevant near the photon sphere and may either increase or decrease the shadow radius depending on the sign and magnitude of $\beta$. Hence, the shadow provides a useful optical probe of the combined effects of nonlinear electrodynamics, Lorentz symmetry violation, and the surrounding dark matter distribution.

\section{Thermodynamics}\label{s4}

This section studies the thermodynamic behaviour of the solution discussed in
the previous section. It is useful to introduce the notation
\begin{equation}
 a\equiv 1-\alpha .
 \label{eq:a-def}
\end{equation}
Then, by solving the horizon condition for the mass, one obtains
\begin{equation}
\begin{split}
 M(r_h)
 &=
 \frac{e^{-\lmm}}{2r_h(\alpha-1)^2}
 \Bigg[
 Q^2
 -e^{\lmm} r_h(\alpha-1)
 \bigg(
 r_h+\beta\ln\frac{r_h}{|\beta|}
 \bigg)
 \Bigg] .
\end{split}
 \label{eq:mass-original}
\end{equation}

\begin{figure*}[ht!]
\centering

\begin{minipage}[t]{0.48\textwidth}
    \centering
    \includegraphics[height=4.8cm]{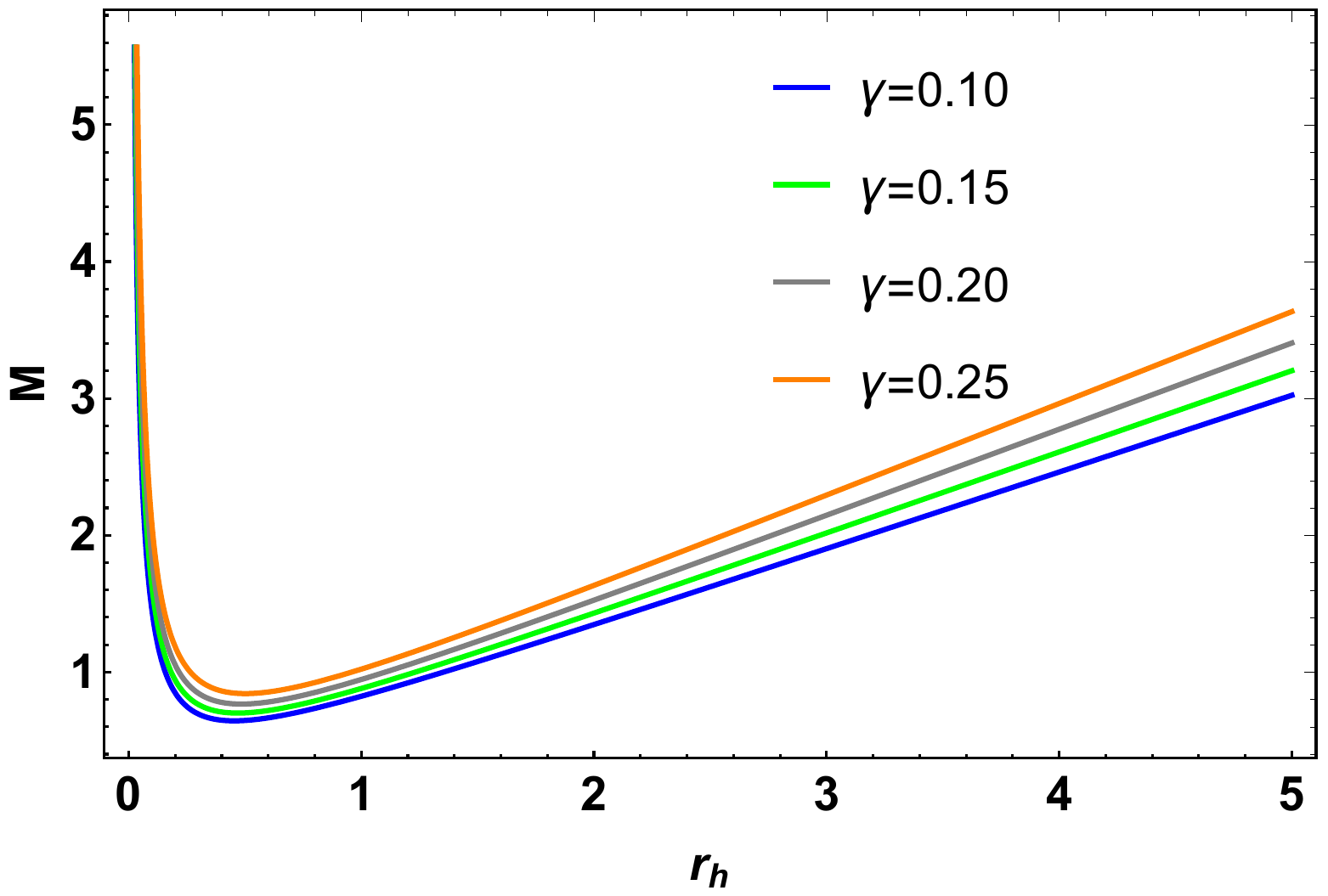}

    \vspace{0.15cm}
    \parbox{0.95\textwidth}{\centering\footnotesize
    (a) Variation of the KR/LV parameter, with the remaining parameters fixed.}
\end{minipage}
\hfill
\begin{minipage}[t]{0.48\textwidth}
    \centering
    \includegraphics[height=4.8cm]{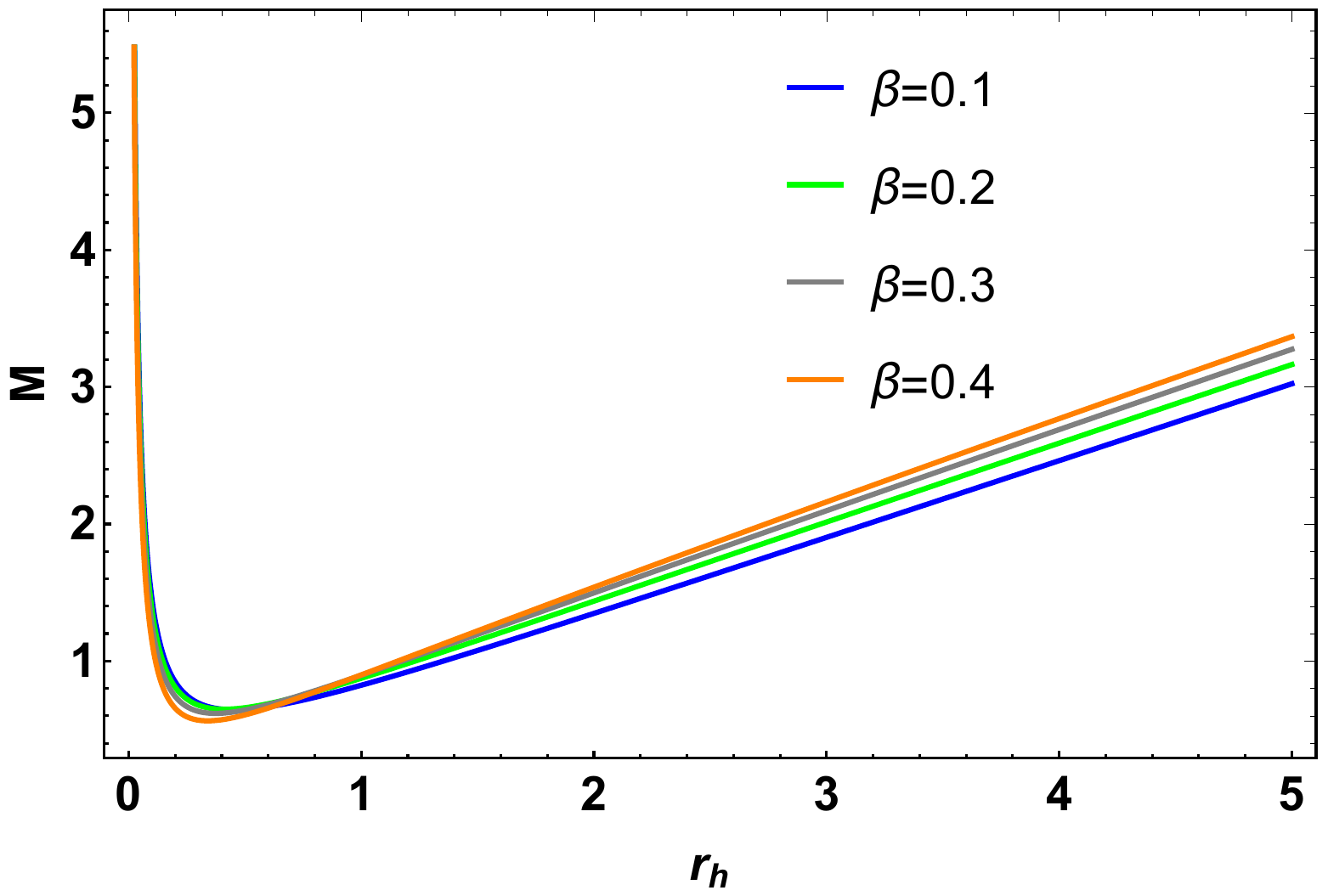}

    \vspace{0.15cm}
    \parbox{0.95\textwidth}{\centering\footnotesize
    (b) Variation of the PFDM parameter, with the remaining parameters fixed.}
\end{minipage}

\vspace{0.45cm}

\begin{minipage}[t]{0.48\textwidth}
    \centering
    \includegraphics[height=4.8cm]{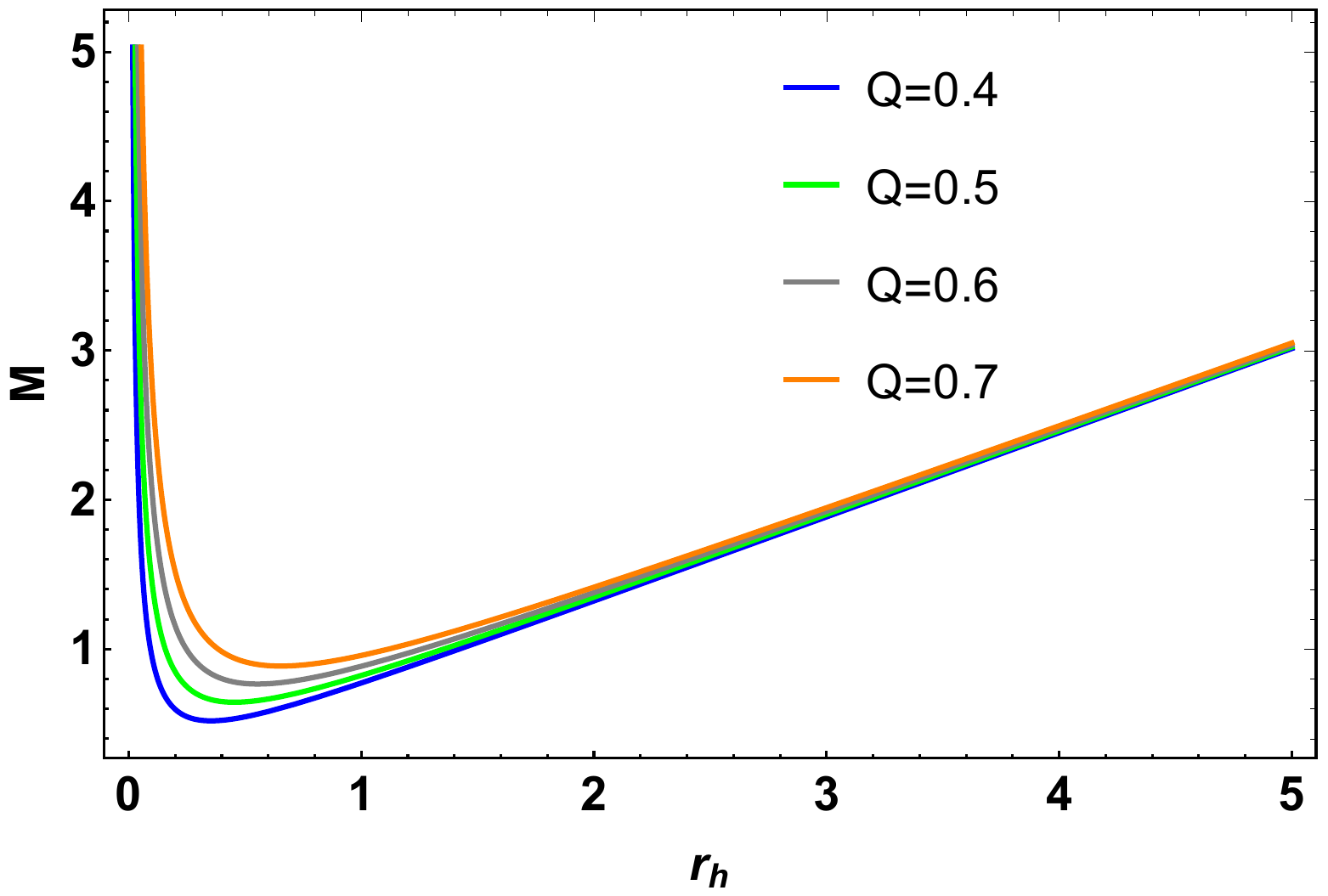}

    \vspace{0.15cm}
    \parbox{0.95\textwidth}{\centering\footnotesize
    (c) Variation of the electric charge, with the remaining parameters fixed.}
\end{minipage}
\hfill
\begin{minipage}[t]{0.48\textwidth}
    \centering
    \includegraphics[height=4.8cm]{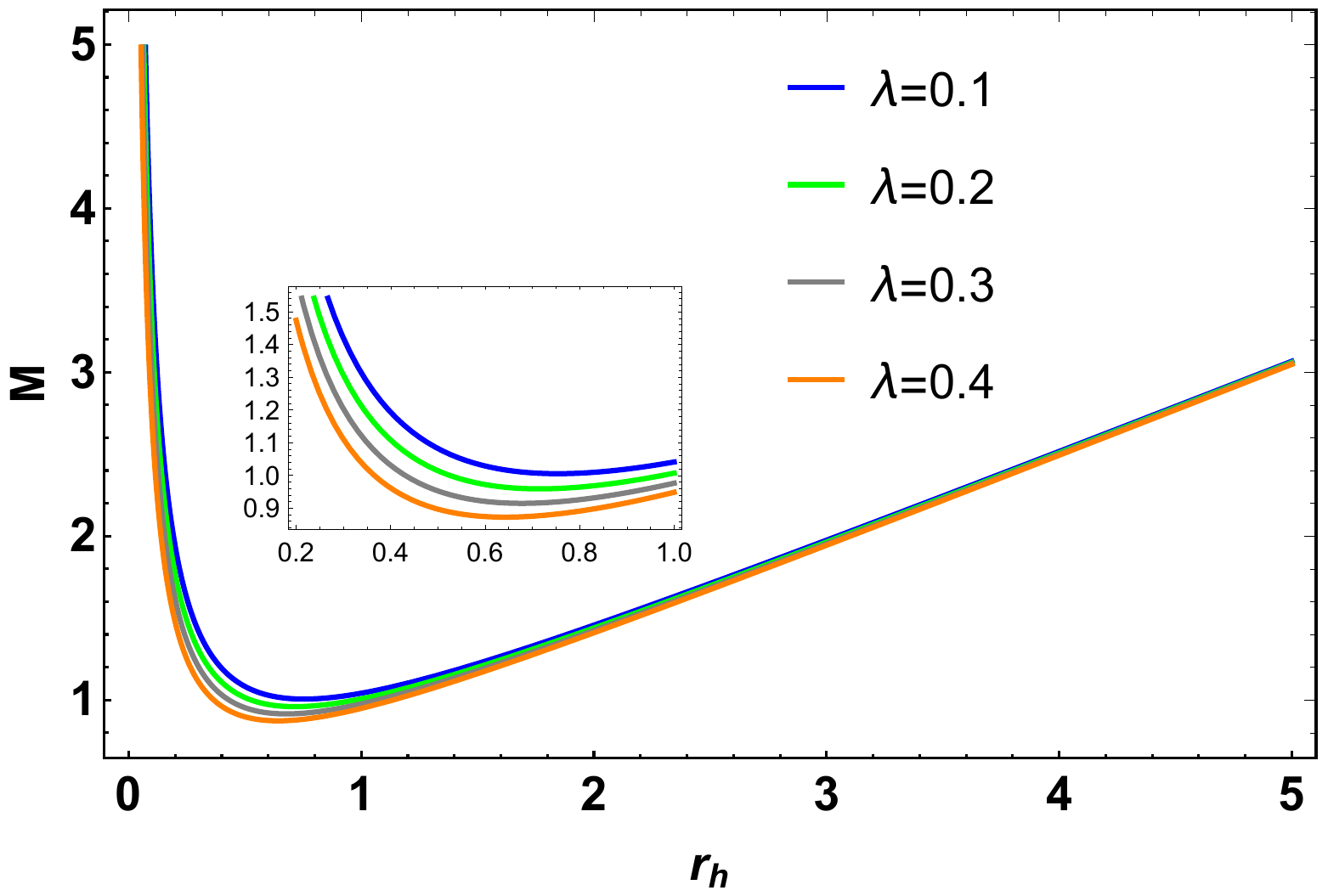}

    \vspace{0.15cm}
    \parbox{0.95\textwidth}{\centering\footnotesize
    (d) Variation of the ModMax parameter, with the remaining parameters fixed.}
\end{minipage}

\vspace{-0.2cm}

\caption{Behavior of the mass function $M(r_h)$ under variations of the model parameters. Panel (a) displays the effect of the KR/LV parameter, panel (b) the effect of the PFDM parameter, panel (c) the effect of the electric charge, and panel (d) the effect of the ModMax parameter. In the graphical legends, $\gamma\equiv\alpha$ and $\lambda\equiv\lambda_{\rm MM}$.}
\label{M}
\end{figure*}

Equivalently, this expression assumes the clearer form
\begin{equation}
 M(r_h)=
 \frac{r_h+\beta\ln(r_h/|\beta|)}{2a}
 +\frac{Q^2e^{-\lmm}}{2a^2r_h}.
 \label{eq:mass-simplified}
\end{equation}
This form makes the physics clear, where the first term contains the
geometrical contribution together with the dark-matter logarithmic correction,
whereas the second term is the effective charged contribution. The ModMax
parameter appears through $e^{-\lmm}$, so increasing $\lmm$ screens the
electric sector and reduces the charge contribution to the mass. On the other
hand, increasing $\alpha$ decreases $a=1-\alpha$, thereby enhancing both the
geometrical and charged parts of the mass. This explains why the mass curves in
the plotted profiles rise as the LV parameter is increased.

The dependence on $\beta$ is more subtle because the dark-matter term is
logarithmic. The combination $\beta\ln(r_h/|\beta|)$ can either increase or
decrease the mass depending on the relative size of $r_h$ and $\beta$. For
sufficiently large horizons, $r_h>\mathrm{e}\beta$, increasing $\beta$ raises
the mass. For very small horizons the logarithmic term can compete with the
charge term, although the $Q^2/r_h$ contribution dominates as $r_h\to0$
whenever $Q\neq0$. Therefore the charged solutions generally display a lower
bound in the mass profile. The condition $dM/dr_h=0$ implies
$a r_h(r_h+\beta)=Q^2e^{-\lmm}$, which marks the minimum-mass configuration.
In a thermodynamically consistent description where the horizon relation is
imposed before analyzing the temperature, this same point corresponds to the
extremal limit.

Figure~\ref{M} explicitly illustrates the parameter dependence of the horizon mass. The KR/LV parameter and the electric charge increase the mass scale required to support a horizon of fixed radius in the plotted range, while the PFDM contribution introduces a logarithmic deformation of the mass profile. The ModMax parameter acts through the screened charge $\Qeff$, so increasing $\lmm$ lowers the effective charged contribution and shifts the small-radius behavior of $M(r_h)$ accordingly.

Because the geometry is not asymptotically flat in the usual sense, since $f(r\to\infty)=1/(1-\alpha)$, the timelike Killing vector must be normalized before defining the surface gravity. With $a=1-\alpha$, the normalized surface gravity gives
\begin{equation}
 T_H=\frac{\sqrt{a}}{4\pi}f'(r_h),
 \label{eq:temperature-def}
\end{equation}
which gives
\begin{equation}
 T_H
 =
 \frac{\sqrt{a}}{4\pi}
 \left[
 \frac{2M}{r_h^2}
 -\frac{2Q^2e^{-\lmm}}{a^2r_h^3}
 +\frac{\beta}{a r_h^2}
 \left(1-\ln\frac{r_h}{|\beta|}\right)
 \right].
 \label{eq:H-temperature}
\end{equation}
This expression shows the usual competition between the gravitational part,
which tends to heat the black hole, and the charged part, which tends to cool
it. The term proportional to \(Q^2\) is negative in the numerator; therefore,
larger electric charge lowers the Hawking temperature and pushes the physical
positive-temperature branch to larger horizon radii. This behavior is the same
qualitative mechanism found in Reissner-Nordström-like black holes, where
charge supports an extremal remnant and suppresses Hawking emission.
\begin{figure*}[tbhp]
\centering

\begin{minipage}[t]{0.48\textwidth}
    \centering
    \includegraphics[height=4.8cm]{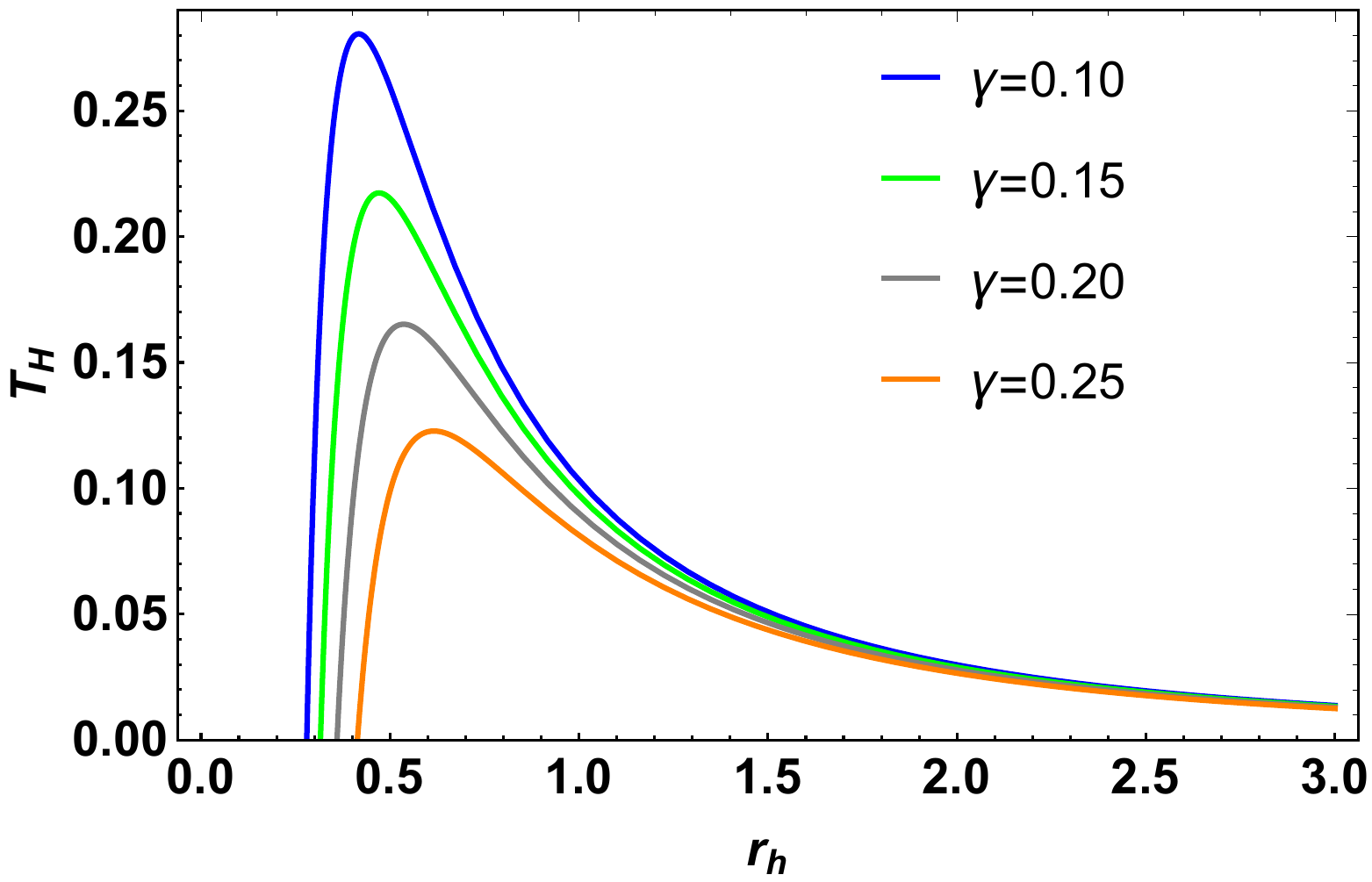}

    \vspace{0.15cm}
    \parbox{0.95\textwidth}{\centering\footnotesize
    (a) Variation of the KR/LV parameter, with the remaining parameters fixed.}
\end{minipage}
\hfill
\begin{minipage}[t]{0.48\textwidth}
    \centering
    \includegraphics[height=4.8cm]{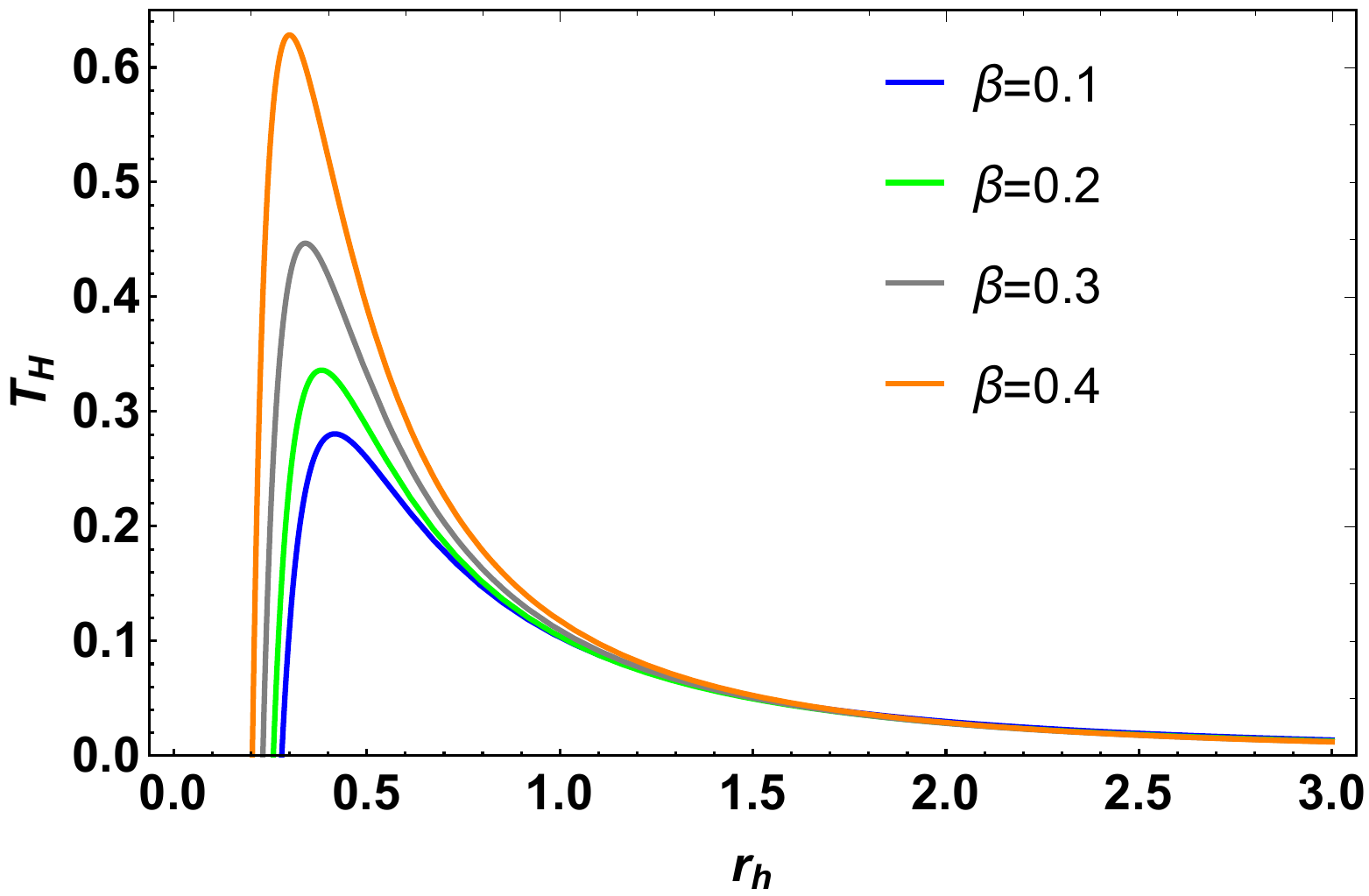}

    \vspace{0.15cm}
    \parbox{0.95\textwidth}{\centering\footnotesize
    (b) Variation of the PFDM parameter, with the remaining parameters fixed.}
\end{minipage}

\vspace{0.45cm}

\begin{minipage}[t]{0.48\textwidth}
    \centering
    \includegraphics[height=4.8cm]{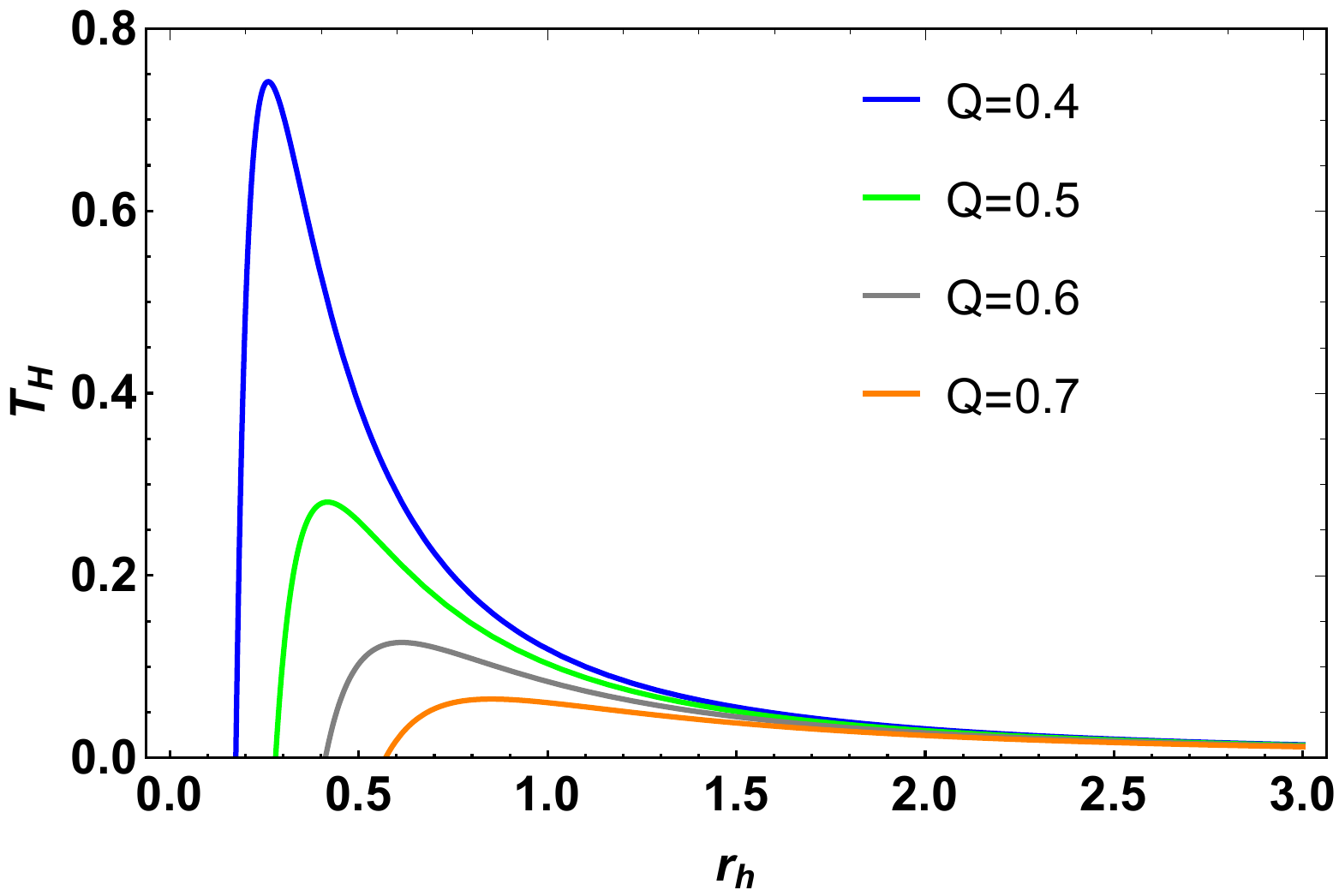}

    \vspace{0.15cm}
    \parbox{0.95\textwidth}{\centering\footnotesize
    (c) Variation of the electric charge, with the remaining parameters fixed.}
\end{minipage}
\hfill
\begin{minipage}[t]{0.48\textwidth}
    \centering
    \includegraphics[height=4.8cm]{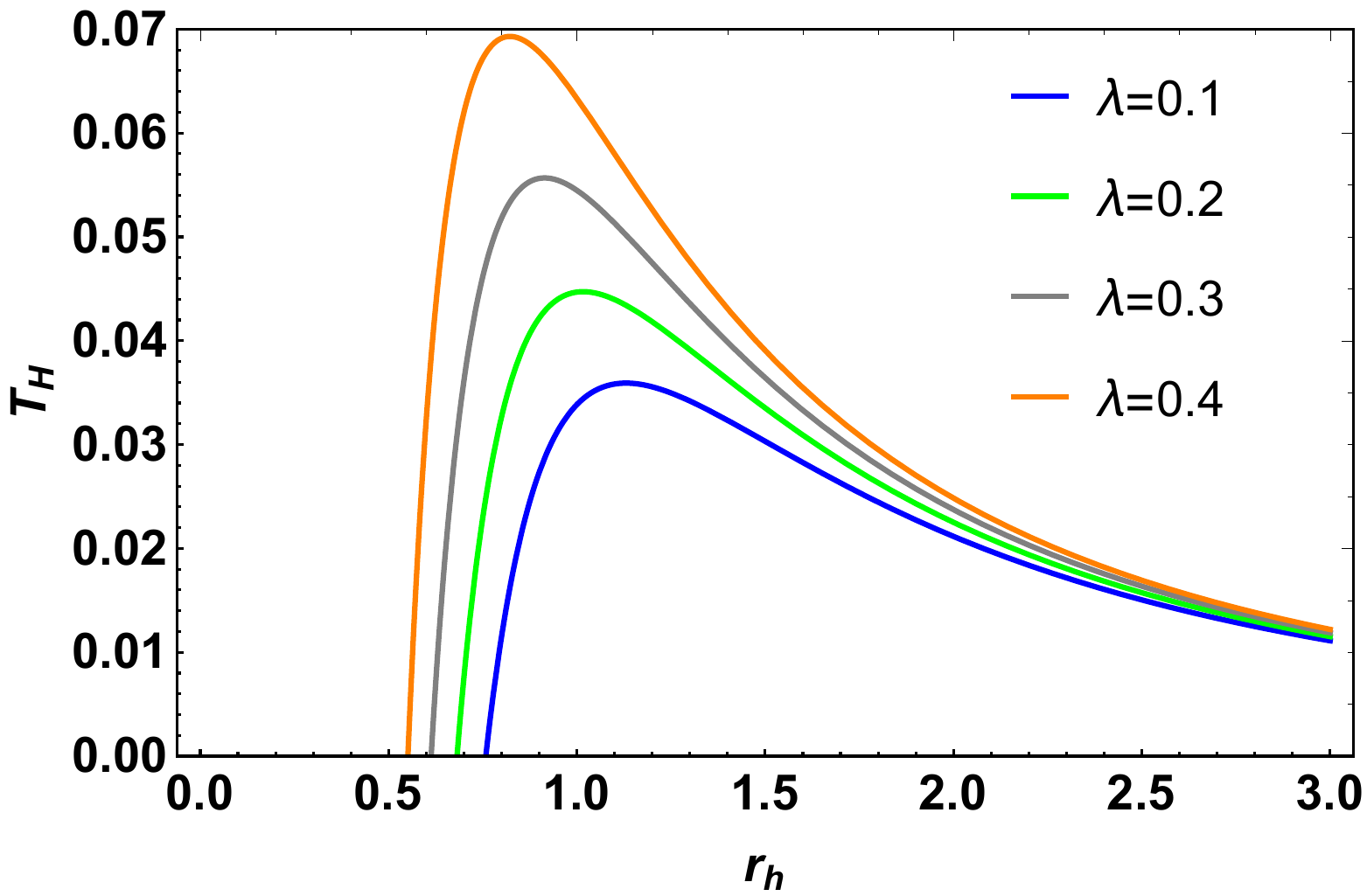}

    \vspace{0.15cm}
    \parbox{0.95\textwidth}{\centering\footnotesize
    (d) Variation of the ModMax parameter, with the remaining parameters fixed.}
\end{minipage}

\vspace{-0.2cm}

\caption{Behavior of the Hawking temperature $T_H$ under variations of the model parameters. Panel (a) displays the effect of the KR/LV parameter, panel (b) the effect of the PFDM parameter, panel (c) the effect of the electric charge, and panel (d) the effect of the ModMax parameter. In the graphical legends, $\gamma\equiv\alpha$ and $\lambda\equiv\lambda_{\rm MM}$.}
\label{TH}
\end{figure*}

If one substitutes the horizon mass \eqref{eq:mass-simplified} into
\eqref{eq:H-temperature}, the temperature reduces to
\begin{equation}
 T_H(r_h)=
 \frac{a r_h(r_h+\beta)-Q^2e^{-\lmm}}
 {4\pi a^{3/2} r_h^3},
\label{eq:temperature-simplified}
\end{equation}
with $a=1-\alpha$. This compact expression is useful for interpretation. The physical branch requires
\begin{equation}
 a r_h(r_h+\beta)>Q^2e^{-\lmm}.
 \label{eq:positive-temperature}
\end{equation}
At equality the temperature vanishes and the black hole becomes extremal. The
critical radius is
\begin{equation}
 r_{\rm ext}
 =
 \frac{-\beta+\sqrt{\beta^2+4Q^2e^{-\lmm}/a}}
 {2}.
 \label{eq:extremal-radius}
\end{equation}
Thus larger $Q$ and larger $\alpha$ increase the extremal radius, while larger
$\lmm$ decreases it because the effective charge is exponentially screened.
The PFDM parameter $\beta$ also shifts the onset of the positive-temperature
branch and modifies the height and position of the temperature maximum.

The temperature curves show that the black hole does not behave
as a simple Schwarzschild object. Instead, the system typically begins near a
cold or unphysical branch at small $r_h$, enters a positive-temperature region,
reaches a maximum temperature, and then cools as $r_h$ becomes large. For large
horizon radius, the leading behavior is
\begin{equation}
 T_H(r_h)\sim \frac{1}{4\pi\sqrt{1-\alpha}\,r_h},
 \qquad r_h\rightarrow\infty,
 \label{eq:large-r-temperature}
\end{equation}
up to subleading dark-matter and charge corrections. The evaporation is
therefore slowed for large black holes, while near extremality the temperature
vanishes and evaporation tends to stop.

Figure~\ref{TH} shows the nonmonotonic thermal profile associated with the charged configuration. The temperature vanishes at the extremal point, reaches a maximum at an intermediate horizon radius, and then decreases for larger black holes. Increasing $Q$ pushes the positive-temperature branch to larger $r_h$ and suppresses the emission, whereas increasing $\lmm$ weakens the effective charge and therefore shifts the temperature curves in the opposite direction. The KR/LV and PFDM sectors modify both the position and the height of the temperature maximum.

The heat capacity is the main diagnostic of local thermodynamic stability. It is
defined as follows
\begin{equation}
 C_Q= \frac{dM/dr_h}{dT_H/dr_h},
 \label{eq:heat-capacity-def}
\end{equation}
and the explicit result is
\begin{equation}
 C_Q=
 \frac{2\pi r_h^2
 \left[a r_h(r_h+\beta)-Q^2e^{-\lmm}\right]}
 {\sqrt{a}\,\left[-a r_h^2-2a\beta r_h+3Q^2e^{-\lmm}\right]} .
 \label{eq:heat-capacity}
\end{equation}
\begin{figure*}[tbhp]
\centering

\begin{minipage}[t]{0.48\textwidth}
    \centering
    \includegraphics[height=4.8cm]{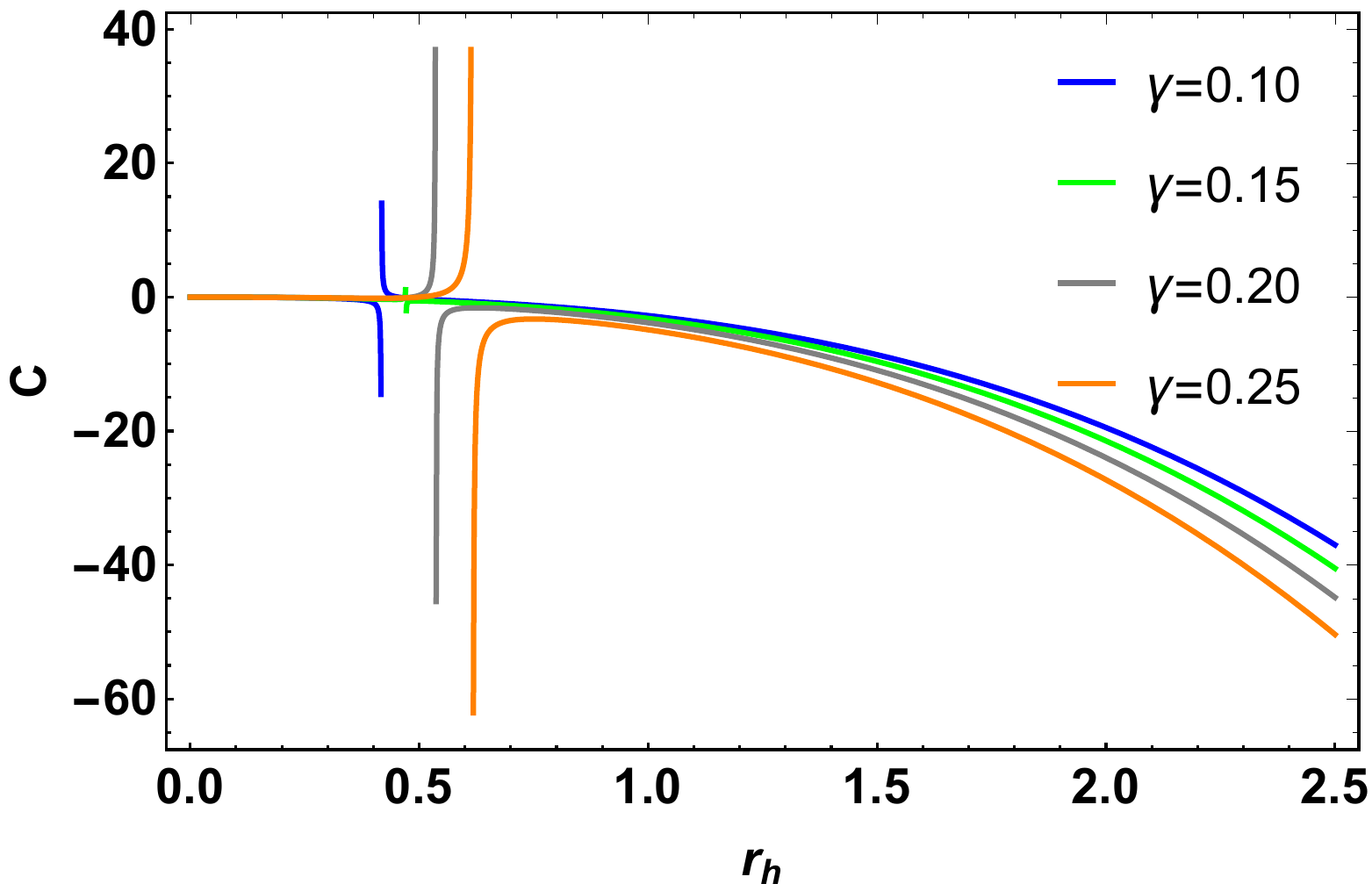}

    \vspace{0.15cm}
    \parbox{0.95\textwidth}{\centering\footnotesize
    (a) Variation of the KR/LV parameter, with the remaining parameters fixed.}
\end{minipage}
\hfill
\begin{minipage}[t]{0.48\textwidth}
    \centering
    \includegraphics[height=4.8cm]{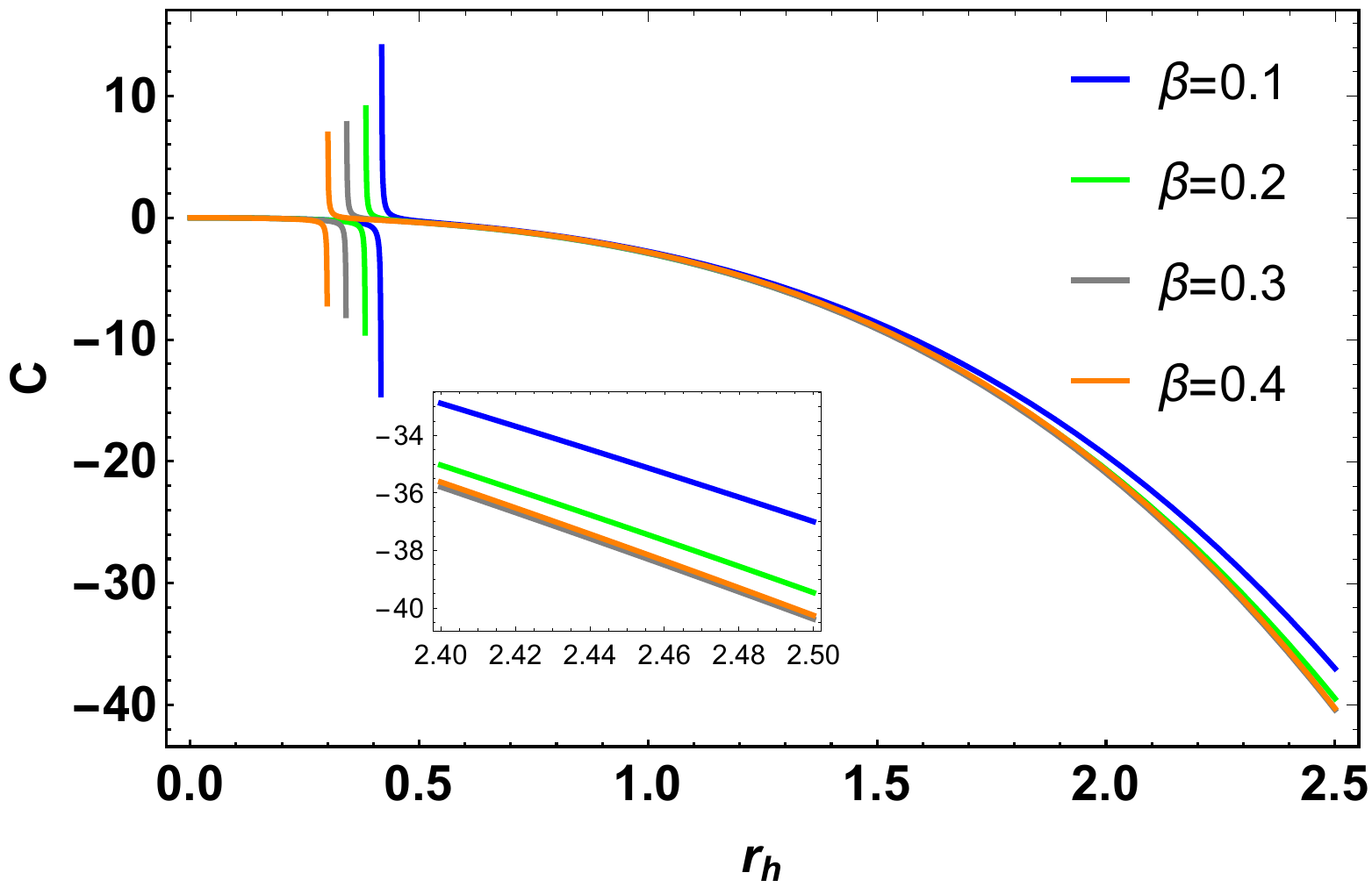}

    \vspace{0.15cm}
    \parbox{0.95\textwidth}{\centering\footnotesize
    (b) Variation of the PFDM parameter, with the remaining parameters fixed.}
\end{minipage}

\vspace{0.45cm}

\begin{minipage}[t]{0.48\textwidth}
    \centering
    \includegraphics[height=4.8cm]{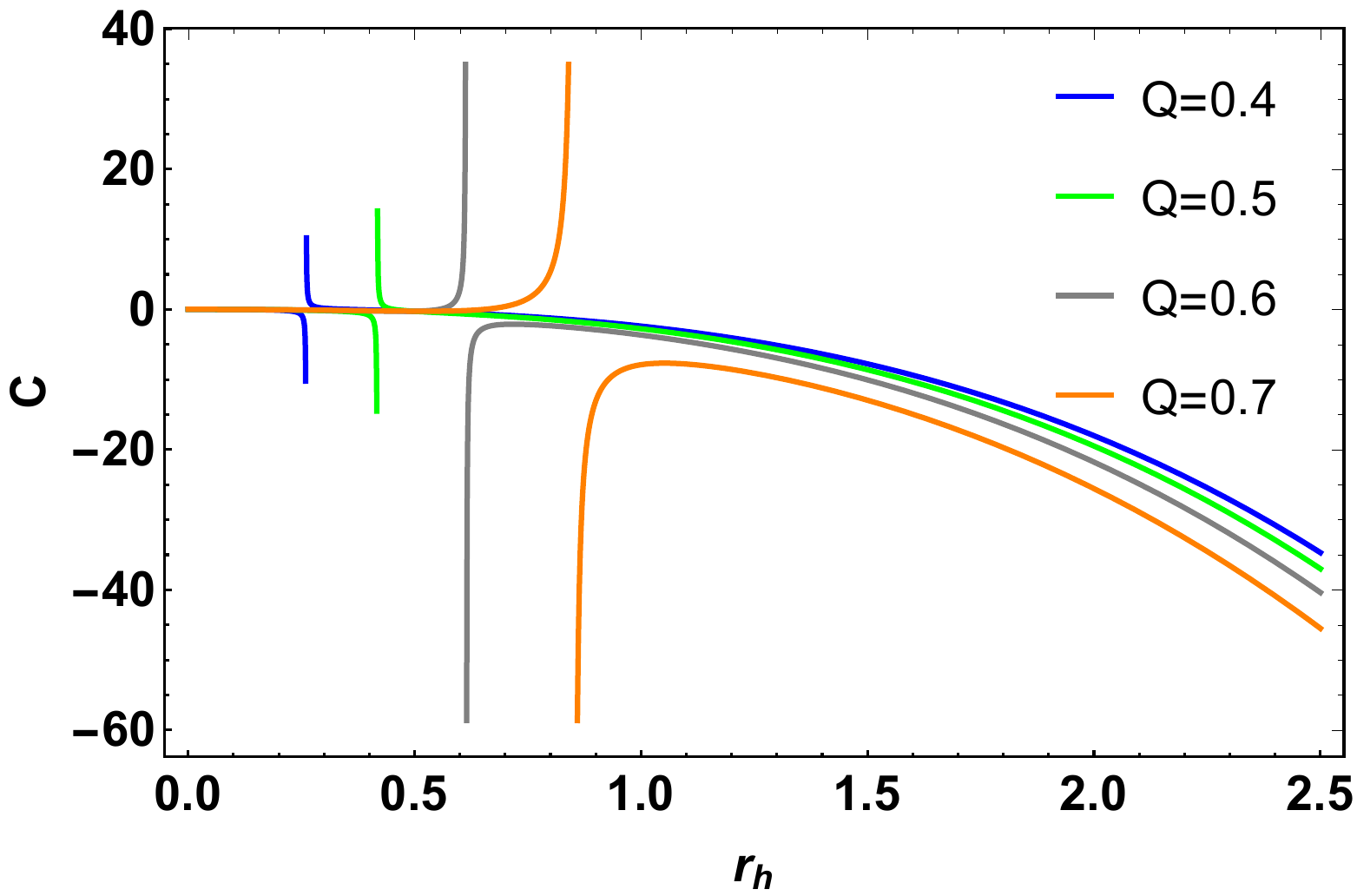}

    \vspace{0.15cm}
    \parbox{0.95\textwidth}{\centering\footnotesize
    (c) Variation of the electric charge, with the remaining parameters fixed.}
\end{minipage}
\hfill
\begin{minipage}[t]{0.48\textwidth}
    \centering
    \includegraphics[height=4.8cm]{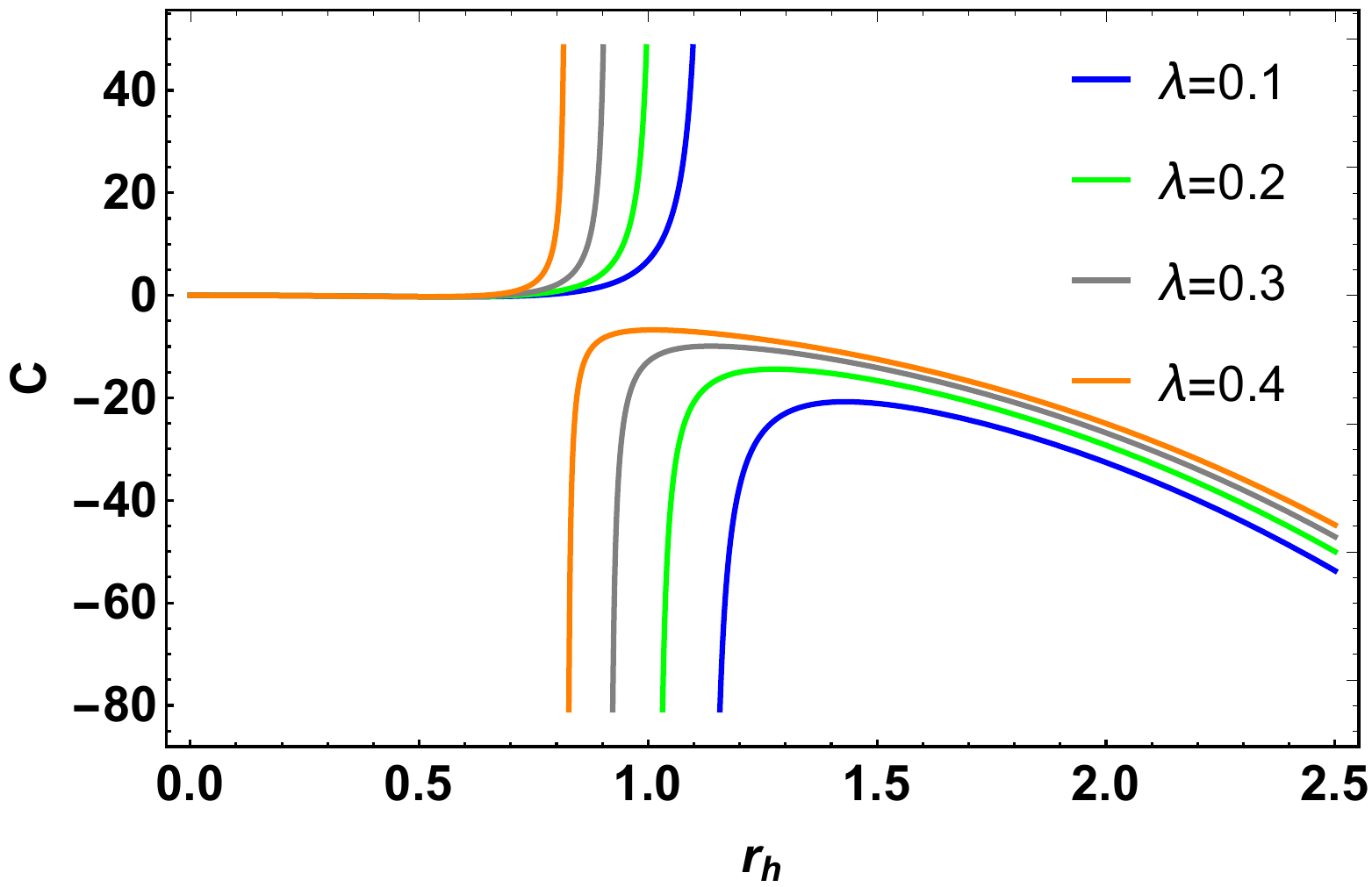}

    \vspace{0.15cm}
    \parbox{0.95\textwidth}{\centering\footnotesize
    (d) Variation of the ModMax parameter, with the remaining parameters fixed.}
\end{minipage}

\vspace{-0.2cm}

\caption{Behavior of the heat capacity $C_Q$ under variations of the model parameters. Panel (a) displays the effect of the KR/LV parameter, panel (b) the effect of the PFDM parameter, panel (c) the effect of the electric charge, and panel (d) the effect of the ModMax parameter. In the graphical legends, $\gamma\equiv\alpha$ and $\lambda\equiv\lambda_{\rm MM}$.}
\label{C}
\end{figure*}

\begin{figure*}[tbhp]
\centering

\begin{minipage}[t]{0.48\textwidth}
    \centering
    \includegraphics[height=4.8cm]{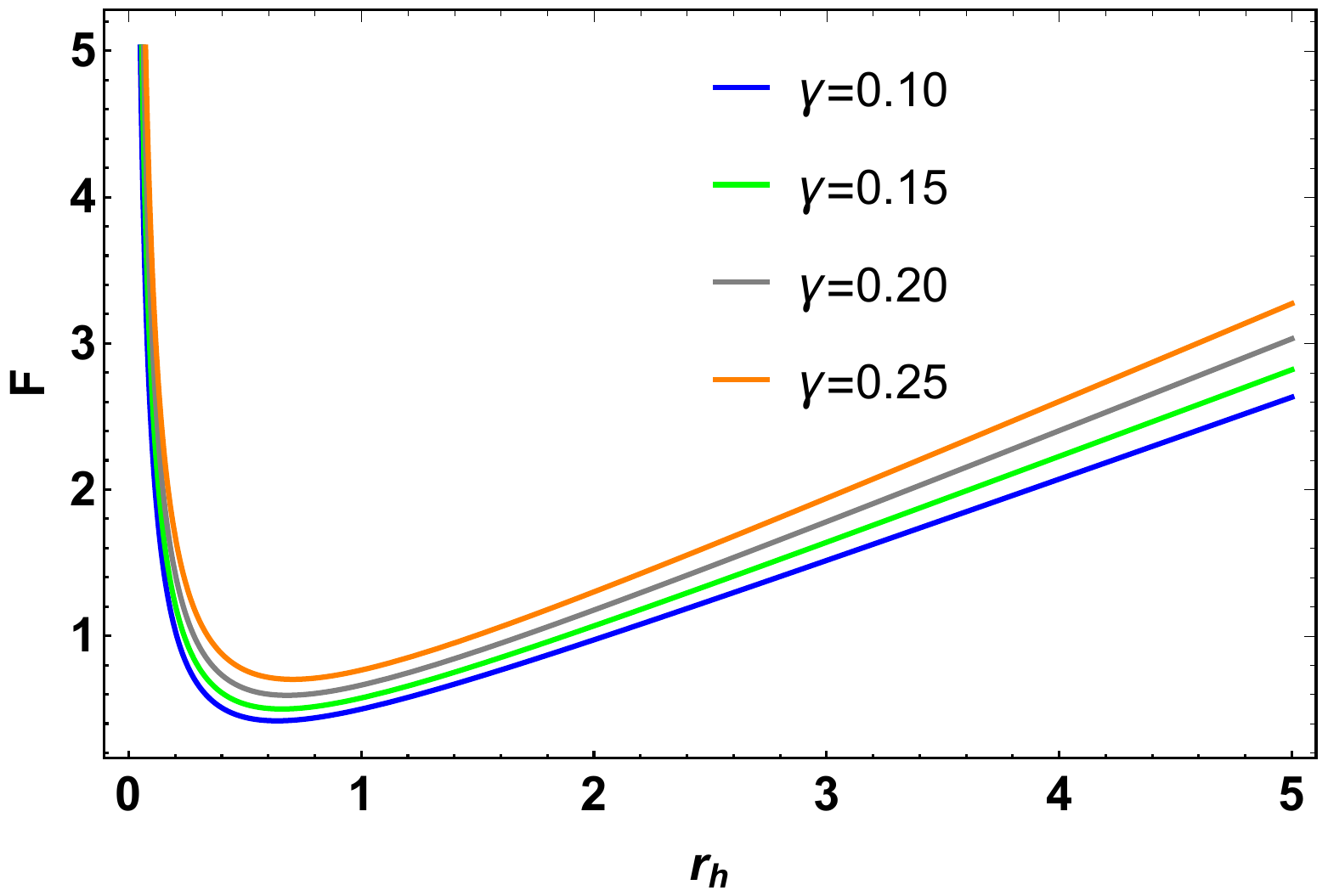}

    \vspace{0.15cm}
    \parbox{0.95\textwidth}{\centering\footnotesize
    (a) Variation of the KR/LV parameter, with the remaining parameters fixed.}
\end{minipage}
\hfill
\begin{minipage}[t]{0.48\textwidth}
    \centering
    \includegraphics[height=4.8cm]{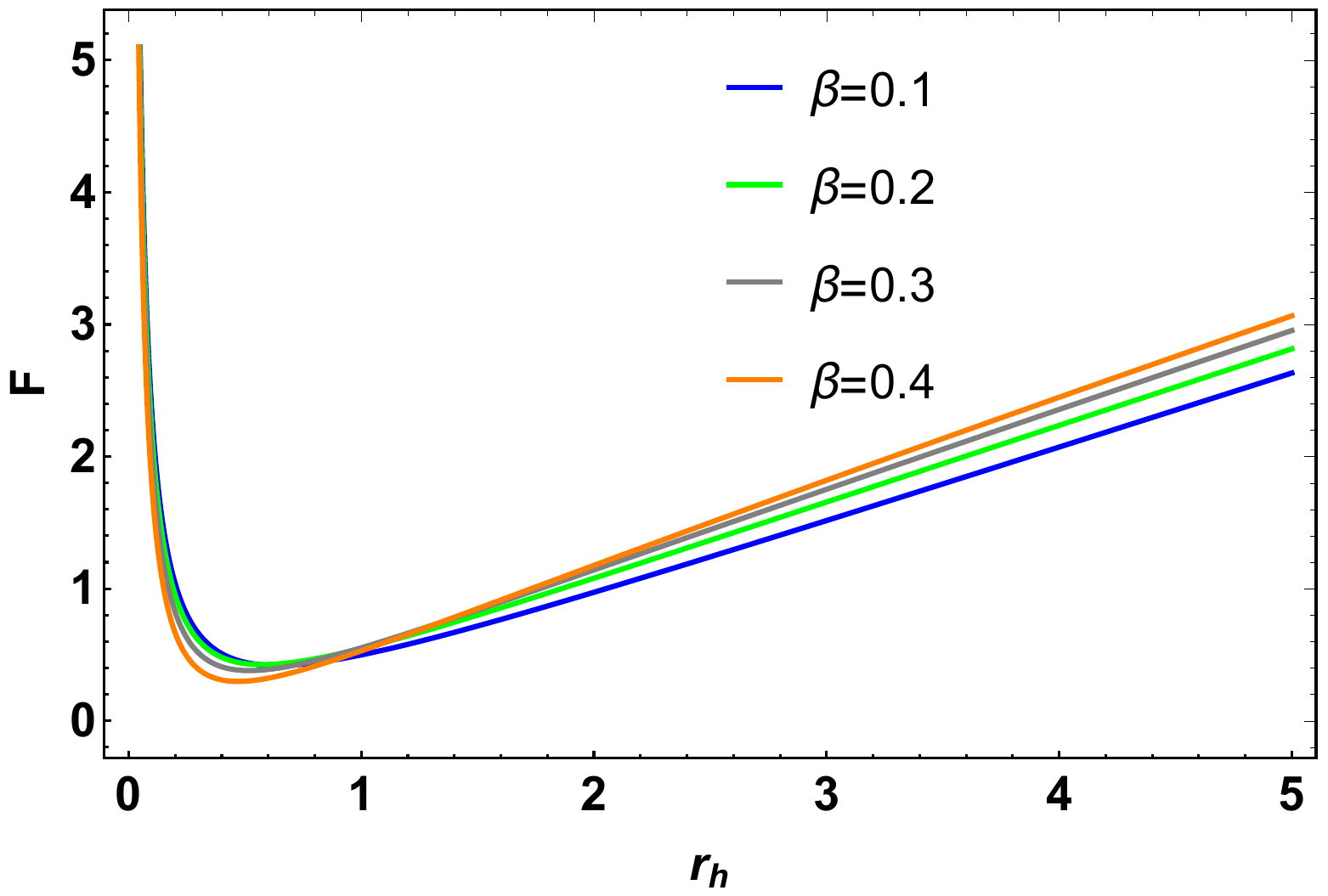}

    \vspace{0.15cm}
    \parbox{0.95\textwidth}{\centering\footnotesize
    (b) Variation of the PFDM parameter, with the remaining parameters fixed.}
\end{minipage}

\vspace{0.45cm}

\begin{minipage}[t]{0.48\textwidth}
    \centering
    \includegraphics[height=4.8cm]{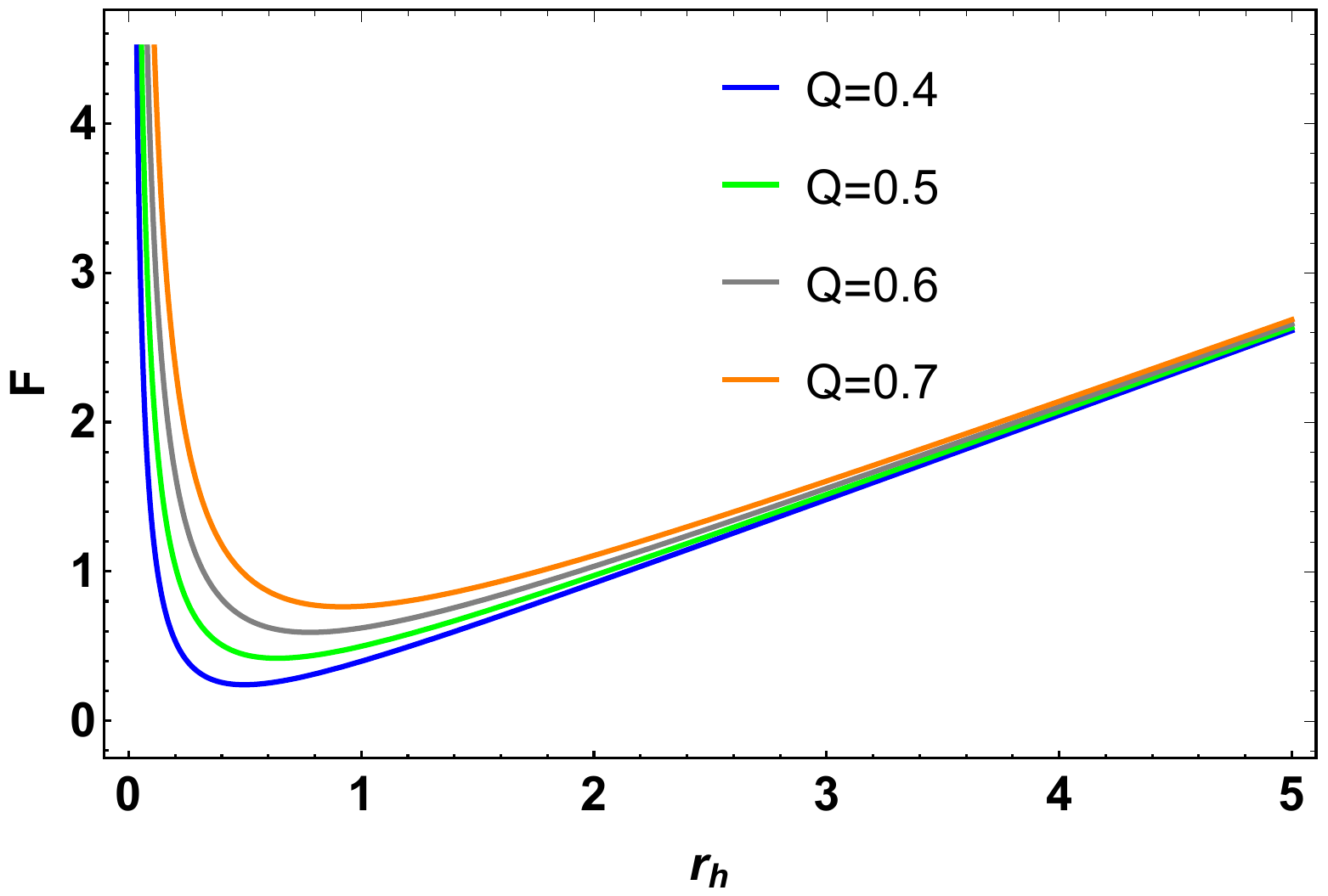}

    \vspace{0.15cm}
    \parbox{0.95\textwidth}{\centering\footnotesize
    (c) Variation of the electric charge, with the remaining parameters fixed.}
\end{minipage}
\hfill
\begin{minipage}[t]{0.48\textwidth}
    \centering
    \includegraphics[height=4.8cm]{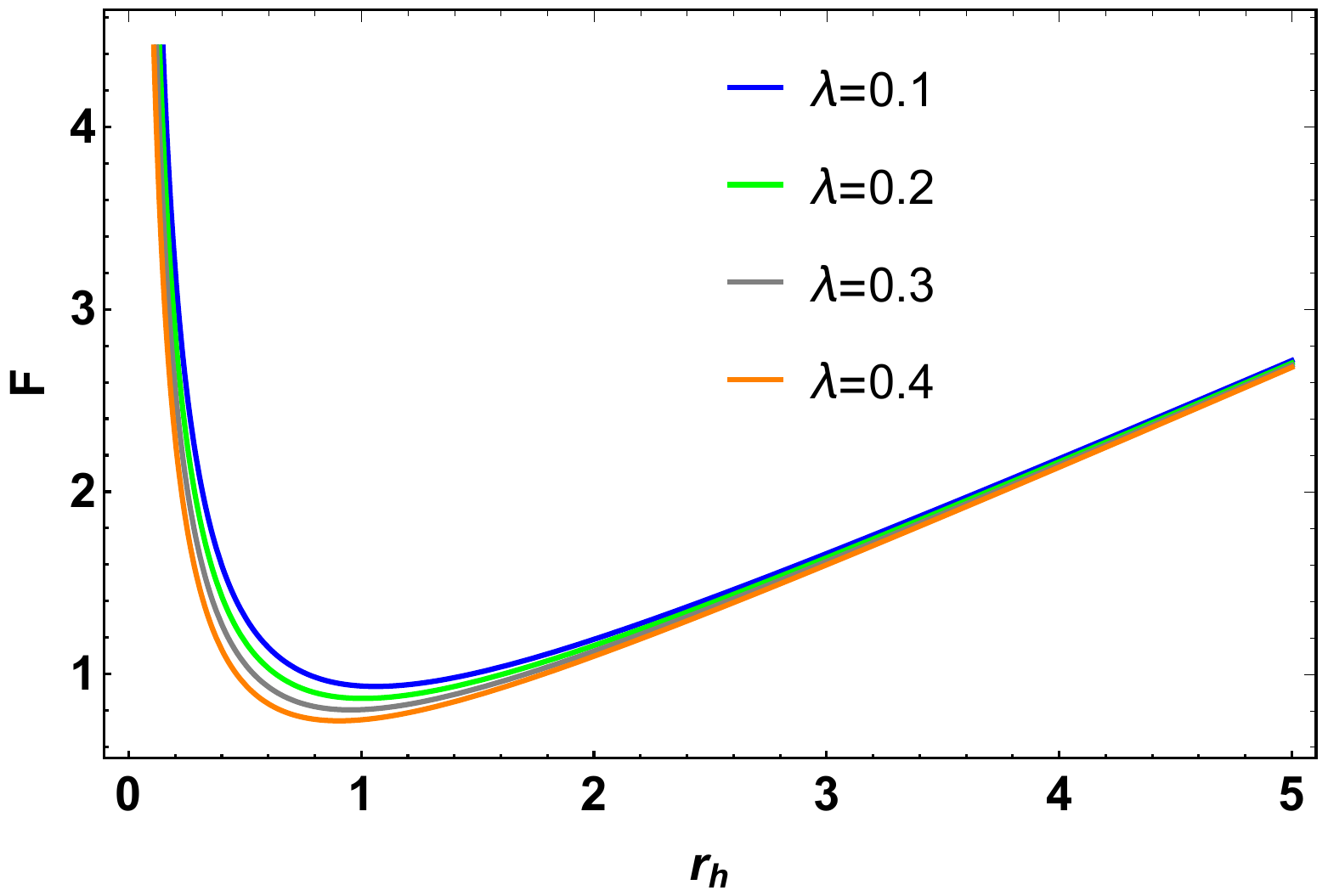}

    \vspace{0.15cm}
    \parbox{0.95\textwidth}{\centering\footnotesize
    (d) Variation of the ModMax parameter, with the remaining parameters fixed.}
\end{minipage}

\vspace{-0.2cm}

\caption{Behavior of the Helmholtz free energy $F$ under variations of the model parameters. Panel (a) displays the effect of the KR/LV parameter, panel (b) the effect of the PFDM parameter, panel (c) the effect of the electric charge, and panel (d) the effect of the ModMax parameter. In the graphical legends, $\gamma\equiv\alpha$ and $\lambda\equiv\lambda_{\rm MM}$.}
\label{F}
\end{figure*}
The sign of $C_Q$ separates locally stable and unstable thermodynamic branches.
When $C_Q>0$, the black hole heats up as energy is added and can be locally
stable in a canonical ensemble. When $C_Q<0$, the system has the usual
self-gravitating instability, so that adding energy lowers the temperature,
while losing energy raises it. The zeros of $C_Q$ occur when $dM/dr_h=0$, which
is associated with the extremal or minimum-mass configuration. The divergences
of $C_Q$ occur when
\begin{equation}
 -a r_h^2-2a\beta r_h+3Q^2e^{-\lmm}=0 .
 \label{eq:heat-capacity-pole}
\end{equation}
Equivalently, the positive critical radius is $r_\ast=-\beta+\sqrt{\beta^2+3Q^2e^{-\lmm}/a}$.
Such poles indicate possible second-order phase transitions, since the
temperature has an extremum there and the response function changes sign. The
plots of $C_Q$ in Fig.~\ref{C} should be read together with the temperature
plots. A peak of $T_H$ generally corresponds to a pole of the heat capacity. On
one side of the pole the black hole belongs to an unstable branch, while on the
other side it may enter a locally stable branch. Increasing the charge usually
shifts the pole to larger $r_h$, because the electric field delays the onset of
the thermally allowed region. Increasing $\lmm$ has the opposite tendency,
since it weakens the effective electric charge. Increasing $\alpha$ amplifies
the deformation through powers of $1/(1-\alpha)$, so the stability intervals
are displaced and the heat-capacity magnitude becomes more sensitive to the
horizon radius.

The entropy follows from the first law after the same normalization is used in the temperature. One obtains
\begin{equation}
 S=\frac{\pi r_h^2}{\sqrt{1-\alpha}},
 \label{eq:entropy-normalized}
\end{equation}
which reduces to the Bekenstein--Hawking area law when the KR parameter vanishes. At fixed charge and fixed external parameters $(\alpha,\beta,\lmm)$, the Helmholtz free energy considered here is therefore
\begin{equation}
 F=M-S T_H.
 \label{eq:free-energy-def}
\end{equation} The explicit
expression obtained reads
\begin{equation}
 F=
 \frac{
 a r_h^2+2a\beta r_h\ln(r_h/|\beta|)-a\beta r_h+3Q^2e^{-\lmm}}
 {4a^2r_h} .
 \label{eq:free-energy}
\end{equation}
The free energy measures global thermodynamic preference in the canonical
ensemble. Lower values of $F$ correspond to more favorable configurations at
fixed external parameters. The charge term is positive and dominant at small
$r_h$, so very small charged black holes are thermodynamically costly. For
large $r_h$, the geometrical contribution grows approximately linearly, with
logarithmic corrections from the PFDM sector. Consequently the free-energy
curve can develop a minimum at an intermediate radius, signalling the most
favourable black hole size within the family of solutions considered.

The parameter dependence of $F$ follows the same physical pattern observed in
the mass and temperature. Larger $Q$ raises the free energy, because more energy
is stored in the electromagnetic field. Larger $\lmm$ suppresses the
effective charge $Q^2e^{-\lmm}$, thereby lowering the free energy,
especially in the small-radius region where the charge term is strongest.
Larger $\alpha$ generally increases the thermodynamic cost of a black hole of
fixed radius, since the factors of $1/(1-\alpha)$ enhance the gravitational and
charged contributions. The PFDM parameter $\beta$ modifies the free energy
through the logarithmic term. Its effect is therefore not merely a vertical
shift since it changes the curvature of $F(r_h)$ and can move the preferred
radius.

Figure~\ref{F} displays the global thermodynamic preference encoded in the Helmholtz free energy. In the plotted range, the free-energy curves exhibit a minimum at finite horizon radius, identifying the most favorable configuration within the fixed-charge ensemble. The electric charge raises the small-radius free-energy cost, while the ModMax parameter reduces this cost by screening the effective charge. The KR/LV and PFDM parameters shift the location and depth of the minimum through the geometric and logarithmic contributions to $F(r_h)$.

Overall, this section shows a consistent physical picture, where the electric
charge produces a cold extremal sector and suppresses the Hawking temperature.
The ModMax parameter $\lmm$ acts as an exponential screening of this electric
contribution, making the black hole effectively less charged as $\lmm$
increases. The LV parameter $\alpha$ strengthens the deformation of the geometry
and enhances the thermodynamic quantities through powers of $1/(1-\alpha)$.
The PFDM parameter $\beta$ introduces a logarithmic correction that is
especially relevant away from the asymptotic large-radius regime. The combined
effect is a thermodynamic phase structure with extremal points, possible
heat-capacity divergences, and preferred intermediate configurations determined
by the minima of the Helmholtz free energy.

\section{Sparsity of the Hawking radiation}\label{s5}

We now analyze the sparsity of the Hawking radiation emitted by the ModMax black hole surrounded by perfect fluid dark matter in Kalb-Ramond gravity. 

Previously, we calculated the Hawking temperature given by \eqref{eq:H-temperature}. To proceed further, it is useful to define the dimensionless horizon factor
\begin{equation}
\Theta_h\equiv1+\frac{\beta}{r_h}-\frac{\Qeff}{\delta r_h^2}.
\label{eq:sparsity_Theta_h}
\end{equation}
Then, we write the Hawking temperature as follows
\begin{equation}
T_H=\frac{\Theta_h}{4\pi\sqrt{\delta}\,r_h}.
\label{eq:sparsity_T_Theta}
\end{equation}
The condition $\Theta_h>0$ corresponds to a positive-temperature black hole. The extremal limit is reached when $\Theta_h=0$, for which the Hawking temperature vanishes. The Hawking spectrum is not exactly blackbody because the emitted particles must propagate through an effective curvature potential outside the horizon. This produces greybody factors, which are crucial in scattering and absorption problems \cite{Sanchez:1976xm,Maldacena:1996ix,Klebanov:1997cx,Boonserm:2021owk}. For massless bosonic quanta, the number flux measured at infinity may be written as
\begin{equation}
\frac{d^2N}{d\tau_\infty d\omega}
=\frac{1}{2\pi}
\sum_{\ell=0}^{\infty}
(2\ell+1)\frac{\Gamma_\ell(\omega)}
{\exp(\omega/T_H)-1},
\label{eq:sparsity_exact_flux}
\end{equation}
where $\Gamma_\ell(\omega)$ is the greybody factor for the partial wave $\ell$. The total number emission rate is therefore
\begin{equation}
\dot{N}=
\frac{dN}{d\tau_\infty}
=\frac{1}{2\pi}
\sum_{\ell=0}^{\infty}
(2\ell+1)
\int_0^\infty
\frac{\Gamma_\ell(\omega)}
{\exp(\omega/T_H)-1}
\,d\omega .
\label{eq:sparsity_total_flux_exact}
\end{equation}

The sparsity of Hawking radiation is measured by comparing two time scales. The first one is the average time gap between consecutive emitted quanta, namely
\begin{equation}
\tau_{\rm gap}
=\frac{1}{\dot{N}}.
\label{eq:sparsity_time_gap}
\end{equation}
The second one is the characteristic oscillation time of a typical emitted quantum, namely
\begin{equation}
\tau_{\rm osc}=\frac{2\pi}{\omega_\ast},
\label{eq:sparsity_oscillation_time}
\end{equation}
where $\omega_\ast$ denotes the characteristic frequency of the Hawking spectrum. The sparsity parameter is then defined as follows
\begin{equation}
\eta\equiv\frac{\tau_{\rm gap}}{\tau_{\rm osc}}
=\frac{\omega_\ast}{2\pi\dot{N}}.
\label{eq:sparsity_parameter_definition}
\end{equation}
The radiation is sparse when $\eta\gg 1$, meaning that the typical time interval between two emitted particles is much larger than the oscillation period of each quantum. On the other hand, $\eta\lesssim 1$ would describe an almost continuous flux. A transparent analytic estimate can be obtained in the geometric-optics approximation. In this regime, the absorption cross section is controlled by the critical impact parameter of null geodesics. Since the shadow radius of the present spacetime is given by
\begin{equation}
R_{\rm sh}=\frac{r_{\rm ph}}{\sqrt{\delta A(r_{\rm ph})}},
\label{eq:sparsity_shadow_radius}
\end{equation}
the high-frequency absorption cross section is approximated by
\begin{equation}
\sigma_{\rm geo}=\pi R_{\rm sh}^2.
\label{eq:sparsity_geo_cross_section}
\end{equation}
The photon-sphere radius $r_{\rm ph}$ is determined by
\begin{equation}
r_{\rm ph}A'(r_{\rm ph})-2A(r_{\rm ph})=0,
\label{eq:sparsity_photon_condition}
\end{equation}
which gives
\begin{equation}
2r_{\rm ph}^2
-6M\delta r_{\rm ph}
+\frac{4\Qeff}{\delta}
+\beta r_{\rm ph}
\left[
3\log\!\left(\frac{r_{\rm ph}}{|\beta|}\right)-1
\right]
=0.
\label{eq:sparsity_photon_equation}
\end{equation}

In the geometric-optics approximation, the number emission rate becomes
\begin{equation}
\dot{N}_{\rm geo}=g\,\frac{\zeta(3)}{\pi}
R_{\rm sh}^2T_H^3,
\label{eq:sparsity_number_flux_geo}
\end{equation}
where $g$ denotes the number of internal degrees of freedom of the emitted massless field. For example, $g=1$ for a real scalar mode. For a bosonic number spectrum, the peak frequency is $\omega_\ast=x_NT_H$, where $x_N=2+W_0(-2e^{-2})
\simeq 1.59362$, and $W_0$ is the principal branch of the Lambert function. Hence the geometric-optics sparsity parameter is given by
\begin{equation}
\eta_{\rm geo}=\frac{x_N}{2g\zeta(3)R_{\rm sh}^2T_H^2}.
\label{eq:sparsity_eta_geo_general}
\end{equation}
This expression already displays the most important physical point: for fixed temperature, a larger optical cross section increases the number flux and therefore decreases the sparsity. Conversely, for fixed optical size, colder black holes radiate more sparsely. Using Eqs.~\eqref{eq:sparsity_T_Theta} and \eqref{eq:sparsity_shadow_radius}, we can write the sparsity parameter explicitly in terms of the horizon radius and photon-sphere radius. First, define
\begin{equation}
D_{\rm ph}
\equiv
\delta A(r_{\rm ph})
=1
-\frac{2M\delta}{r_{\rm ph}}
+\frac{\Qeff}{\delta r_{\rm ph}^2}
+\frac{\beta}{r_{\rm ph}}
\log\!\left(\frac{r_{\rm ph}}{|\beta|}\right).
\label{eq:sparsity_Dph}
\end{equation}
Then, we have
\begin{equation}
R_{\rm sh}^2
=\frac{r_{\rm ph}^2}{D_{\rm ph}}.
\label{eq:sparsity_Rshadow_square}
\end{equation}
Substitution into Eq.~\eqref{eq:sparsity_eta_geo_general} yields
\begin{equation}
\eta_{\rm geo}
=\frac{8\pi^2x_N}{g\zeta(3)}
\,\frac{
\delta r_h^2D_{\rm ph}
}{r_{\rm ph}^2\Theta_h^2}.
\label{eq:sparsity_eta_explicit_Dph}
\end{equation}
Equivalently, we can write
\begin{equation}
\eta_{\rm geo}
=\frac{8\pi^2x_N}{g\zeta(3)}
\,\frac{
(1-\alpha) r_h^2D_{\rm ph}
}{r_{\rm ph}^2
\left[1+\dfrac{\beta}{r_h}
-\dfrac{Q^2e^{-\lmm}}{(1-\alpha)r_h^2}
\right]^2}.
\label{eq:sparsity_eta_explicit_original}
\end{equation}

The quantity $D_{\rm ph}$ can be simplified by using the photon-sphere equation. From Eq.~\eqref{eq:sparsity_photon_equation}, one finds
\begin{equation}
D_{\rm ph}=\frac{1}{3}
\left[1+\frac{\beta}{r_{\rm ph}}
-\frac{\Qeff}{\delta r_{\rm ph}^2}
\right].
\label{eq:sparsity_Dph_simplified}
\end{equation}
Then if we define
\begin{equation}
\Theta_{\rm ph}\equiv 1+\frac{\beta}{r_{\rm ph}}
-\frac{\Qeff}{\delta r_{\rm ph}^2},
\label{eq:sparsity_Theta_ph}
\end{equation}
we obtain
\begin{equation}
R_{\rm sh}^2=\frac{3r_{\rm ph}^2}{\Theta_{\rm ph}},
\label{eq:sparsity_Rshadow_Theta}
\end{equation}
and the sparsity parameter becomes
\begin{equation}
\eta_{\rm geo}=\frac{8\pi^2x_N}{3g\zeta(3)}
\,\delta
\left(\frac{r_h}{r_{\rm ph}}\right)^2
\frac{\Theta_{\rm ph}}{\Theta_h^2}.
\label{eq:sparsity_eta_final}
\end{equation}
In terms of the original physical parameters, this reads
\begin{equation}
\eta_{\rm geo}=\frac{8\pi^2x_N}{3g\zeta(3)}
(1-\alpha)
\left(\frac{r_h}{r_{\rm ph}}\right)^2
\frac{
1+\dfrac{\beta}{r_{\rm ph}}
-\dfrac{Q^2e^{-\lmm}}{(1-\alpha)r_{\rm ph}^2}
}{\left[
1+\dfrac{\beta}{r_h}
-\dfrac{Q^2e^{-\lmm}}{(1-\alpha)r_h^2}
\right]^2}.
\label{eq:sparsity_eta_final_original}
\end{equation}

Equation~\eqref{eq:sparsity_eta_final_original} is the central result of this section. It shows that the sparsity is controlled by three ingredients: the ratio between the horizon and photon-sphere radii, the optical size of the black hole, and the square of the horizon temperature factor. Since $\eta_{\rm geo}\propto \Theta_h^{-2}$, the radiation becomes increasingly sparse as the black hole approaches the extremal configuration.

Let us verify the Schwarzschild limit. For $\alpha=0$, $Q=0$, and
$\beta=0$, one has
\begin{align}
&r_h=2M,\\
&r_{\rm ph}=3M,\\
&T_H=\frac{1}{8\pi M},\\
&R_{\rm sh}=3\sqrt{3}\,M.
\label{eq:sparsity_schwarzschild_quantities}
\end{align}
With this, we obtain
\begin{equation}
\Theta_h=\Theta_{\rm ph}=1.
\label{eq:sparsity_schwarzschild_theta}
\end{equation}
Therefore, Eq.~\eqref{eq:sparsity_eta_final} gives
\begin{equation}
\eta_{\rm geo}^{\rm Schw}=\frac{32\pi^2x_N}{27g\zeta(3)}.
\label{eq:sparsity_schwarzschild_eta}
\end{equation}
For a single scalar degree of freedom, $g=1$, this gives $\eta_{\rm geo}^{\rm Schw}\simeq 15.5$. Thus, even in the Schwarzschild limit, the Hawking flux is sparse rather than continuous. Greybody suppression generally reduces the number flux further and therefore increases the actual sparsity relative to the geometric-optics estimate.

Let us examine the near-extremal behavior. In this case, the extremal limit is determined by
\begin{equation}
\Theta_h=1+\frac{\beta}{r_h}
-\frac{\Qeff}{\delta r_h^2}=0.
\label{eq:sparsity_extremal_theta}
\end{equation}
Close to extremality, $\Theta_h\ll 1$, and therefore
\begin{equation}
T_H\propto \Theta_h,
\qquad
\dot{N}_{\rm geo}\propto T_H^3\propto \Theta_h^3,
\qquad
\eta_{\rm geo}\propto \frac{1}{T_H^2}\propto \frac{1}{\Theta_h^2}.
\label{eq:sparsity_near_extremal_scaling}
\end{equation}
Hence, $\lim_{\Theta_h\rightarrow 0}\eta_{\rm geo}\rightarrow \infty$.
This means that the radiation becomes infinitely sparse as the black hole approaches a cold extremal state. Physically, the average time gap between emitted quanta grows much faster than the characteristic oscillation time of each quantum.

The next step is to discuss the physical influence of the black hole parameters. The ModMax parameter $\lmm$ enters through the effective squared charge $\Qeff=Q^2e^{-\lmm}$. Thus, increasing $\lmm$ weakens the electromagnetic contribution. At fixed $r_h$ and $r_{\rm ph}$, this increases both $\Theta_h$ and $\Theta_{\rm ph}$. Since the sparsity depends on $\Theta_h^{-2}$, the dominant effect is usually a reduction of $\eta_{\rm geo}$: the black hole becomes hotter and the Hawking flux becomes less sparse. Conversely, increasing the electric charge $Q$ tends to lower the temperature and drives the system toward a more sparse emission regime.

On the other hand, the LV parameter $\alpha$ acts through $\delta=1-\alpha$. Its influence is twofold. First, it modifies the overall normalization of the Hawking temperature through the factor $1/\sqrt{\delta}$. Second, it changes the effective strength of the charge contribution through the combination $\Qeff/\delta$. As a result, the dependence of the sparsity on $\alpha$ is not purely monotonic in general. Away from extremality, the normalization effect may enhance the temperature, while near extremality the charge-sensitive term may dominate and strongly increase the sparsity.

The perfect fluid dark matter parameter $\beta$ contributes through the following terms
\begin{equation}
\frac{\beta}{r_h},
\qquad
\frac{\beta}{r_{\rm ph}},
\qquad
\beta\log\!\left(\frac{r}{|\beta|}\right).
\label{eq:sparsity_beta_terms}
\end{equation}
Therefore, $\beta$ affects both the thermal sector, through $\Theta_h$, and the optical sector, through $r_{\rm ph}$ and $R_{\rm sh}$. Positive values of $\beta$ tend to increase the horizon factor $\Theta_h$ when the horizon radius is held fixed, which can reduce the sparsity. However, the logarithmic correction also shifts the photon sphere and the shadow radius, so the complete effect must be determined from the coupled equations for $r_h$ and $r_{\rm ph}$. This is one of the main differences between perfect fluid dark matter and ordinary charge corrections: the dark matter term decays slowly and can remain relevant near the photon sphere.

Finally, the relation $\eta_{\rm geo}\propto\frac{1}{R_{\rm sh}^2T_H^2}$ summarizes the physical content of the result. A larger shadow radius increases the effective absorption cross section and therefore makes the Hawking flux less sparse. A lower Hawking temperature, on the other hand, strongly increases the sparsity. In the present geometry, the observed sparsity is therefore the result of a competition between the optical size of the black hole and the thermal suppression induced by charge, LV corrections, and the surrounding perfect fluid dark matter.

Moreover, we should note that the geometric-optics expression \eqref{eq:sparsity_eta_final_original} should be regarded as an analytic estimate. The exact sparsity requires the greybody factors $\Gamma_\ell(\omega)$ obtained from the wave equation of the emitted field in the black-hole background. Since greybody factors satisfy $0\leq \Gamma_\ell(\omega)\leq 1$, they suppress the total number flux relative to a perfect blackbody emitter. Consequently, we have $\dot{N}_{\rm exact}<\dot{N}_{\rm geo}$, which implies $\eta_{\rm exact}>\eta_{\rm geo}$, at least when the geometric-optics cross section provides the dominant high-frequency estimate. Thus, Eq.~\eqref{eq:sparsity_eta_final_original} gives a useful lower estimate for the sparsity. The true Hawking emission is expected to be even more intermittent, especially for low-frequency modes, where the effective potential barrier strongly suppresses transmission.

Therefore, with the analysis we performed here, we see that the Hawking radiation emitted by the ModMax black hole surrounded by perfect fluid dark matter in Kalb-Ramond gravity is generically sparse. The sparsity increases near extremality, grows when the effective charge contribution cools the black hole, and is modified by the LV and dark matter parameters through both the temperature and the shadow radius. 

\section{Scalar perturbations, absorption and greybody factor}\label{s6}

In this section, we will calculate and analyze the absorption cross section and greybody factor for massless scalar perturbation by considering the geometry described by \eqref{BH sol}.

For a massless scalar perturbation, after the usual separation of variables and the introduction of the tortoise coordinate
\begin{equation}
    \frac{dr_*}{dr}=\frac{1}{f(r)},
\end{equation}
the radial equation can be written in Schr\"odinger-like form, namely
\begin{equation}
\label{eq:schrodinger}
    \frac{d^2\Psi_{\omega\ell}}{dr_*^2}
    +\left[\omega^2-V_0(r)\right]\Psi_{\omega\ell}=0 .
\end{equation}
The spin-zero effective potential is given by
\begin{equation}
\label{eq:V0_general}
    V_0(r) =f(r)\left[
        \frac{\ell(\ell+1)}{r^2}
        +\frac{f'(r)}{r}\right],
\end{equation}
where
\begin{equation}
\label{eq:fprime}
    f'(r)=
    \frac{2M}{r^2}
    -\frac{2Q^2 e^{-\lmm}}{(1-\alpha)^2r^3}
    +\frac{\beta}{(1-\alpha)r^2}
      \left[
        1-\ln\!\left(\frac{r}{|\beta|}\right)
      \right].
\end{equation}

Thus, the potential is controlled by two pieces. The first one is the centrifugal barrier $\ell(\ell+1)f(r)/r^2$ and the second one is the curvature contribution $f(r)f'(r)/r$. For the plotted mode $\ell=1$, the centrifugal part is already significant, so the potential develops a single positive barrier outside the event horizon. It vanishes at the horizon, because $f(r_h)=0$, and it also decays at large radius as the inverse powers of $r$ dominate over the logarithmic correction. The numerical behavior is physically transparent. Increasing $\alpha$ raises and narrows the potential barrier. This happens because the LV factor increases the overall strength of the metric function and simultaneously reduces the outer horizon radius in the chosen parameter range. The scattering region is then effectively compressed closer to the black hole, and the maximum of $V_0$ becomes higher.

The PFDM parameter $\beta$ has a similar net effect in the plots. Although the derivative contribution in Eq.~\eqref{eq:fprime} contains the combination $1-\ln(r/\beta)$, the full potential depends on both $f$ and $f'$. For the displayed values, increasing $\beta$ raises the barrier and shifts its peak inward. In physical terms, the surrounding matter distribution modifies the exterior geometry in such a way that the scalar wave sees a stronger obstacle before reaching the horizon.

\begin{figure*}[tbhp]
\centering

\begin{minipage}[t]{0.48\textwidth}
    \centering
    \includegraphics[height=4.8cm]{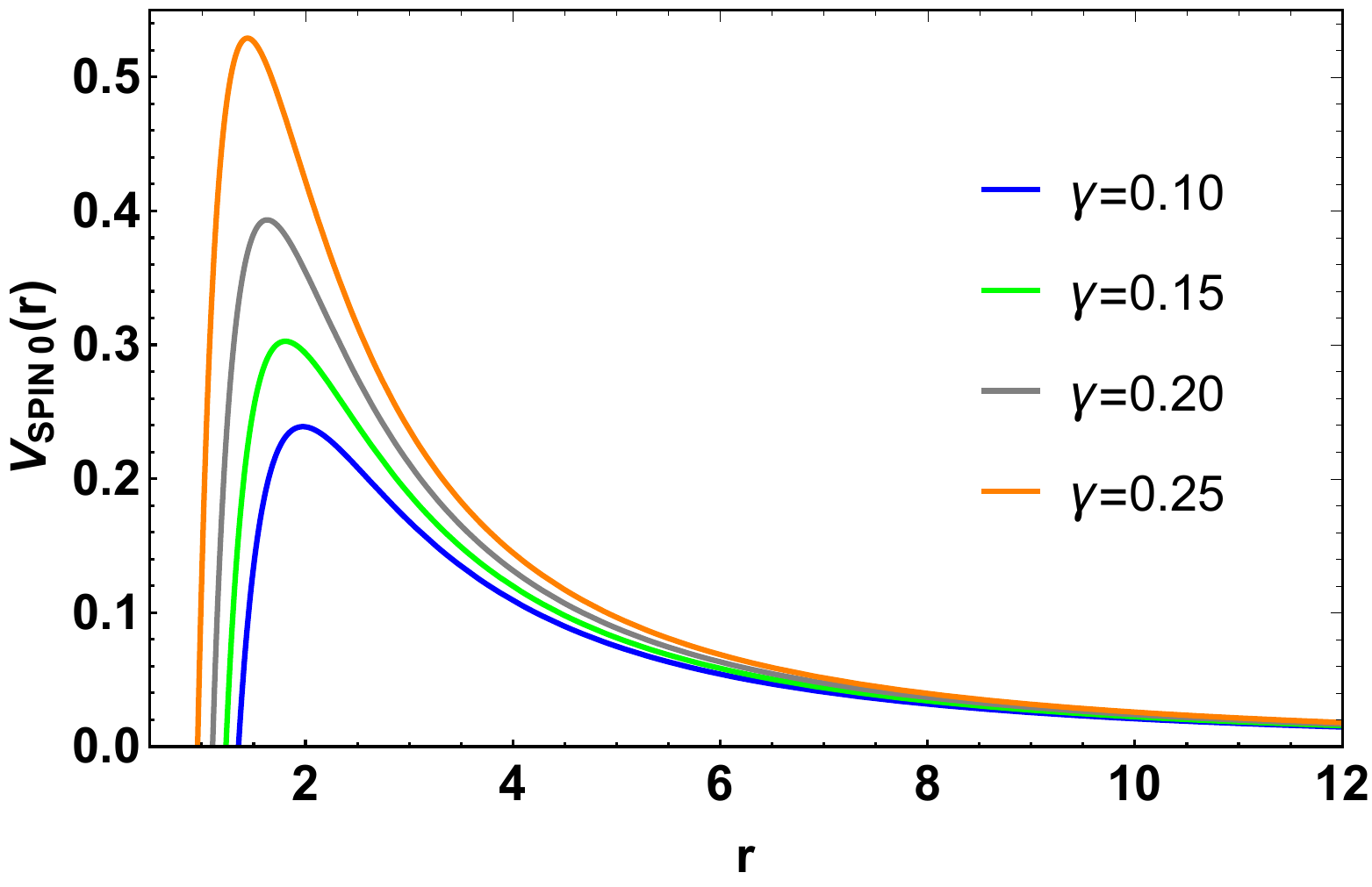}

    \vspace{0.15cm}
    \parbox{0.95\textwidth}{\centering\footnotesize
    (a) Variation of the KR/LV parameter, with the remaining parameters fixed.}
\end{minipage}
\hfill
\begin{minipage}[t]{0.48\textwidth}
    \centering
    \includegraphics[height=4.8cm]{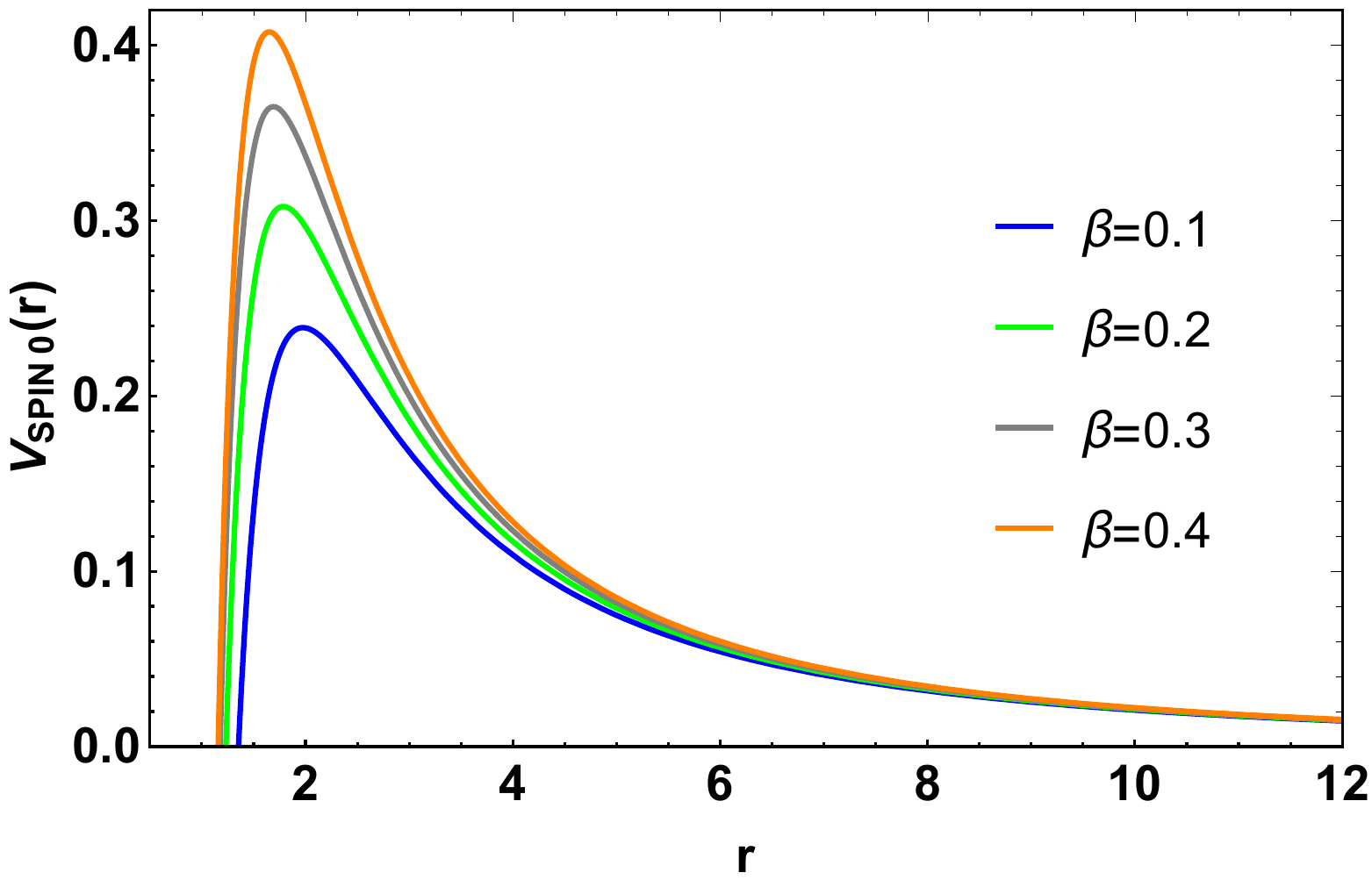}

    \vspace{0.15cm}
    \parbox{0.95\textwidth}{\centering\footnotesize
    (b) Variation of the PFDM parameter, with the remaining parameters fixed.}
\end{minipage}

\vspace{0.45cm}

\begin{minipage}[t]{0.48\textwidth}
    \centering
    \includegraphics[height=4.8cm]{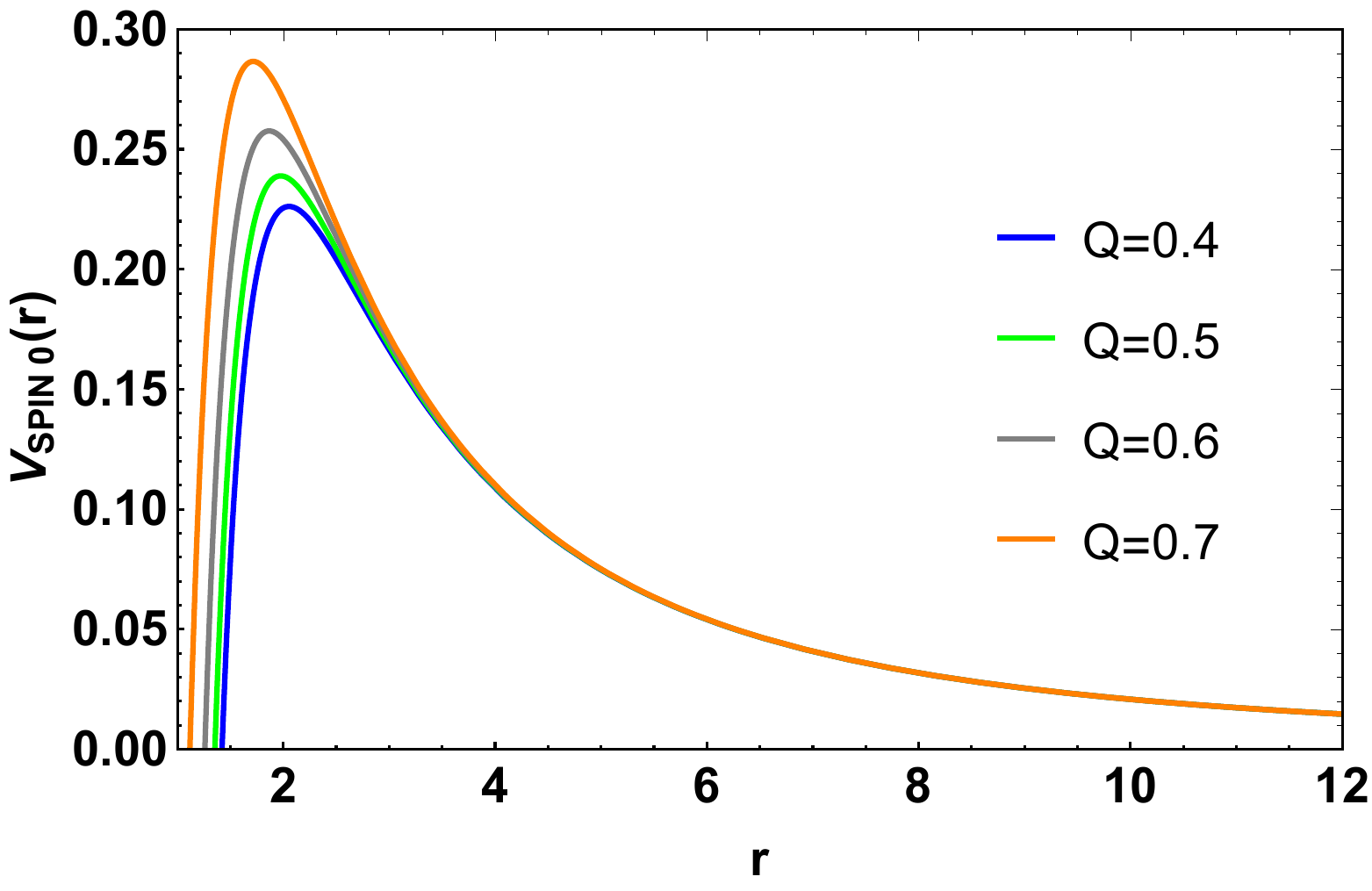}

    \vspace{0.15cm}
    \parbox{0.95\textwidth}{\centering\footnotesize
    (c) Variation of the electric charge, with the remaining parameters fixed.}
\end{minipage}
\hfill
\begin{minipage}[t]{0.48\textwidth}
    \centering
    \includegraphics[height=4.8cm]{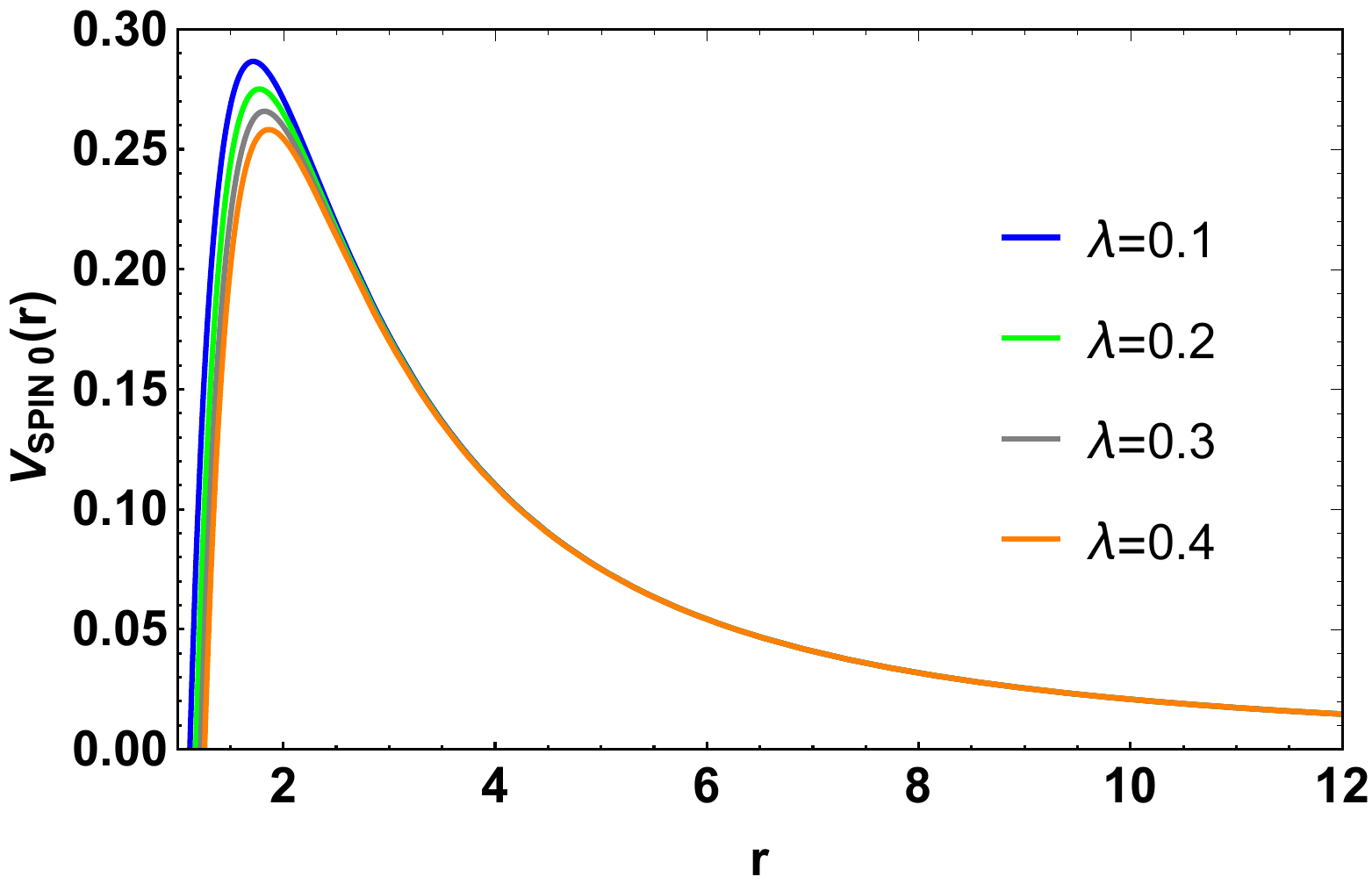}

    \vspace{0.15cm}
    \parbox{0.95\textwidth}{\centering\footnotesize
    (d) Variation of the ModMax parameter, with the remaining parameters fixed.}
\end{minipage}

\vspace{-0.2cm}

\caption{Behavior of the scalar effective potential $V_0(r)$ under variations of the model parameters. Panel (a) displays the effect of the KR/LV parameter, panel (b) the effect of the PFDM parameter, panel (c) the effect of the electric charge, and panel (d) the effect of the ModMax parameter. In the graphical legends, $\gamma\equiv\alpha$ and $\lambda\equiv\lambda_{\rm MM}$.}
\label{V}
\end{figure*}

The charge $Q$ also increases the height of the effective barrier. This is the same qualitative behavior familiar from charged black holes, where a larger charge term strengthens the near-horizon structure and produces a more reflective exterior potential. The ModMax parameter $\lmm$, however, works oppositely when it is positive. Since $Q$ appears through $Q^2e^{-\lmm}$, increasing $\lmm$ reduces the effective electromagnetic contribution. Consequently, the peak of the scalar potential decreases and shifts outward. In short, $Q$ enhances the barrier, whereas positive $\lmm$ screens the charge and lowers it. These features are displayed explicitly in Fig.~\ref{V}, where the changes in the height and position of the scalar effective-potential barrier are shown for variations of the KR/LV parameter, the PFDM parameter, the electric charge, and the ModMax parameter.

\begin{figure*}[tbhp]
\centering

\begin{minipage}[t]{0.48\textwidth}
    \centering
    \includegraphics[height=4.8cm]{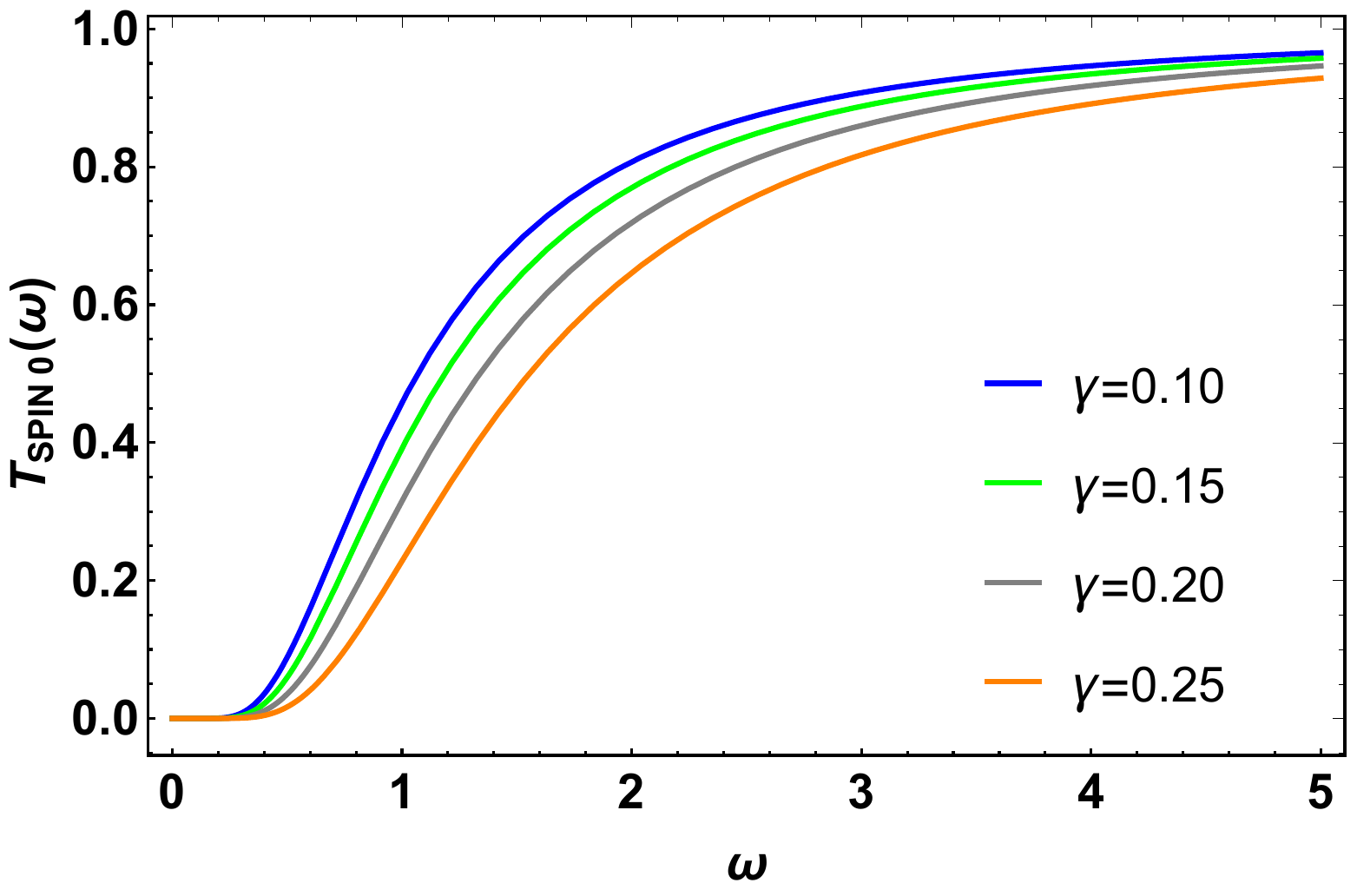}

    \vspace{0.15cm}
    \parbox{0.95\textwidth}{\centering\footnotesize
    (a) Variation of the KR/LV parameter, with the remaining parameters fixed.}
\end{minipage}
\hfill
\begin{minipage}[t]{0.48\textwidth}
    \centering
    \includegraphics[height=4.8cm]{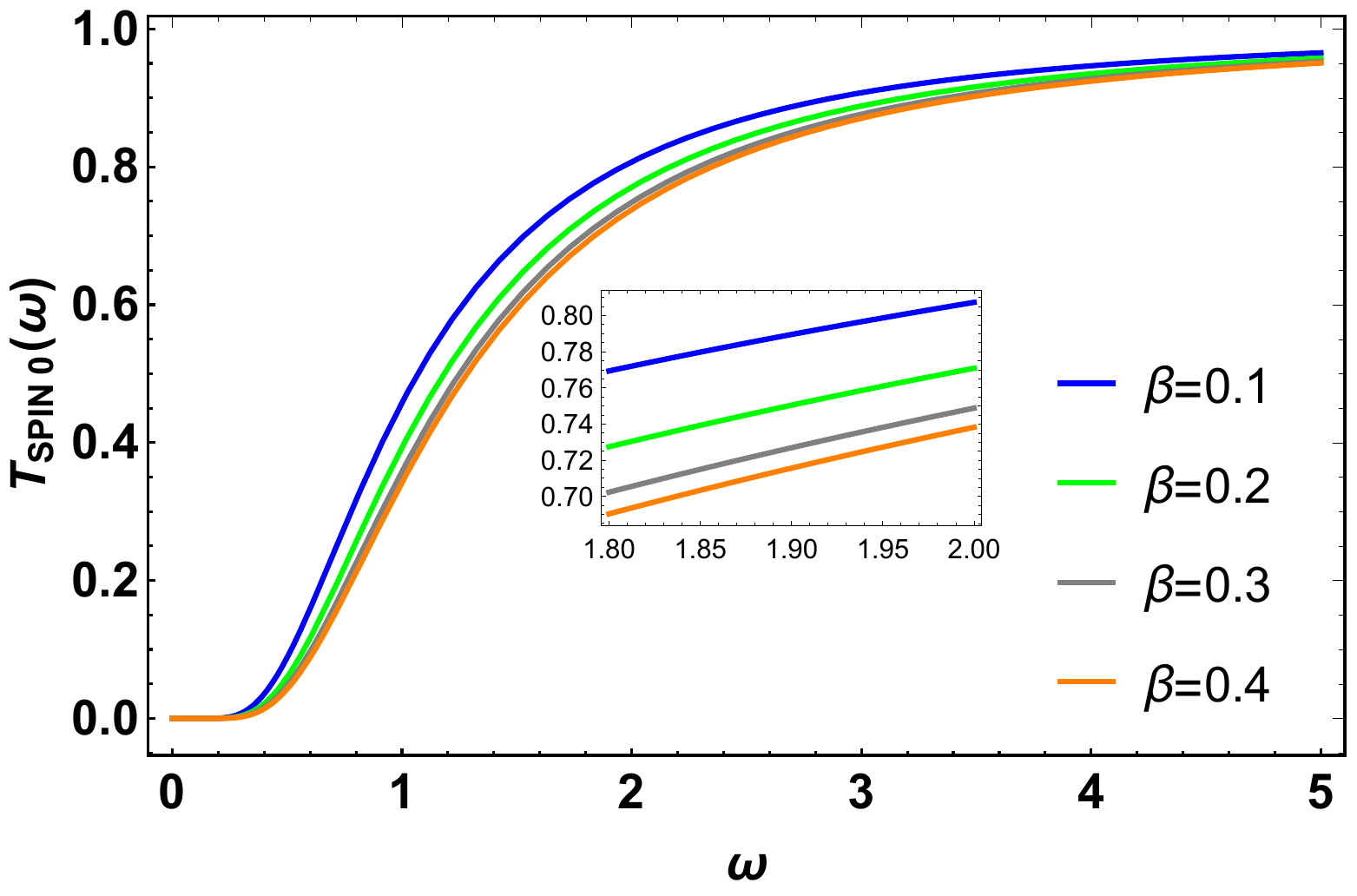}

    \vspace{0.15cm}
    \parbox{0.95\textwidth}{\centering\footnotesize
    (b) Variation of the PFDM parameter, with the remaining parameters fixed.}
\end{minipage}

\vspace{0.45cm}

\begin{minipage}[t]{0.48\textwidth}
    \centering
    \includegraphics[height=4.8cm]{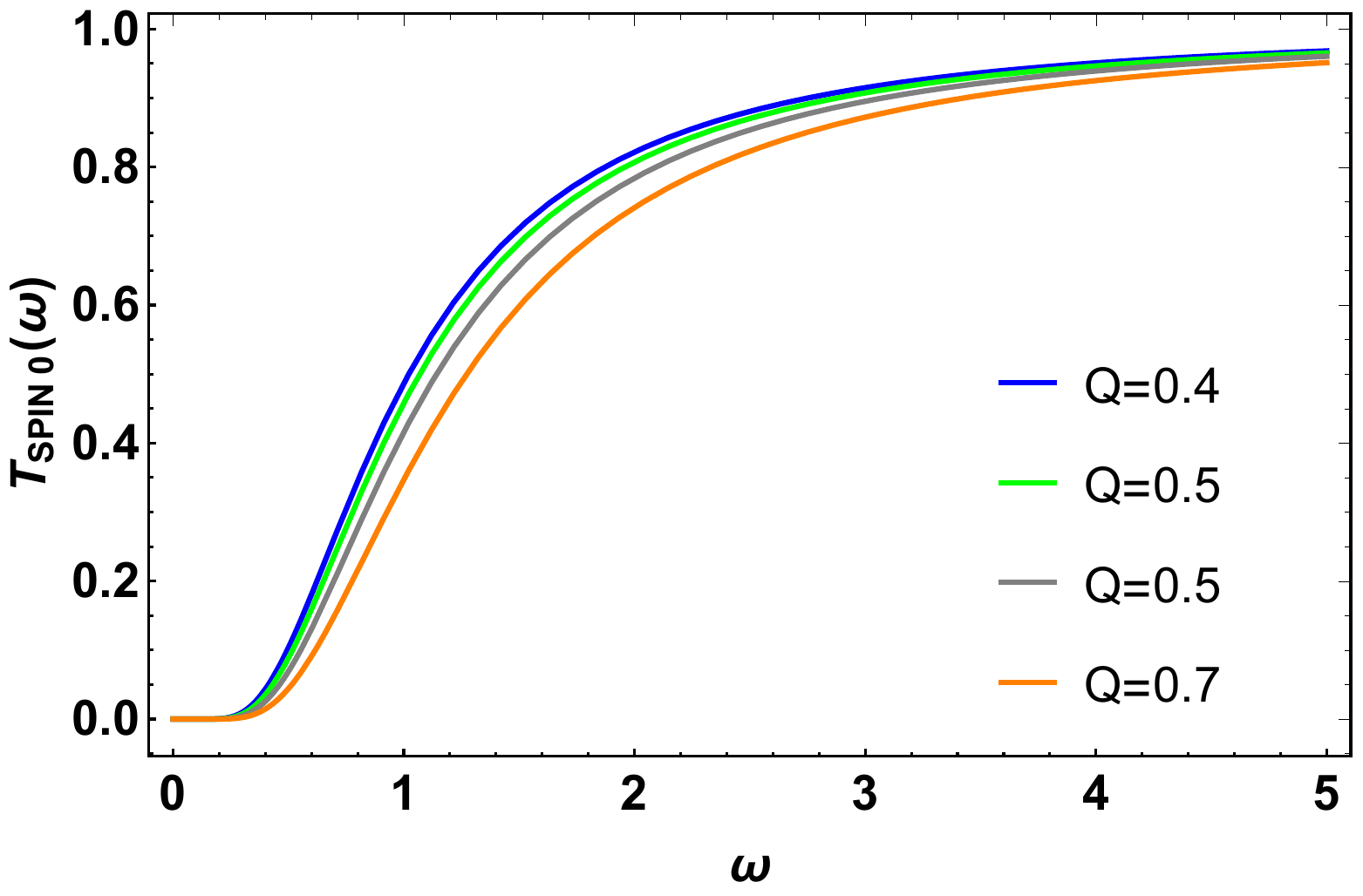}

    \vspace{0.15cm}
    \parbox{0.95\textwidth}{\centering\footnotesize
    (c) Variation of the electric charge, with the remaining parameters fixed.}
\end{minipage}
\hfill
\begin{minipage}[t]{0.48\textwidth}
    \centering
    \includegraphics[height=4.8cm]{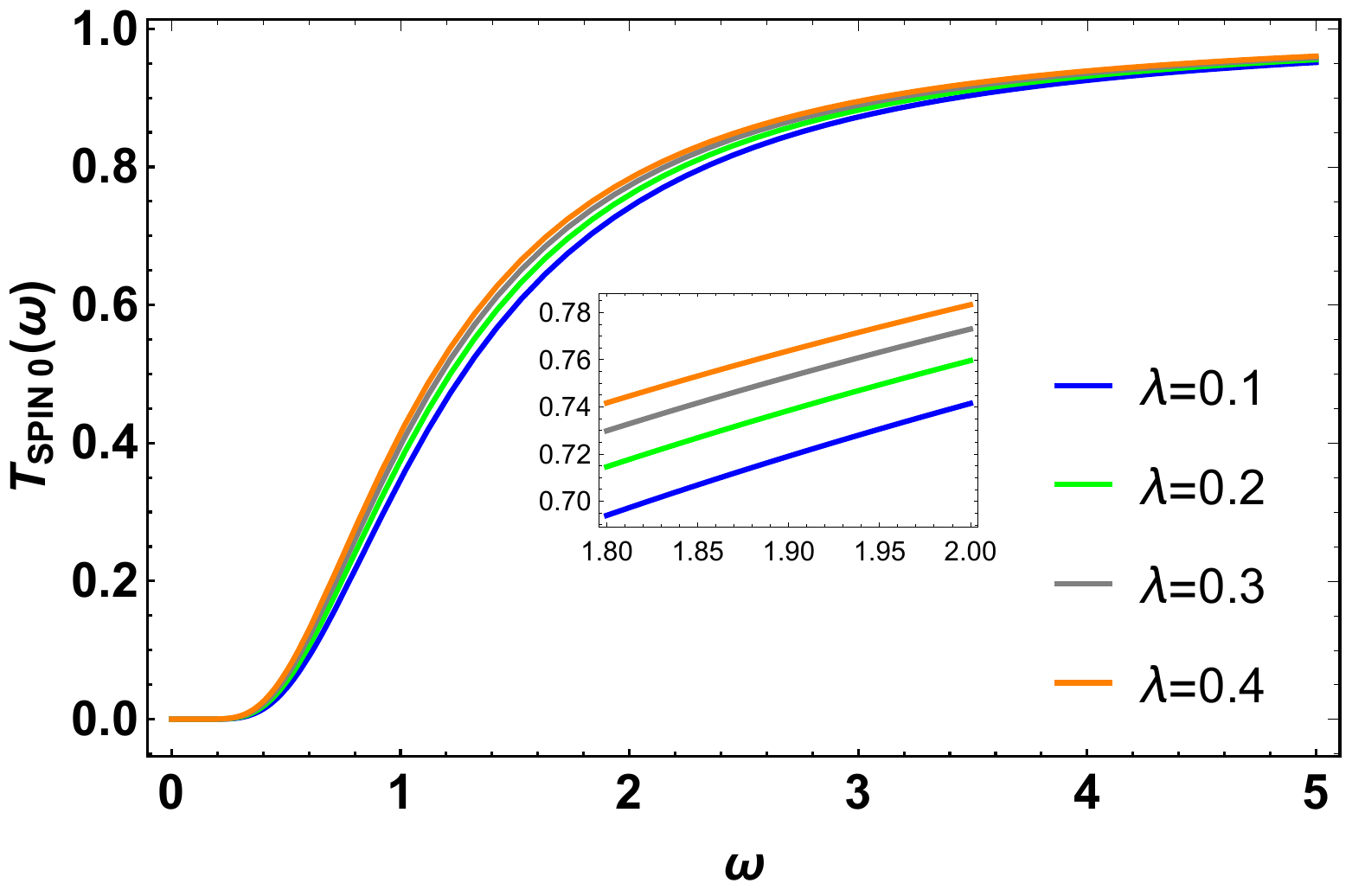}

    \vspace{0.15cm}
    \parbox{0.95\textwidth}{\centering\footnotesize
    (d) Variation of the ModMax parameter, with the remaining parameters fixed.}
\end{minipage}

\vspace{-0.2cm}

\caption{Behavior of the scalar transmission factor, namely the greybody lower bound $\mathcal{T}_{\ell}(\omega)$, under variations of the model parameters. Panel (a) displays the effect of the KR/LV parameter, panel (b) the effect of the PFDM parameter, panel (c) the effect of the electric charge, and panel (d) the effect of the ModMax parameter. In the graphical legends, $\gamma\equiv\alpha$ and $\lambda\equiv\lambda_{\rm MM}$.}
\label{T}
\end{figure*}

The greybody factor measures the probability that an outgoing Hawking quantum escapes from the near-horizon region to infinity, or equivalently that an incoming wave from infinity is transmitted through the effective potential barrier. This analysis employs the standard rigorous lower bound of the form
\begin{equation}
\label{eq:greybody_bound}
    T_\ell(\omega)
    \geq
    \operatorname{sech}^2
    \left[
        \frac{1}{2\omega}
        \int_{r_h}^{\infty}
        \left(
            \frac{\ell(\ell+1)}{r^2}
            +\frac{f'(r)}{r}
        \right)dr
    \right],
\end{equation}
where we used $dr_*=dr/f(r)$ and the potential in Eq.~\eqref{eq:V0_general}. For the present geometry, the integral can be evaluated analytically. Defining
\begin{equation}
\label{eq:I_def}
    \mathcal{I}_\ell(r_h)
    =
    \int_{r_h}^{\infty}
    \left(
        \frac{\ell(\ell+1)}{r^2}
        +\frac{f'(r)}{r}
    \right)dr ,
\end{equation}
one obtains
\begin{align}
\label{eq:I_result}
    \mathcal{I}_\ell(r_h)
   & =
    \frac{\ell(\ell+1)}{r_h}
    +\frac{M}{r_h^2}
    -\frac{2Q^2e^{-\lmm}}{3(1-\alpha)^2r_h^3}
    \notag\\&+\frac{\beta}{4(1-\alpha)r_h^2}
       \left[
          1-2\ln\!\left(\frac{r_h}{|\beta|}\right)
       \right].
\end{align}
Hence, the working expression for the bound is
\begin{equation}
\label{eq:T_final}
T_\ell(\omega)=\operatorname{sech}^2
\left[\frac{\mathcal{I}_\ell(r_h)}{2\omega}
\right],
\end{equation}
where the equality sign should be understood as the analytic expression used for the lower bound. With this convention, Eq.~\eqref{eq:T_final} explains the main features of the numerical curves. At low frequency, $\omega\ll \mathcal{I}_\ell$, the argument of the hyperbolic secant is large and the transmission is exponentially suppressed. At high frequency, $\omega\gg \mathcal{I}_\ell$, the argument tends to zero and $T_\ell(\omega)$ approaches unity. High-energy modes therefore cross the barrier almost freely. The parameter dependence follows from $\mathcal{I}_\ell$. Any parameter choice that increases the effective barrier also increases $\mathcal{I}_\ell$ and lowers the greybody factor at fixed $\omega$. Thus the plots show that increasing $\alpha$, $\beta$, or $Q$ suppresses transmission. On the other hand, increasing $\lmm$ weakens the effective charge contribution and decreases $\mathcal{I}_\ell$, so the greybody factor increases. This is the same physical ordering already visible in the effective potential. Figure~\ref{T} makes this behavior explicit: the scalar transmission factor approaches zero at low frequency, tends to unity at high frequency, and is shifted upward or downward according to whether the corresponding parameter lowers or raises the effective barrier.

\begin{figure*}[tbhp]
\centering

\begin{minipage}[t]{0.48\textwidth}
    \centering
    \includegraphics[height=4.8cm]{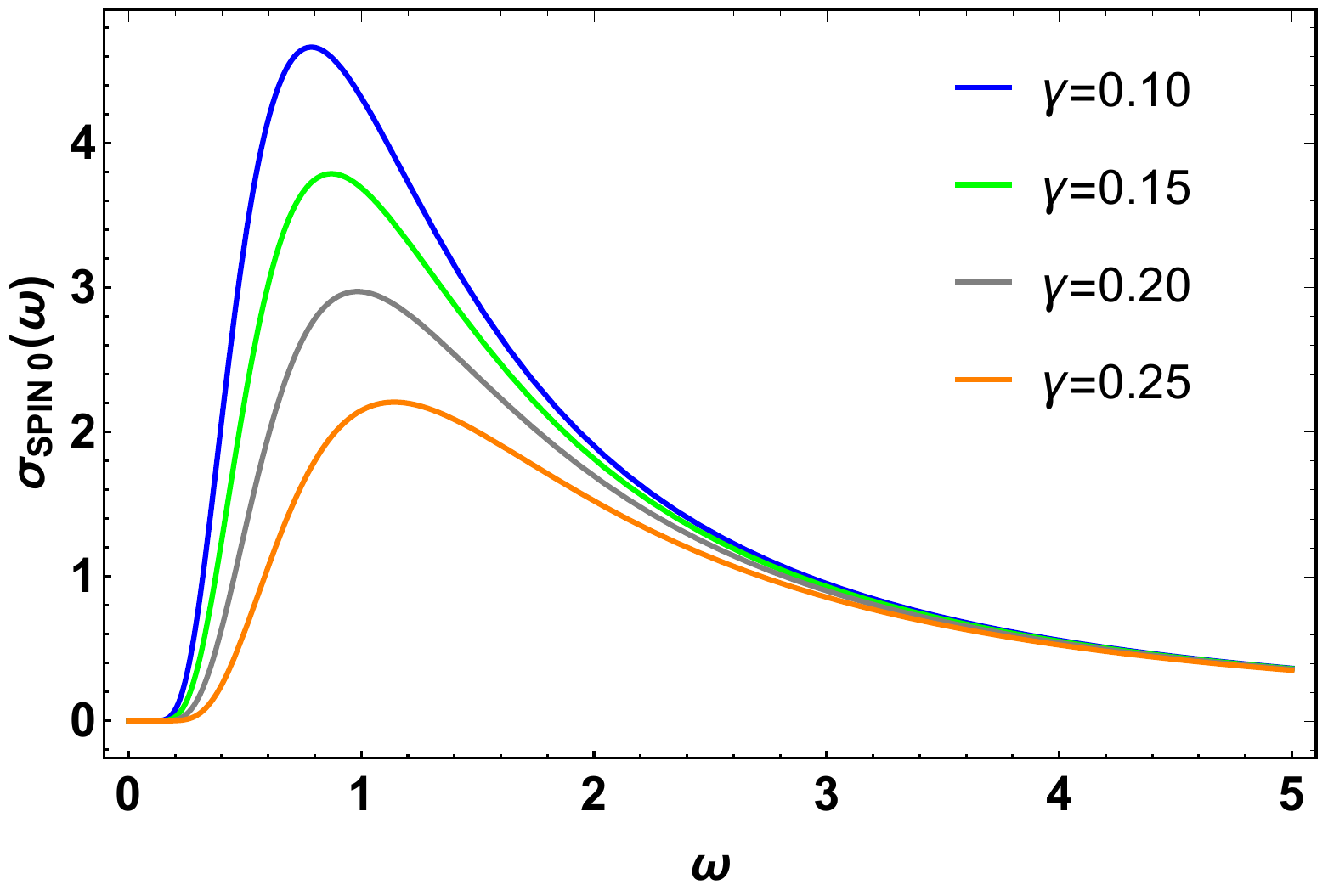}

    \vspace{0.15cm}
    \parbox{0.95\textwidth}{\centering\footnotesize
    (a) Variation of the KR/LV parameter, with the remaining parameters fixed.}
\end{minipage}
\hfill
\begin{minipage}[t]{0.48\textwidth}
    \centering
    \includegraphics[height=4.8cm]{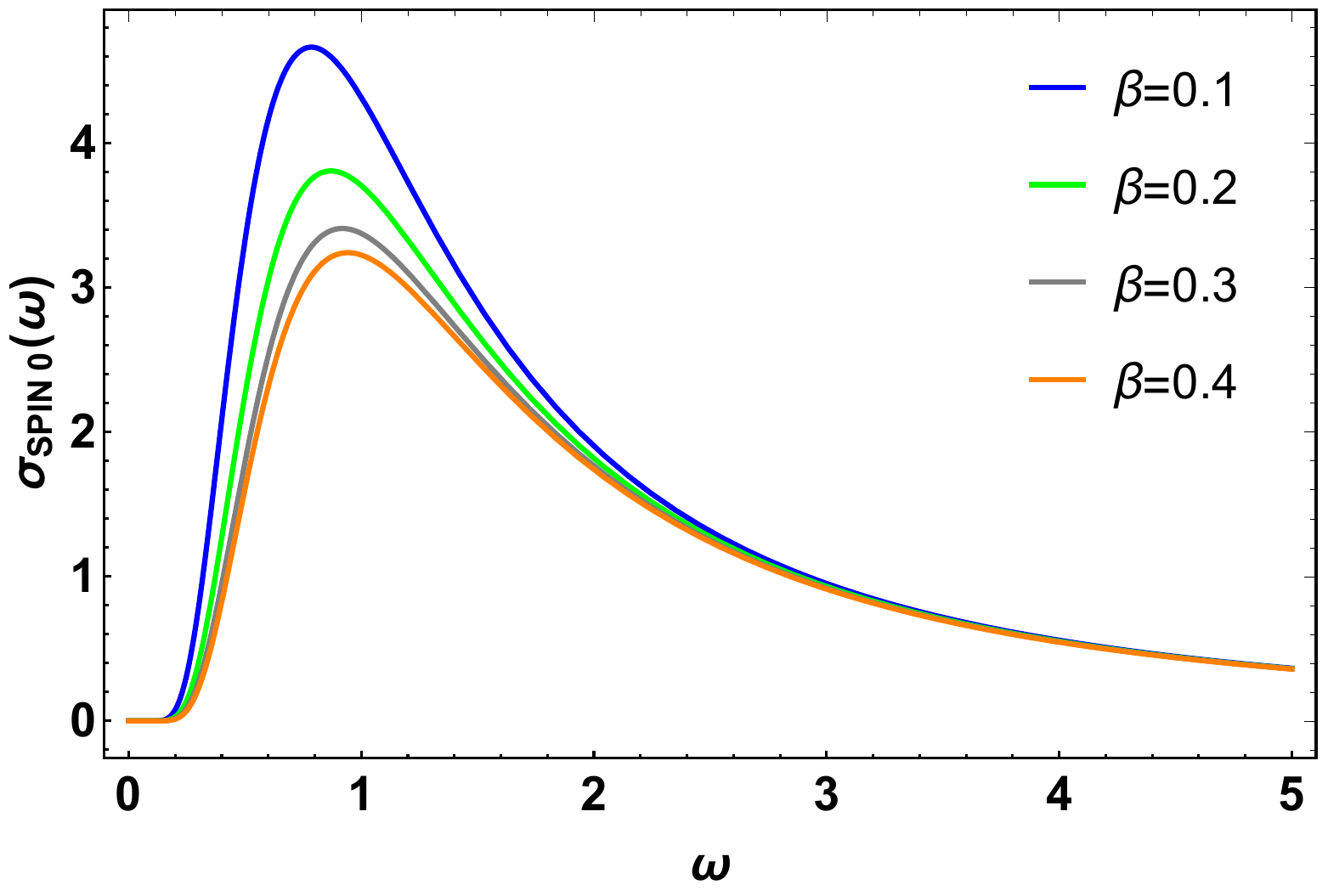}

    \vspace{0.15cm}
    \parbox{0.95\textwidth}{\centering\footnotesize
    (b) Variation of the PFDM parameter, with the remaining parameters fixed.}
\end{minipage}

\vspace{0.45cm}

\begin{minipage}[t]{0.48\textwidth}
    \centering
    \includegraphics[height=4.8cm]{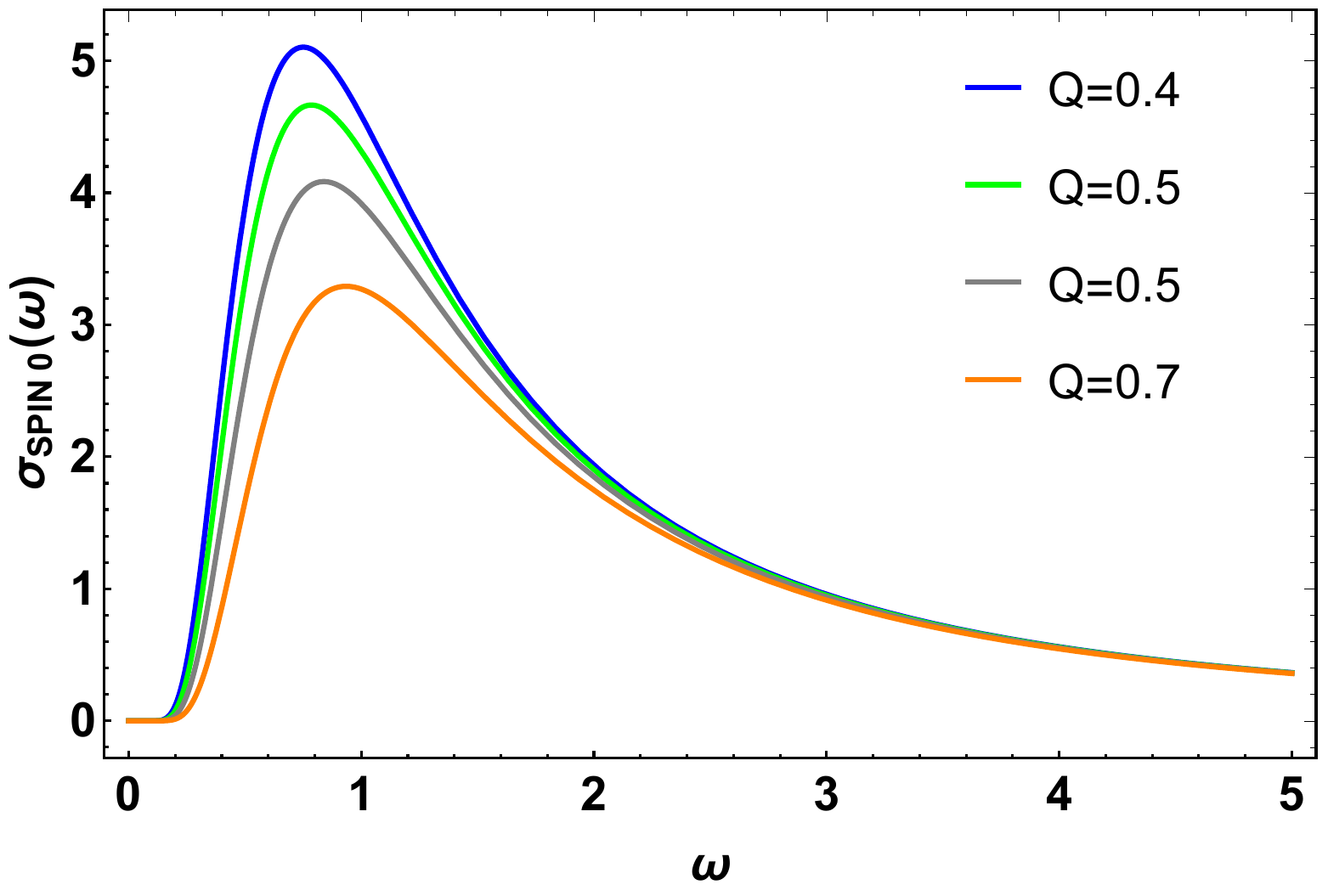}

    \vspace{0.15cm}
    \parbox{0.95\textwidth}{\centering\footnotesize
    (c) Variation of the electric charge, with the remaining parameters fixed.}
\end{minipage}
\hfill
\begin{minipage}[t]{0.48\textwidth}
    \centering
    \includegraphics[height=4.8cm]{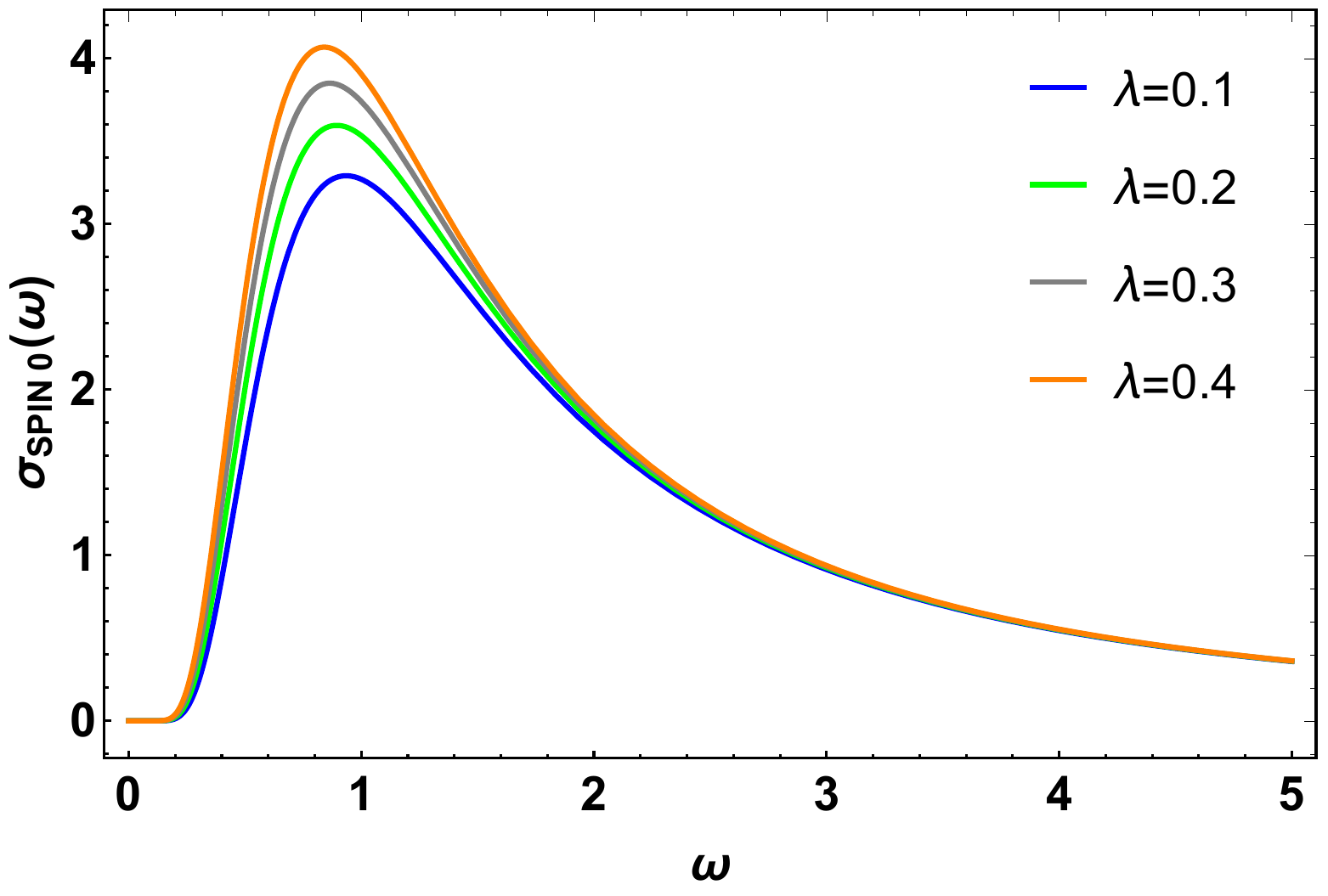}

    \vspace{0.15cm}
    \parbox{0.95\textwidth}{\centering\footnotesize
    (d) Variation of the ModMax parameter, with the remaining parameters fixed.}
\end{minipage}

\vspace{-0.2cm}

\caption{Behavior of the partial scalar absorption cross section $\sigma_{\ell}(\omega)$ under variations of the model parameters. Panel (a) displays the effect of the KR/LV parameter, panel (b) the effect of the PFDM parameter, panel (c) the effect of the electric charge, and panel (d) the effect of the ModMax parameter. In the graphical legends, $\gamma\equiv\alpha$ and $\lambda\equiv\lambda_{\rm MM}$.}
\label{ABS}
\end{figure*}

The partial absorption cross section is given by
\begin{equation}
\label{eq:sigma_abs}
    \sigma_\ell(\omega)
    =
    \frac{\pi(2\ell+1)}{\omega^2}\,
    T_\ell(\omega).
\end{equation}
For the plotted scalar mode $\ell=1$, this becomes
\begin{equation}
    \sigma_1(\omega)
    =
    \frac{3\pi}{\omega^2}
    \operatorname{sech}^2
    \left[
        \frac{\mathcal{I}_1(r_h)}{2\omega}
    \right].
\end{equation}
Note that this expression contains two competing frequency effects. The greybody factor grows with $\omega$, because the wave more easily crosses the barrier, while the geometrical prefactor $1/\omega^2$ decreases with $\omega$. As a result, the partial cross section is suppressed at very small frequency by the exponential behavior of $T_\ell$, rises in an intermediate region, and then falls again approximately as $1/\omega^2$ once $T_\ell\simeq1$. The same trend seen in the transmission coefficient appears in the absorption curves. Larger $\alpha$, $\beta$, and $Q$ lower the absorption probability at a fixed frequency because they make the scalar barrier more difficult to penetrate. Larger positive $\lmm$ increases the absorption probability because it screens the electric charge sector and reduces the barrier. In this sense, the ModMax parameter acts as a transparency parameter in the studied range, while the LV parameter, the dark-matter strength, and the electric charge act as opacity parameters. Figure~\ref{ABS} summarizes this behavior for the partial scalar absorption cross section: the curves vanish at very low frequency, peak at intermediate frequency, and decrease at high frequency due to the overall $1/\omega^2$ factor.

With this analysis, this section provides a consistent picture of scalar-wave propagation in the ModMax-Kalb-Ramond black hole with a dark-matter logarithmic correction. The geometry determines an effective barrier outside the horizon. The LV parameter $\alpha$, the dark-matter parameter $\beta$, and the charge $Q$ make this barrier stronger for the parameter intervals considered. They therefore reduce the scalar greybody factor and the partial absorption cross section at a fixed frequency. The ModMax parameter $\lmm$ has the opposite role, so that positive $\lmm$ weakens the effective charge through $\Qeff=Q^2e^{-\lmm}$, lowers the potential barrier, and enhances transmission.

Physically, this means that the Hawking spectrum observed at infinity is not determined only by the near-horizon temperature. It is filtered by the exterior geometry. A stronger exterior barrier removes a larger fraction of low- and intermediate-frequency scalar modes before they reach infinity. Conversely, when the ModMax sector screens the charge, the exterior becomes more transparent and the emitted scalar flux is less suppressed. The PFDM term is especially relevant because it modifies the geometry over a relatively broad radial region, rather than only very close to the horizon.

\section{Final remarks}\label{s7}

In this work, we investigated the background of a charged black hole with ModMax electrodynamics surrounded by perfect fluid dark matter within the Lorentz-violating framework of Kalb-Ramond. Such a geometry incorporates two corrections to the standard charged black hole scenario. The first one is associated with the nonlinear electrodynamic contribution of the ModMax sector, while the second one is due to the logarithmic correction induced by the surrounding perfect fluid dark matter distribution. This framework provides a useful setting in which to analyse how quantum-electrodynamic nonlinearities and dark matter effects modify the scattering properties of black holes.

Our analysis regarding the thermodynamic behavior provides a coherent physical interpretation. The electric charge gives rise to a cold extremal sector and tends to lower the Hawking temperature, as expected for charged black holes. The ModMax parameter $\lmm$ effectively screens the electric contribution through an exponential factor, so that larger values of $\lmm$ make the geometry behave as if the black hole carried a weaker effective charge. The LV parameter $\alpha$ amplifies the geometric deformation and enhances the thermodynamic response through the factors involving powers of $(1-\alpha)^{-1}$. Meanwhile, the PFDM parameter $\beta$ contributes a logarithmic correction, whose influence is most pronounced at finite and intermediate horizon radii rather than in the asymptotically large-radius regime. All these effects generate a rich phase structure characterized by extremal configurations, possible divergences of the heat capacity, and thermodynamically favored intermediate states associated with the minima of the Helmholtz free energy.

Upon analyzing the greybody factor and absorption cross section in this model, we observe that the parameters $\alpha$, $\beta$, and $Q$ tend to make the black hole optically darker to scalar perturbations, while positive $\lmm$ tends to make it optically brighter. This conclusion follows simultaneously from the metric function, the effective potential, the analytic greybody bound, and the absorption cross section computed above.\\

\section*{Acknowledgments}

\hspace{0.5cm} The author Fernando Belchior would like to express gratitude to the Conselho Nacional de Desenvolvimento Cient\'{i}fico e Tecnol\'{o}gico CNPq for grant No. 151845/2025-5. The author Edilberto Silva acknowledges the support from Conselho Nacional de Desenvolvimento Cient\'{i}fico e Tecnol\'{o}gico (CNPq) (grants 306308/2022-3), Funda\c{c}\~{a}o de Amparo \`{a} Pesquisa e ao Desenvolvimento Cient\'{i}fico e Tecnol\'{o}gico do Maranh\~{a}o (FAPEMA) (grants UNIVERSAL-06395/22), and Coordena\c{c}\~{a}o de Aperfei\c{c}oamento de Pessoal de N\'{i}vel Superior (CAPES) - Brazil (Code 001). The author Faizuddin Ahmed acknowledges the Inter University Centre for Astronomy and Astrophysics (IUCAA), Pune, India for granting visiting associateship.

\bibliographystyle{apsrev4-2}
\bibliography{references}

\end{document}